\documentclass[useAMS, usenatbib]{mnras}
\usepackage[final]{graphicx}
\usepackage{amssymb}
\usepackage{amsmath}
\usepackage{breqn}
\usepackage{color}
\usepackage{blindtext}
\usepackage[usenames,dvipsnames]{xcolor}
\usepackage{enumitem}

\usepackage [english]{babel}
\usepackage [autostyle, english = american]{csquotes}
\MakeOuterQuote{"}

\usepackage{tablefootnote}
\usepackage[flushleft]{threeparttable} 
\usepackage{subfig}

\usepackage{makecell}

\usepackage{dblfloatfix}    

\usepackage{titlesec}
 
\titleformat{\paragraph}
{\normalfont\normalsize\it}{\theparagraph}{1em}{}
\titlespacing*{\paragraph}
{0pt}{3.25ex plus 1ex minus .2ex}{1.5ex plus .2ex}

\title[The possible hierarchical scales of observed clumps in high-redshift disc galaxies]{The possible hierarchical scales of observed clumps \\ in high-redshift disc galaxies}
\author[M. Behrendt et al.]{M. ~Behrendt$^{1,2}$\thanks{E-mail: mabe@mpe.mpg.de}, M. ~Schartmann$^{1,2,3}$, A. ~Burkert$^{1,2}$ \\
$^1$Max-Planck Institute for Extraterrestrial Physics, Giessenbachstra${\ss}$e, D-85741 Garching, Germany\\
$^2$University Observatory,  Ludwig-Maximilians-Universit\"at M\"unchen, Scheinerstra\ss e 1, D-81679 M\"unchen, Germany\\
$^3$Excellence Cluster ORIGINS, Ludwig-Maximilians-Universit\"at M\"unchen, Boltzmannstra${\ss}$e 2, D-85748 Garching, Germany}

\begin{document}

\date{\today}

\pagerange{\pageref{firstpage}--\pageref{lastpage}} \pubyear{2018}

\maketitle

\label{firstpage}

\begin{abstract}
Giant clumps on $\sim$ kpc scales and with masses of $10^8-10^9 \ \mathrm{M_{\odot}}$ are ubiquitous in observed high-redshift disc galaxies. Recent simulations and observations with high spatial resolution indicate the existence of substructure within these clumps. We perform high-resolution simulations of a massive galaxy to study the substructure formation within the framework of gravitational disc instability. We focus on an isolated and pure gas disk with an isothermal equation of state with $T=10^4 \ K$ that allows capturing the effects of self-gravity and hydrodynamics robustly. The main mass of the galaxy resides in rotationally supported clumps which grow by merging to a maximum clump mass of $10^8 \ \mathrm{M_{\odot}}$ with diameter $\sim 120$ pc for the dense gas. They group to clump clusters (CCs) within relatively short times ($\ll 50 \ \mathrm{Myr}$), which are present over the whole simulation time. We identify several mass and size scales on which the clusters appear as single objects at the corresponding observational resolution between $\sim 10^8 - 10^9 \ \mathrm{M_{\odot}}$. Most of the clusters emerge as dense groups and for larger beams they are more likely to be open structures represented by a single object. In the high resolution runs higher densities can be reached, and the initial structures can collapse further and fragment to many clumps smaller than the initial Toomre length. In our low resolution runs, the clumps directly form on larger scales $0.3-1$ kpc with $10^8-10^9 \ \mathrm{M_{\odot}}$. Here, the artificial pressure floor which is typically used to prevent spurious fragmentation strongly influences the initial formation of clumps and their properties at very low densities.

\end{abstract}

\begin{keywords}
hydrodynamics -- instabilities -- methods: numerical -- galaxies: evolution -- galaxies: high-redshift -- galaxies: structure.
\end{keywords}

 \begin{table*}
 \setlength{\tabcolsep}{12pt}
 \caption{The main differences of the simulations.}
 \label{tab:Main differences of the simulations}
 \begin{threeparttable}
 \begin{tabular}{llccccccc}
  \hline
  
  Simulation	& Description				& $\Delta x_{\mathrm{min}}$\tnote{a} &	$N_{\mathrm{J}}$\tnote{b}	&	$N_{\mathrm{P}}$\tnote{c} 	& $n_{0}$\tnote{d} 	& $L_{\mathrm{MinJeans}}$\tnote{e}  & $N_{\mathrm{Toomre}}$\tnote{f}\\

   			 	&		      				&  $\mathrm{[pc]}$		&		[cells]			&		[cells]		&$\mathrm{[cm^{-3}]}$  	&	 $\mathrm{[pc]}$	 & [cells]	\\
  \hline
    \textcolor{red}{$\textbf{MS}$}   & Main simulation					& 2.9					&	19					&	7				& $4.5 \times 10^3$ 	&  20.5		& 25-55\\  
    \textcolor{blue}{$\textbf{SR}$}	 & Lower resolution			& 5.9					&	4					&	4				&	$3.4 \times 10^3$ &  23.4		& 4-10\\ 
    \textcolor{Plum}{$\textbf{LR}$}	 & Low resolution			& 46.9					&	4					&	4				&	$53$  &  187.5	 & 4-10\\ 
    \textcolor{PineGreen}{$\textbf{ULR}$}	 & Ultra low resolution			& 93.8					&	4					&	4			&	$13$ 	&  375	& 2-10\\	 				
  \hline
 \end{tabular}
  \begin{tablenotes}
  \item[a] The cell size at maximum resolution.
  \item[b] The number of cells that resolve the Jeans length at all refinement levels except for the maximum resolution.
  \item[c] The number of cells that represent the Jeans length at maximum resolution ensured by the artificial pressure floor.
  \item[d] The critical density where the artificial pressure floor sets in.
  \item[e] The smallest Jeans length that can be resolved. 
  \item[e] The number of cells that represent the initial Toomre length $L_{\mathrm{Toomre}}$.

  \end{tablenotes}
  \end{threeparttable}
\end{table*}

\section{Introduction}
\label{sec:Introduction}
The challenging observations of the distant universe give us the outstanding opportunity to study the galaxy evolution at very early times. They reveal a diversity of galactic systems with physical properties that are different than in the local universe. At redshift $z\sim 2$, gas-rich \citep{2010ApJ...713..686D, 2013ApJ...768...74T} galaxies with an irregular morphology \citep{2005ApJ...631...85E} and high random motions with an underlying coherent rotating disc \citep{2009ApJ...706.1364F, 2012MNRAS.422.3339W} have been found. The star formation rates are a factor of 10 to 100 times larger than in present-day Milky Way type galaxies \citep{2005ApJ...627..632E, 2011ApJ...733..101G}. A key feature are $\sim$ kpc-sized, massive clumps of several $\sim 10^8 -10^{9} \ \mathrm{M_{\odot}}$ with high star formation rate densities containing relatively young stars of 10 Myrs to several 100 Myrs \citep{2011ApJ...739...45F, 2013ApJ...779..135W, 2015Natur.521...54Z, 2018ApJ...853..108G}.\\
The origin of the giant clumps is well explained in the framework of the gravitational disc instability. The observed high gas densities and high random motions can lead to kpc-sized growing perturbations and the discs fragment into massive star-forming clumps if the Toomre Q parameter is below a critical value $Q < Q_{\mathrm{crit}}$ \citep{2008ApJ...687...59G, 2009ApJ...703..785D}. With this approach, theoretical models have been very successful in explaining or simulating many observed properties. In general, cosmological zoom-in simulations find typically a handful of giant clumps which directly form on the mass and size scales which are observed at kpc spatial resolution \citep{2009ApJ...703..785D, Ceverino:2010eh, 2012MNRAS.420.3490C}. Here, the advantage is that the cosmological context can be taken into account, e.g. infalling gas from the cosmic web. A complementary approach is to simulate a galaxy in an isolated box. The advantages are the significant higher resolution that can be reached (initially and finally) and the well-defined conditions to reduce the complexity. At higher resolution a rich substructure is found within giant clumps \citep{2012MNRAS.420.3490C, 2014ApJ...780...57B, Bournaud2016, 2016ApJ...819L...2B, 2017MNRAS.465..952O}. Overall, two possible scenarios may lead to the formation of giant clump sub-structure. In the so-called "top-down" scenario, giant gravitationally bound clumps may form directly and sub-fragment due to growing perturbations on the sub-clump scale enabled by higher resolution simulations. This picture may be supported by cosmological simulations, but at present has not been showcased in any study. In \cite{2017MNRAS.465..952O} they find sub-fragmentation caused by stellar feedback which leads to very short lived ($\sim 20$ Myr) giant clumps. In the second scenario, many gravitationally bound clumps form on smaller scales than the estimated initial growing Toomre length \citep{2018arXiv180802438B}. They later on conglomerate to clusters representing giant clumps with several smaller clumps ("sub-structure") \citep{2015MNRAS.448.1007B, 2016ApJ...819L...2B, 2017MNRAS.468.4792T}. These clusters would appear homogeneous if observed at a typical resolution of 1.6 kpc at $z \sim 2$. These clump clusters have been shown to have similar kinematic properties \citep{2016ApJ...819L...2B} as observed for the giant star forming clumps in \citet{2011ApJ...733..101G}. Which of the two massive clump formation scenarios is reproduced in nature is still unclear. The effect on clump properties due to different observational resolution is evident in the recent study of a strong gravitationally lensed massive galaxy ($\mathrm{M_{\star}} \approx 4 \times 10^{10} \ \mathrm{M_{\odot}}$) "the cosmic snake" \citep{2018NatAs...2...76C} and the comparison to its counter image. With the $\mathrm{SFR \thicksim 30 \ M_{\odot} \ yr^{-1}}$ it is comparable to galaxies on the main sequence at $z \sim 1-2$. Furthermore, the identified clump masses and sizes tend to correlate with the different magnifications of the images, giving  $10^7 \ \mathrm{M_{\odot}}$ at high magnification ($\sim 30$ pc resolution) and several $10^8 \ \mathrm{M_{\odot}}$ at low magnification ($\sim 300$ pc resolution). Currently, for observations at $z \sim 2$ the spatial resolution is limited to $\sim 1-2$ kpc. One hint for a possible substructure on sub-kpc scales of a bright clump has been found in the seminal work of \citet{2011ApJ...733..101G}, by studying the clump gas kinematics of H$\alpha$ and comparing the individual velocity channel maps. Those galaxies have very high gas fractions and much higher $\mathrm{SFR \sim 120-290 \ M_{\odot} \ yr^{-1}}$ corresponding to the upper range of mass and bolometric luminosity of the $z \sim 2$ star forming galaxies on the main sequence \citep{2009ApJ...706.1364F}. Another possibility to study clump clustering are the extremely rare local clumpy galaxies whose properties are similar to those of high-redshift discs (Fisher et al. 2017), found in a sub-sample of the DYNAMO-HST survey with massive galaxies of $\mathrm{M_{\star}}  \sim 1-5 \times 10^{10} \mathrm{M_{\odot}}$, which correspond to galaxies that dominated the cosmic SFR at $z \sim 1-3$.  In their simulated high-redshift maps (taking blurring, surface brightness dimming, and a sensitivity cut similar to high-z observations with adaptive optics into account) the number of clumps is significantly reduced to 1-4 clumps per galaxy comparable to the amount of clumps detected in high-z surveys \citep{2011ApJ...733..101G, 2012ApJ...760..130S, 2012MNRAS.422.3339W}. The limited sensitivity may also restrict the observations to more massive clumps within a galaxy as was shown in \cite{2017MNRAS.468.4792T} by H$\alpha$ mocks created by radiative transfer postprocessing of a clumpy galaxy simulation. \\
Besides the resolution effects on clump properties, an important numerical issue is the so-called artificial pressure floor (APF) in simulations. Typically, the Jeans length is resolved by several elements to avoid spurious fragmentation of a collapsing clump \citep{1997ApJ...489L.179T}. At the highest refinement level, this can be satisfied by using an artificially induced pressure which is commonly used in clumpy galaxy simulations \citep{2008ApJ...680.1083R, Dekel:2009bn, 2009MNRAS.392..294A, 2010ApJ...720L.149T, 2012MNRAS.420.3490C, 2014ApJ...780...57B, 2016ApJ...833..202K, 2017arXiv170206148H}. The effect on the number of clumps by varying the APF at the same resolution has been studied in \citet{Ceverino:2010eh}. They suggested to investigate its influence on clump properties also for simulations better than 10 pc maximum resolution. \\
\\
 To understand the two fragmentation scenarios from top-down and bottom-up and the clump clustering better we complement the previous studies in this article. Section \ref{sec:Simulation Methods} gives an overview of the used simulation methods and the different runs. Section \ref{sec:Results}
(i) analyses the fragmentation of a Toomre unstable gas disc and gives the detailed clump statistic over time by comparing between simulations of $<100$ pc and $<10$ pc maximum resolution. 
(ii) Furthermore, we investigate the effect of the artificial pressure floor on the clump properties. In the second part of the paper in Section \ref{sec:The effects of beam smearing on the main clump properties} we investigate 
(iii) on which scales the clumps appear as clusters  
(iv) and how their properties are related to low and high beam smearing. In Section \ref{sec:Summary an Discussion} we give the summary and discussion of the findings.

\begin{figure}
\centering
 
  {\includegraphics[width=0.99\linewidth]{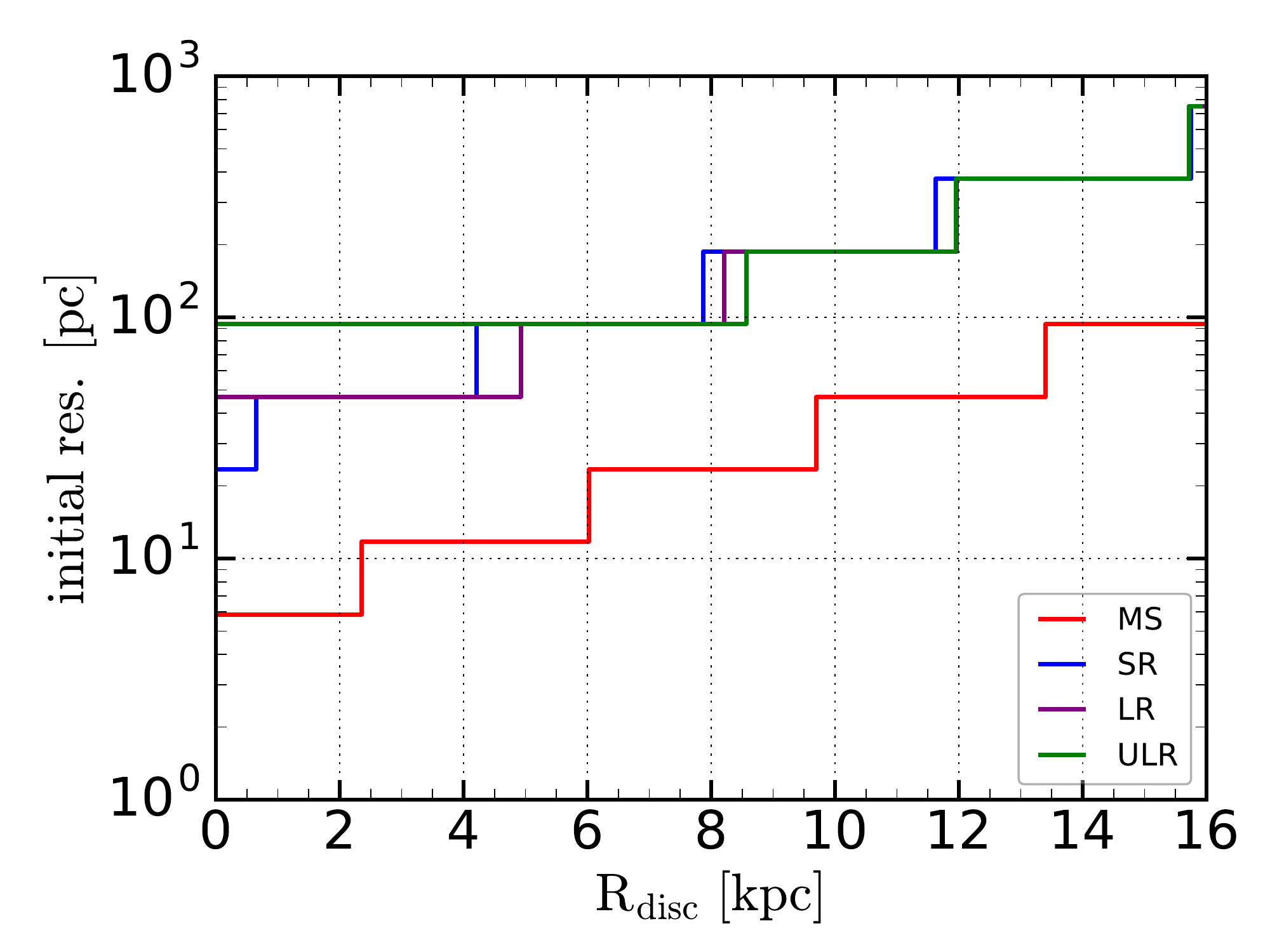}}

\caption{Radial dependence of the initial resolution in the mid-plane of the axisymmetric disc for the different runs.   \label{fig:amr_levels_indisk}}

\end{figure} 

\section{Simulation Methods}
\label{sec:Simulation Methods}
\subsection{Disc model}
\label{subsec:Gas disc setup}
For the disc model, we use the same setup as in \citet{2015MNRAS.448.1007B} which is initially in hydrostatic equilibrium in vertical direction and in centrifugal equilibrium in radial direction. The main parameters were chosen to resemble a massive high-redshift disc galaxy \citep{2011ApJ...733..101G}. The gas disc has an exponential surface density profile with a total mass of $M_{\mathrm{disc}} = 2.7 \times 10^{10} \ \mathrm{M_{\mathrm{\odot}}}$ with a maximum radius of $R_{\mathrm{disc}} = 16 \ \mathrm{kpc}$ and a scale-length $h = 5.26 \ \mathrm{kpc}$ and is Toomre unstable within the inner $10.5 \ \mathrm{kpc}$. The relatively large scale-length ensures the unstable regime and the relatively large disc radius gives a smoother transition of the low density part to the environment. The dark matter halo follows the \citet{Burkert:1995jr} density profile with a mass of $M_{\mathrm{DM}} = 1.03 \times 10^{11} \ \mathrm{M_{\odot}}$ within the disc radius and a scale length of $r_{\mathrm{s}}=4$ kpc. Efficient cooling is needed to form gravitationally bound clumps, ensured by the isothermal equation of state (EoS), which keeps the temperature at $10^4$ K (mean molecular weight $\mu = 1$) for the initial setup and for higher densities that are not affected by the artificial pressure floor (see Section \ref{subsubsec:The low-resolution simulations}). Here, it represents the typical micro-turbulent pressure floor of the interstellar medium (ISM) which cannot be resolved in our simulations and keeps the initial conditions of the vertical density distribution stable until the growing perturbations begin to transform the disc locally. For the low resolution simulations the temperature is deviating from $10^4$K in the inner parts due to the artificial pressure floor, see Section \ref{subsubsec:The low-resolution simulations}. The model does not include a stellar disc or a bulge component (see Section \ref{sec:Summary an Discussion}).

\begin{figure}
\centering
\includegraphics[width=86mm]{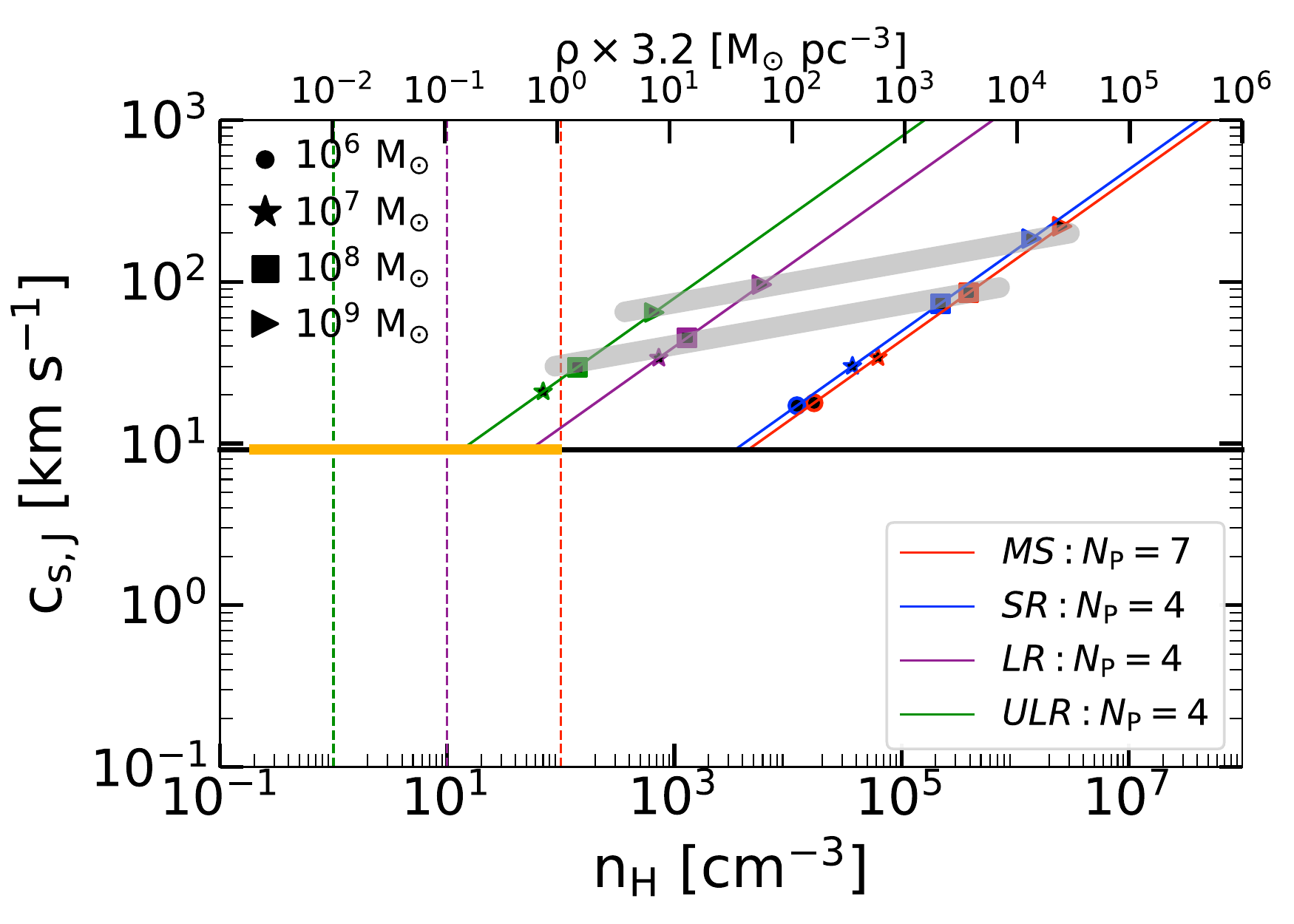}
\caption{Density dependence of the induced artificial pressure floor (Section \ref{sec:Simulation Methods}) expressed as the sound speed $\mathrm{c_{s,J}}$ at the maximum resolution for the different simulations (Table \ref{tab:Main differences of the simulations}). The minimum sound speed in the simulations corresponds to $\sim 10 \ \mathrm{km \ s^{-1}}$ and applies for the initial mid-plane densities  between $0.2 - 93 \ \mathrm{cm^{-3}}$ (orange line). The vertical dashed lines correspond to the density thresholds used by the clump finder in the different simulations ($n_{\mathrm{H}} = 1, 10, 100 \ \mathrm{cm^{-3}}$ ). The different symbols correspond to the typical central densities of the selected clump-masses, defined in Section \ref{subsec:Detailed clump properties}. We measure the maximum density in each mass bin and average over the simulation time. The grey shaded lines illustrate the relation between the clump densities per mass bin for $10^8 \ \mathrm{M_{\odot}}$ and $10^9 \ \mathrm{M_{\odot}}$, the resolution and accordingly the artificial pressure floor.
   \label{fig:pressure_floor}} 
\end{figure}
  
   \begin{figure*}
\centering

  {\includegraphics[width=0.24\linewidth]{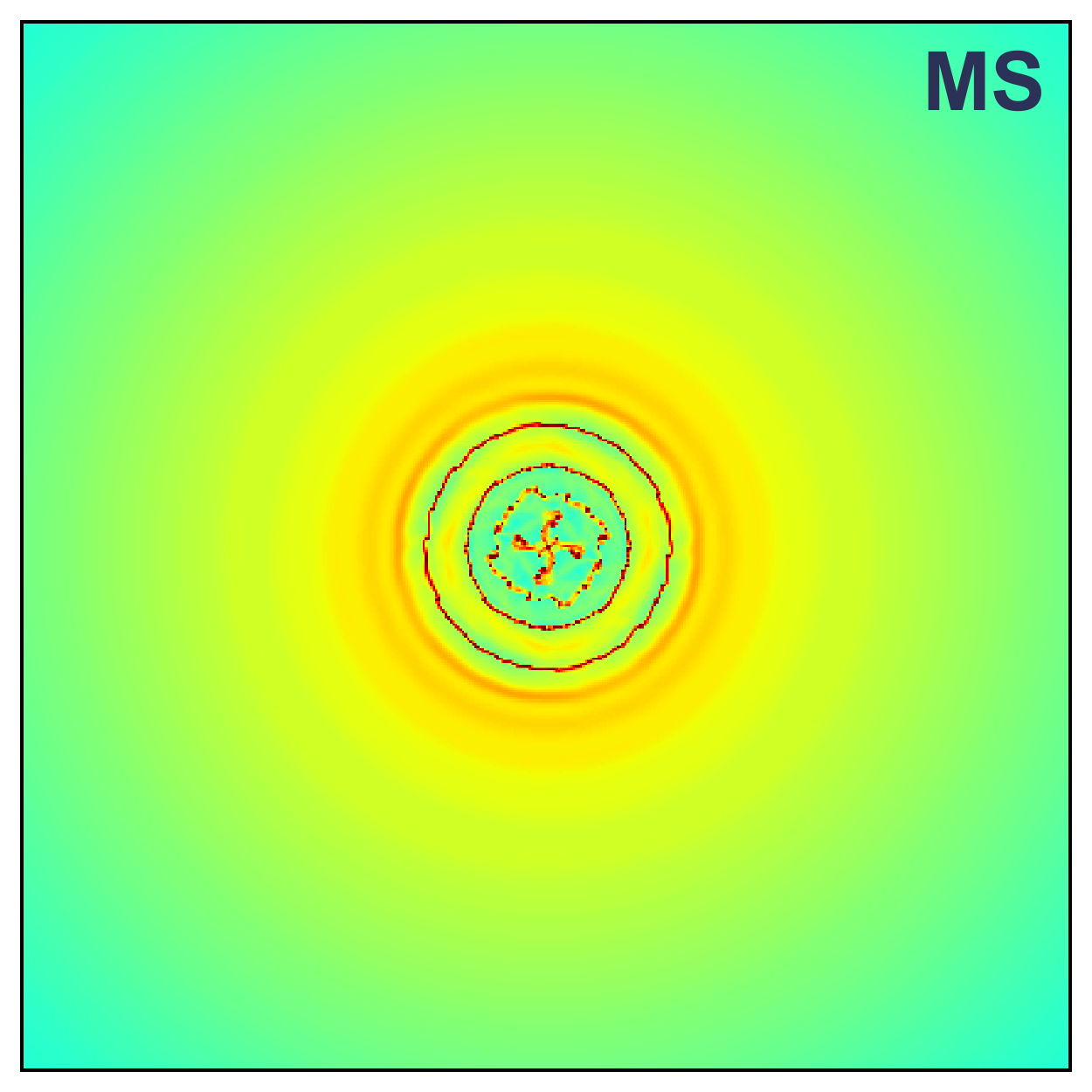}}
  {\includegraphics[width=0.24\linewidth]{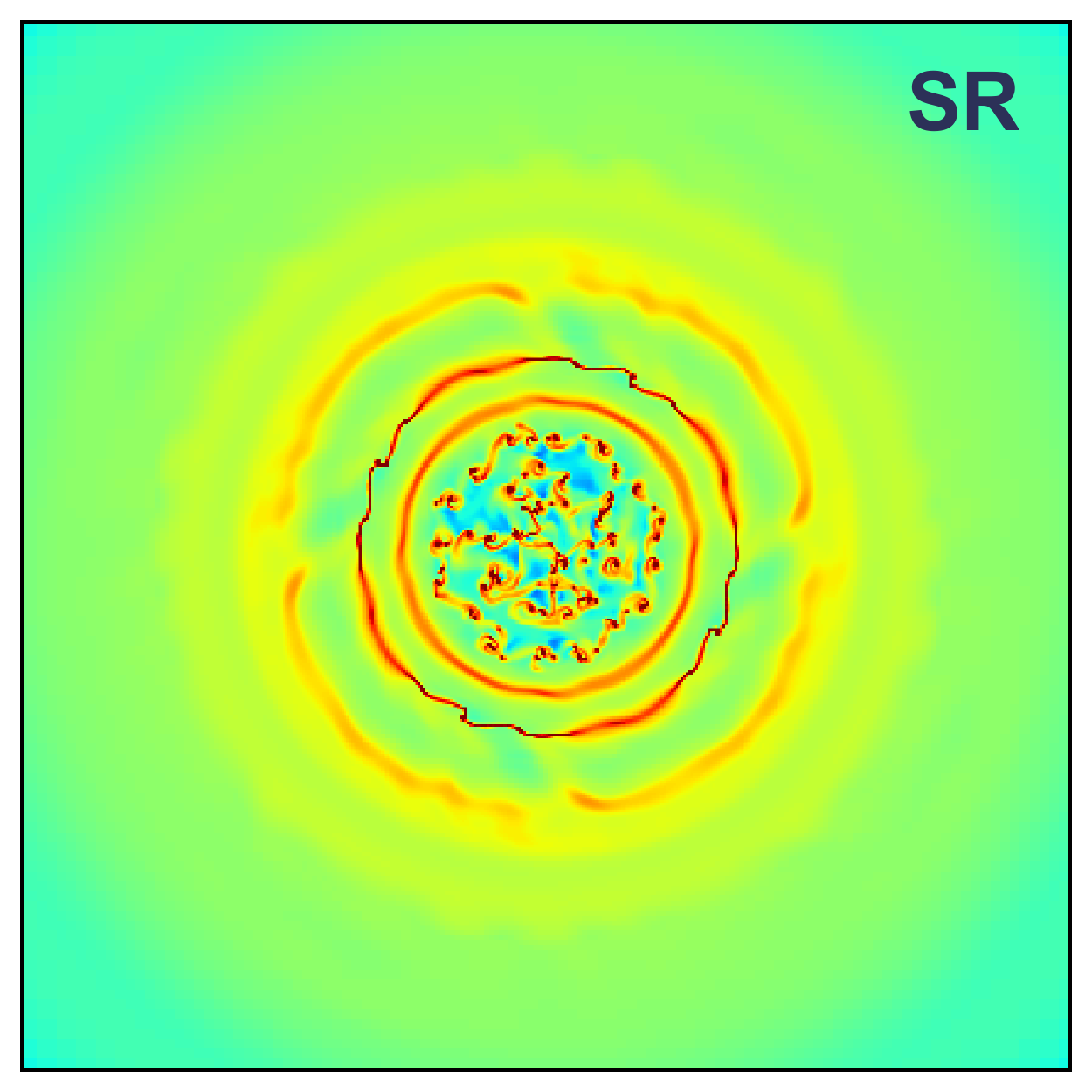}}
  {\includegraphics[width=0.24\linewidth]{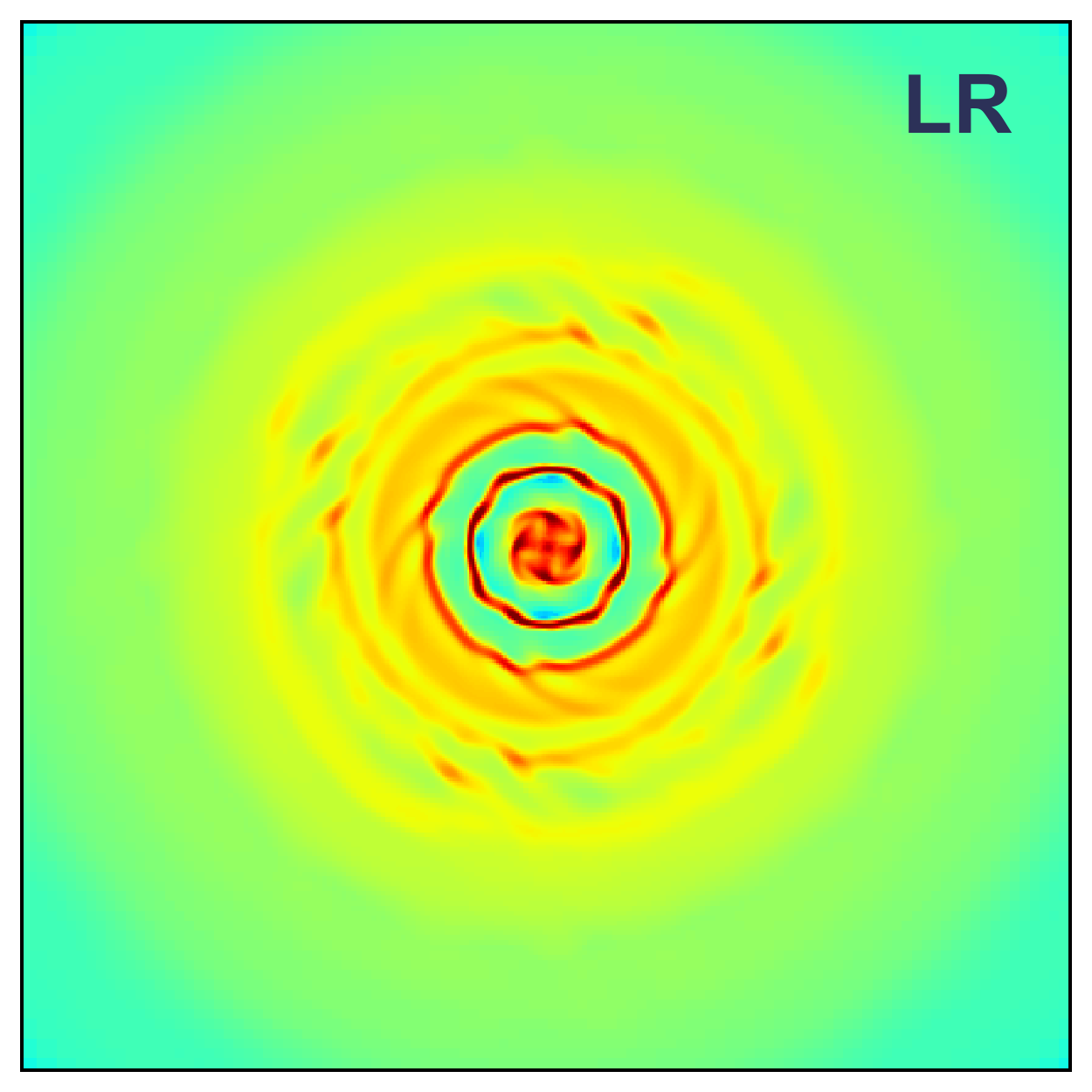}}
  {\includegraphics[width=0.24\linewidth]{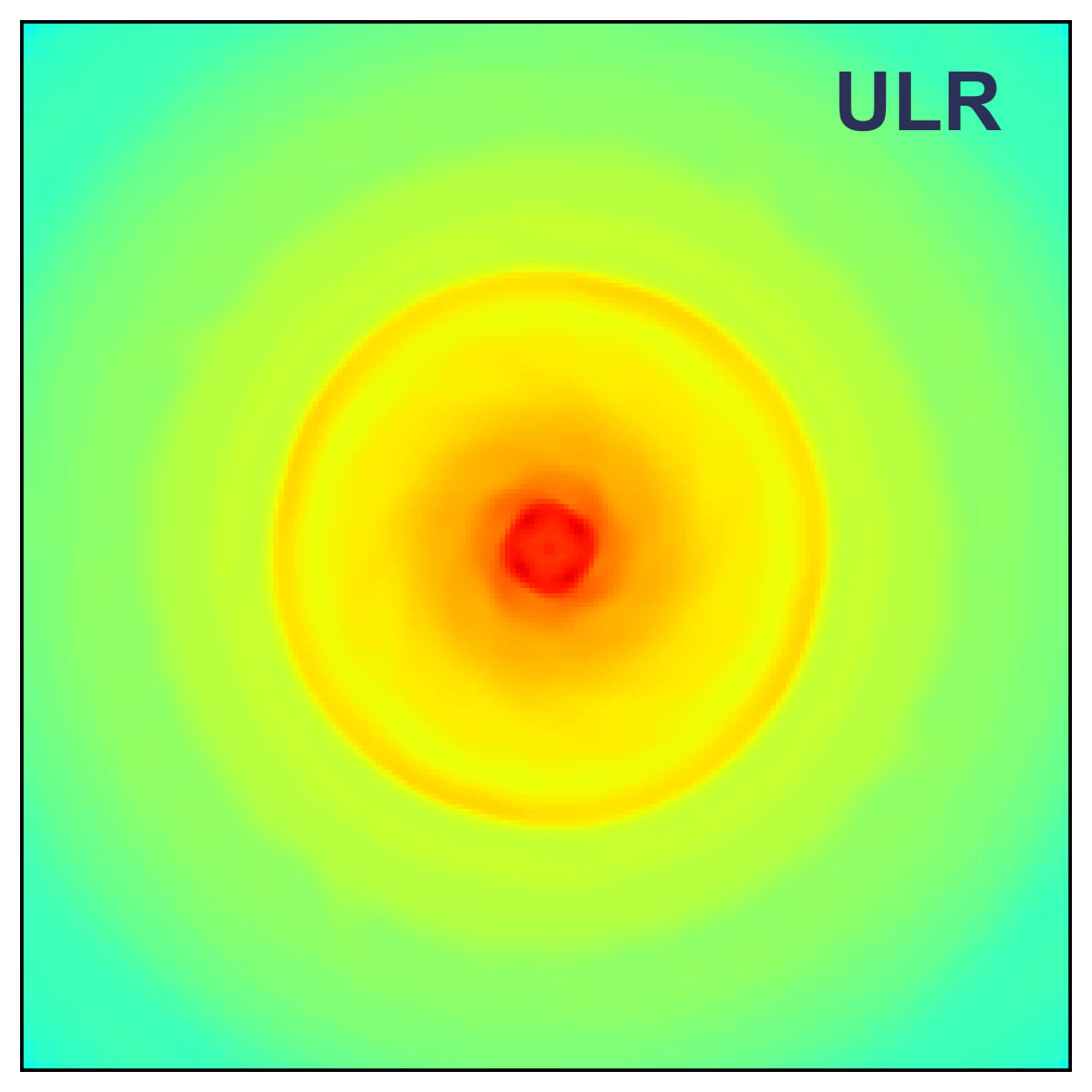}}\hfill

  {\includegraphics[width=0.24\linewidth]{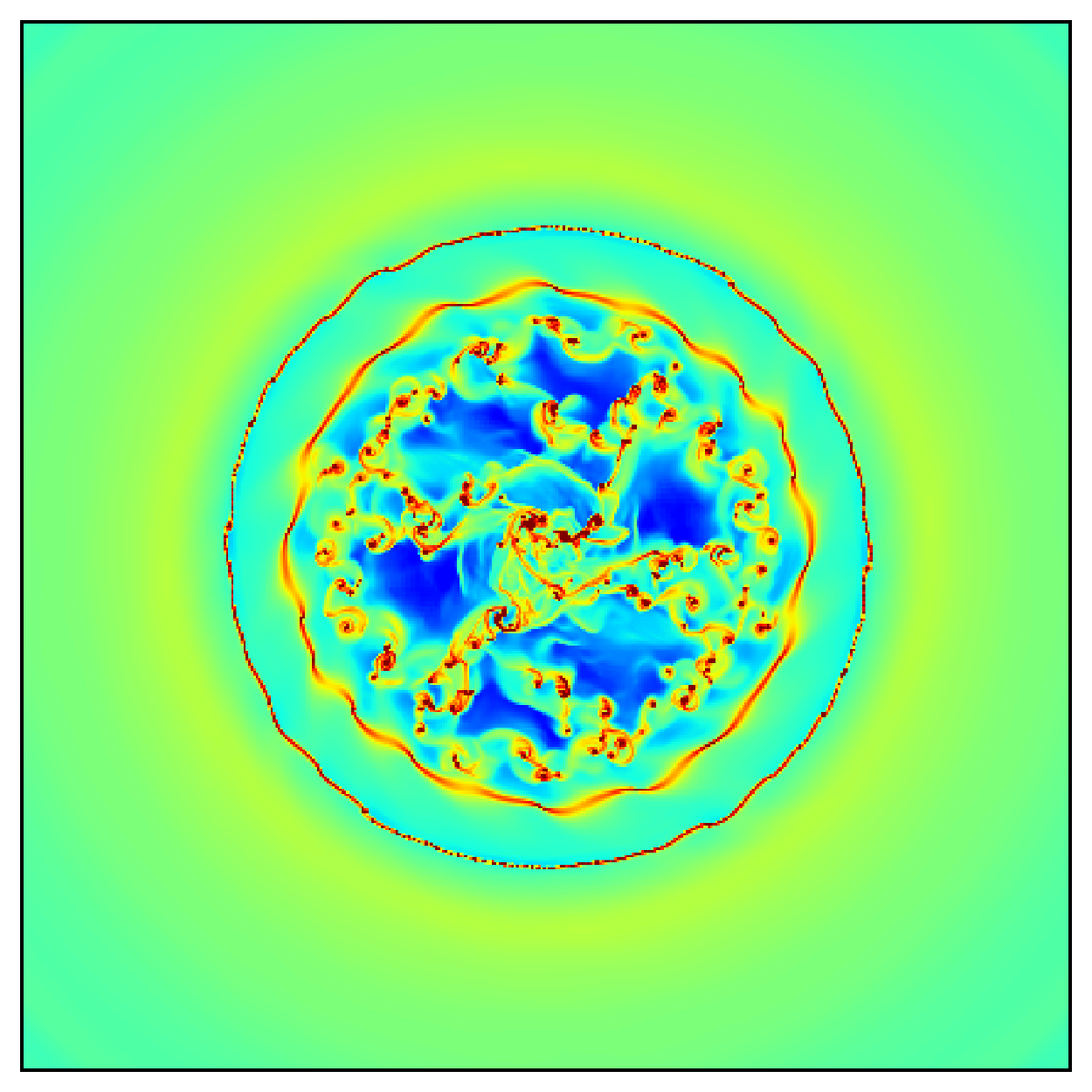}}
  {\includegraphics[width=0.24\linewidth]{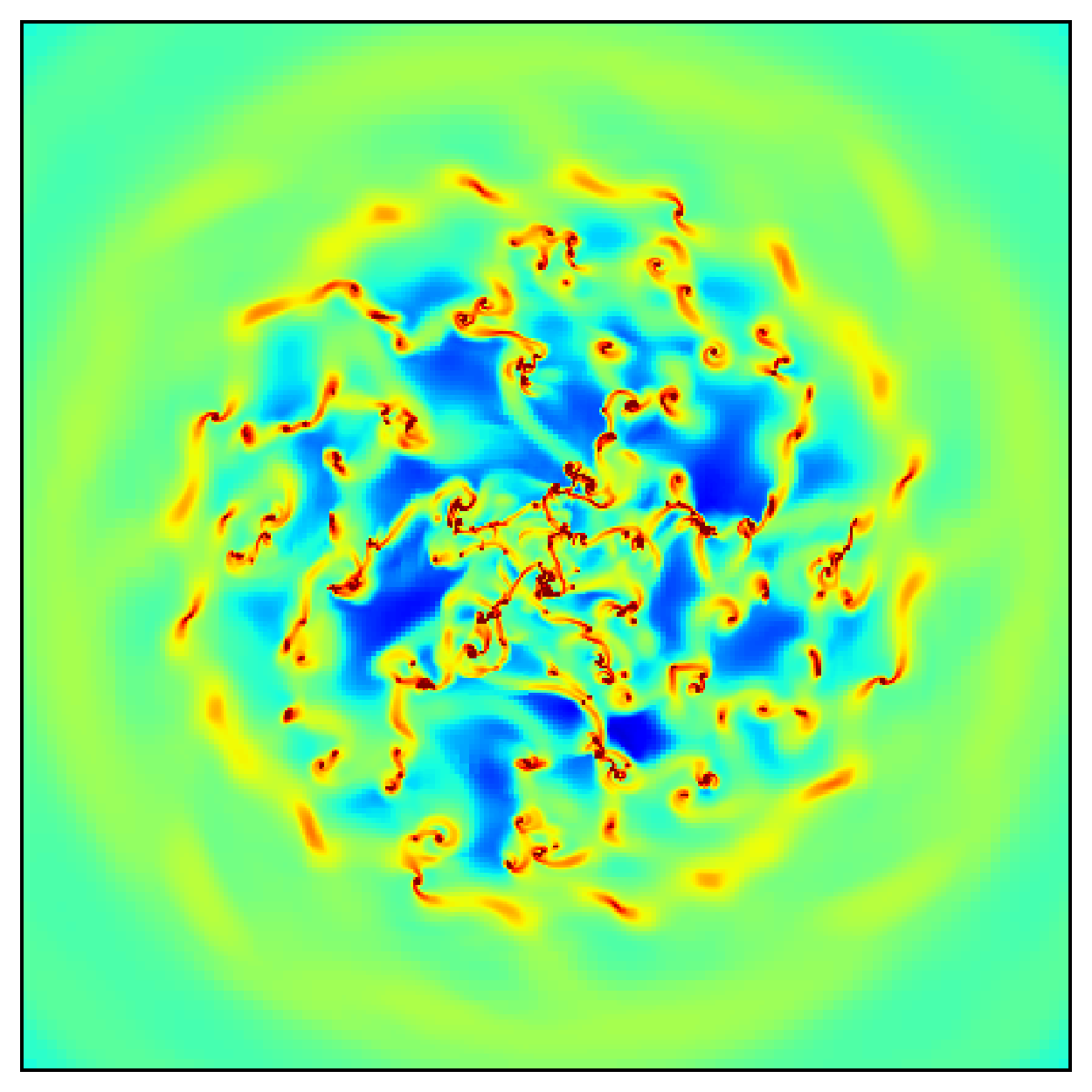}}
  {\includegraphics[width=0.24\linewidth]{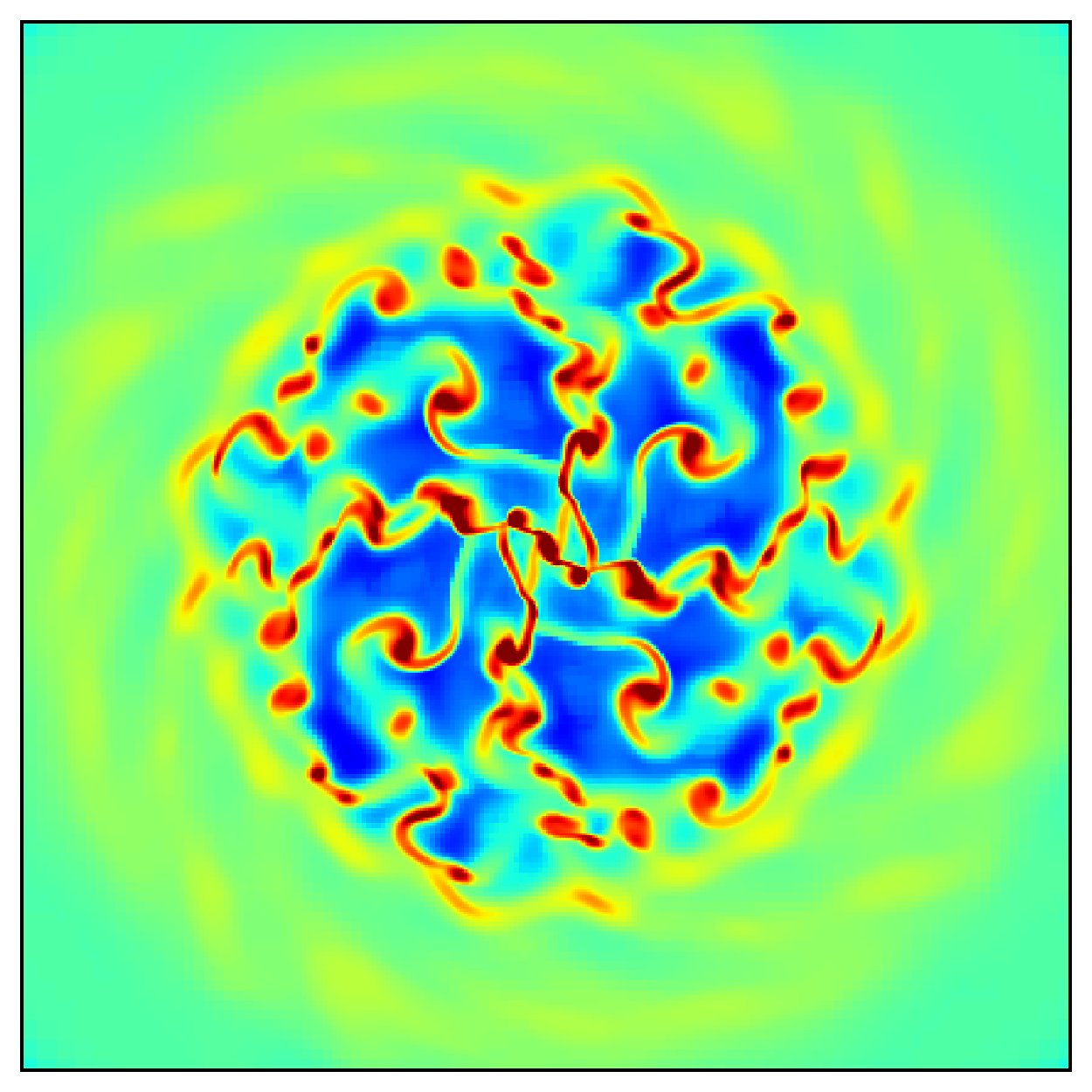}}
  {\includegraphics[width=0.24\linewidth]{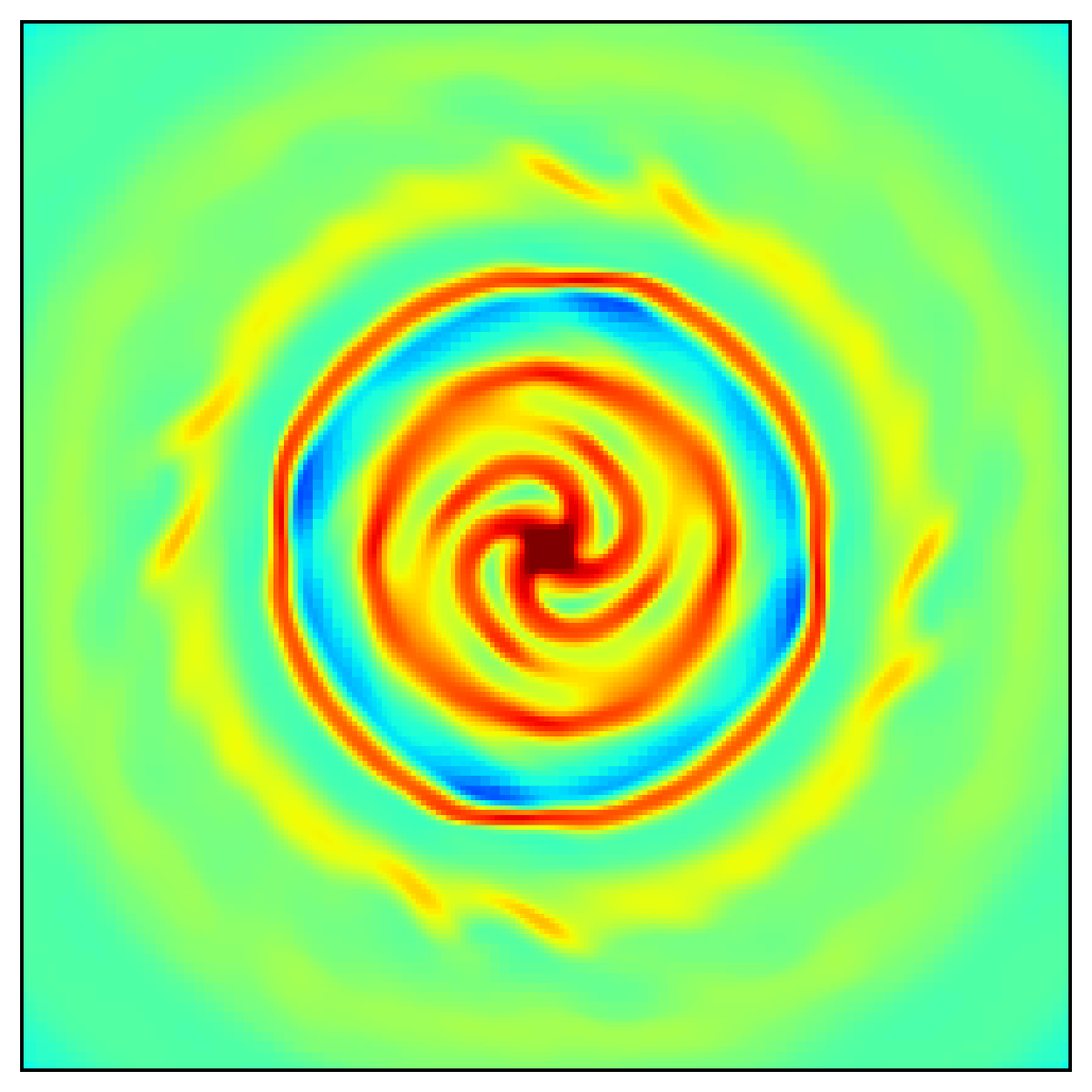}}\hfill

  {\includegraphics[width=0.24\linewidth]{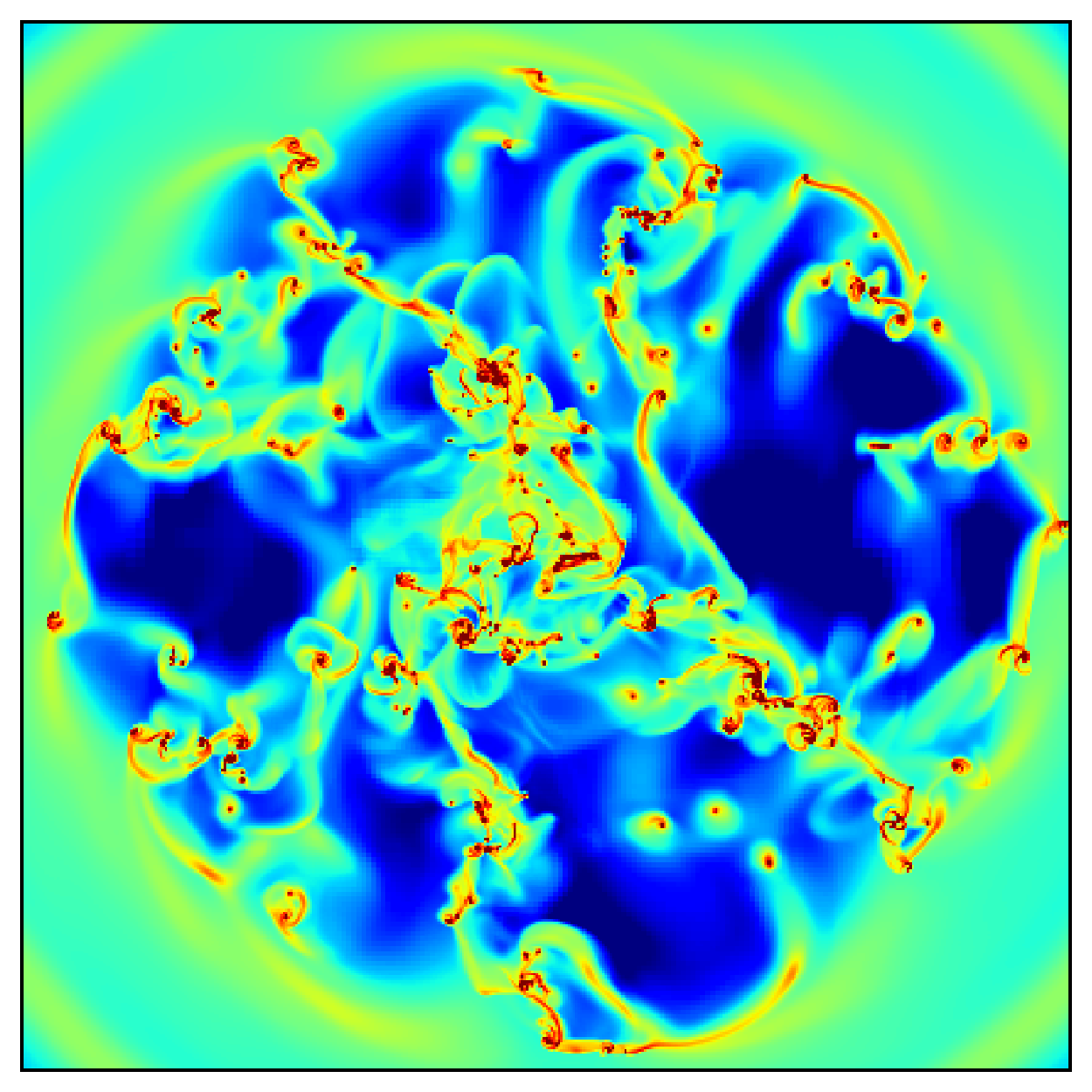}}
  {\includegraphics[width=0.24\linewidth]{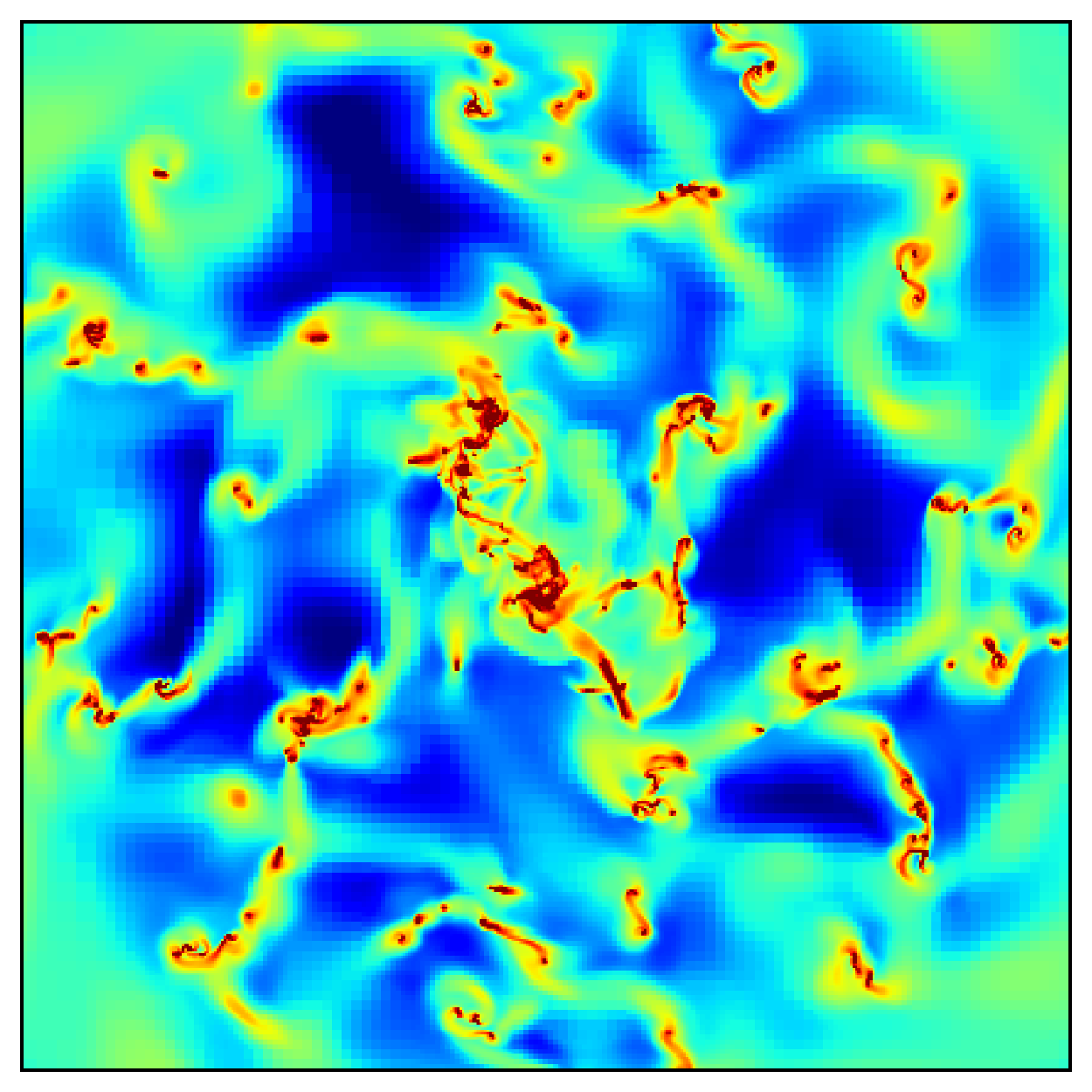}}
  {\includegraphics[width=0.24\linewidth]{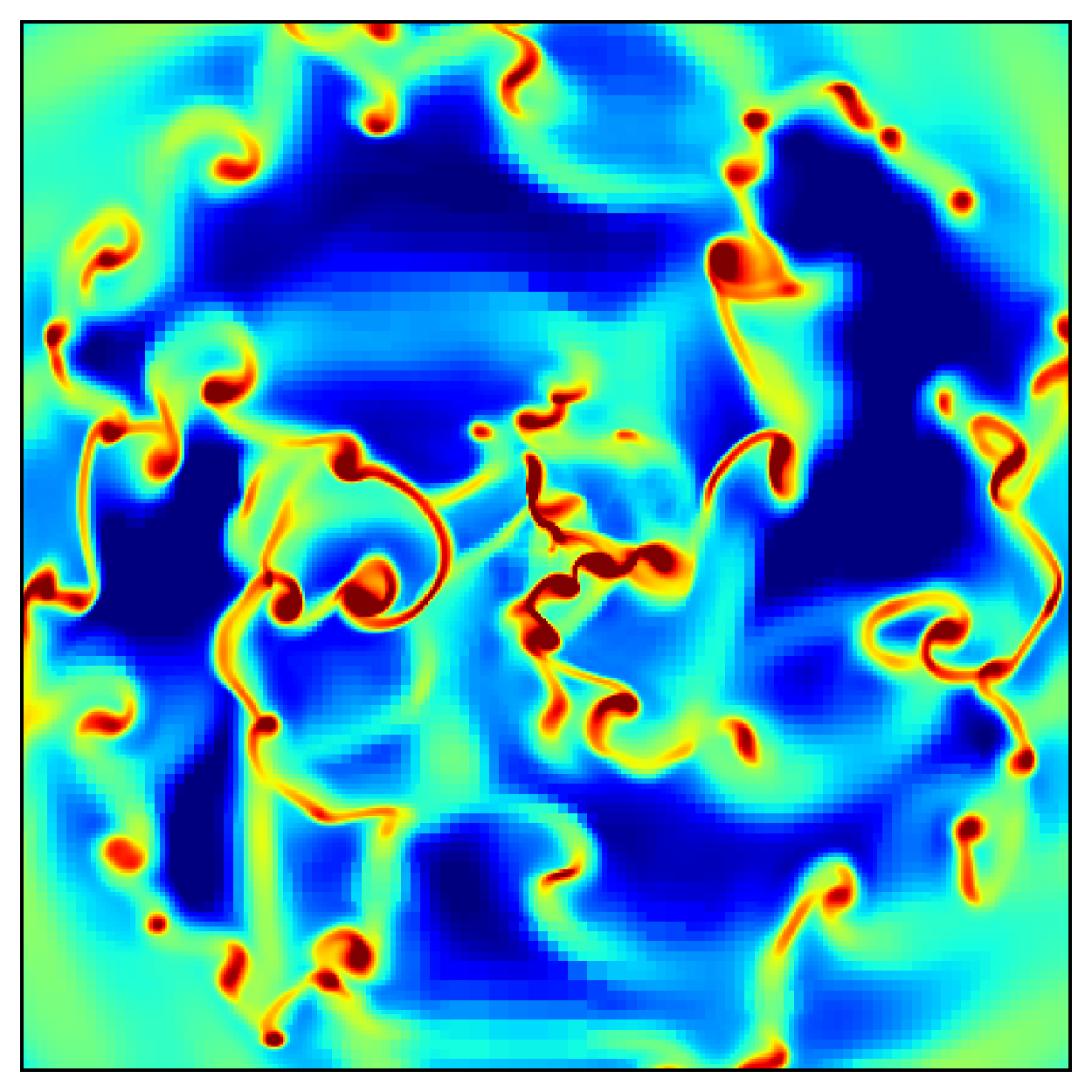}}
  {\includegraphics[width=0.24\linewidth]{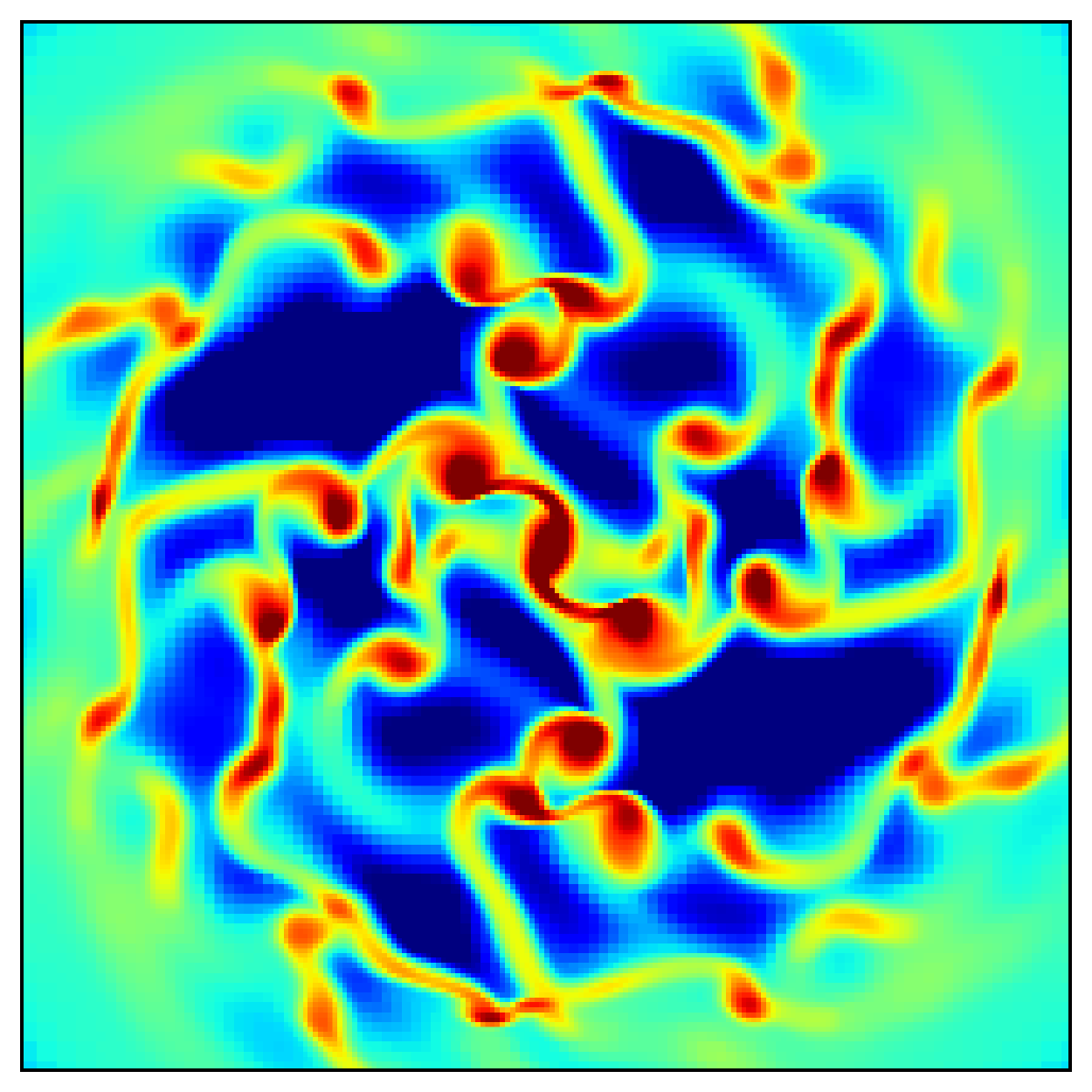}}\hfill

  {\includegraphics[width=0.24\linewidth]{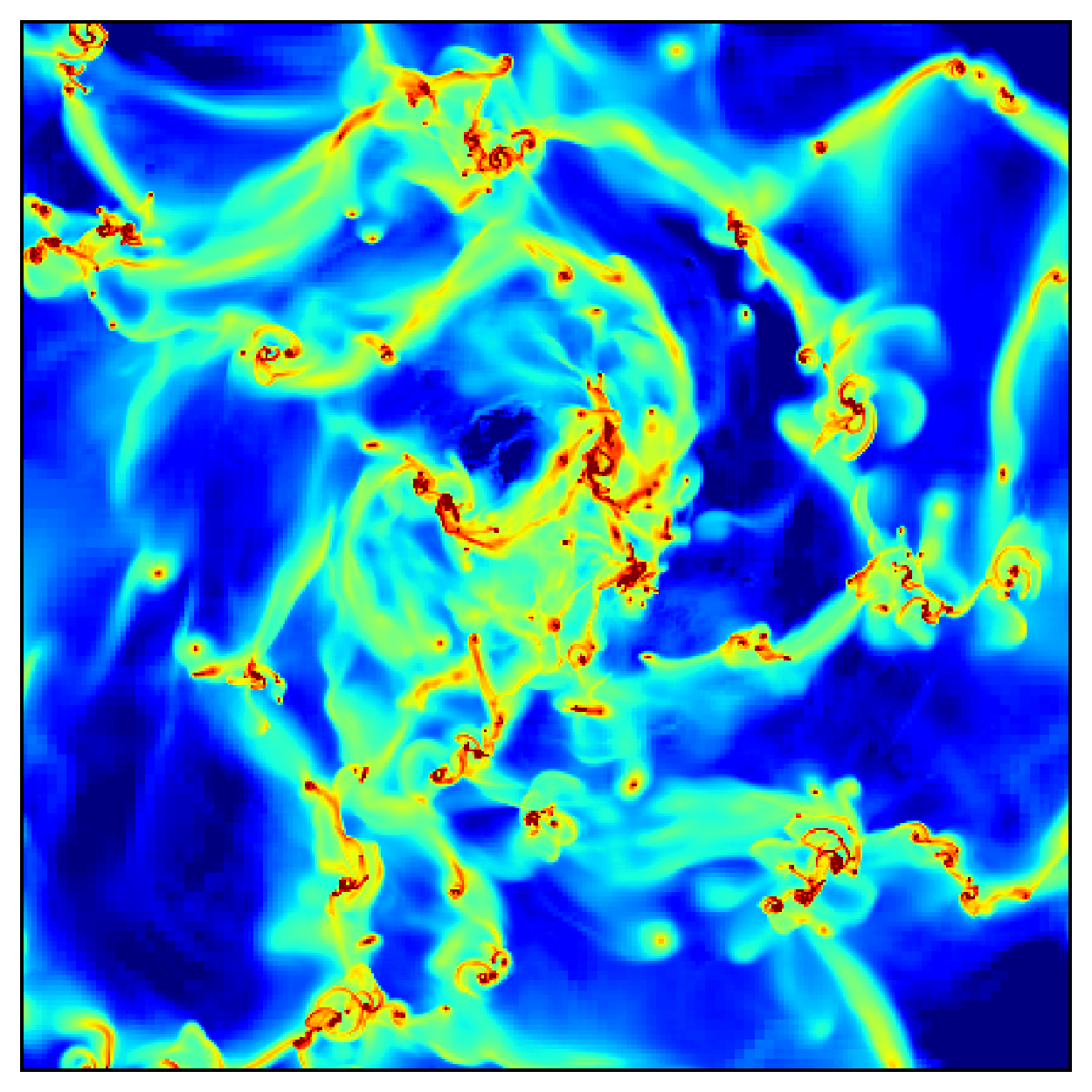}}
  {\includegraphics[width=0.24\linewidth]{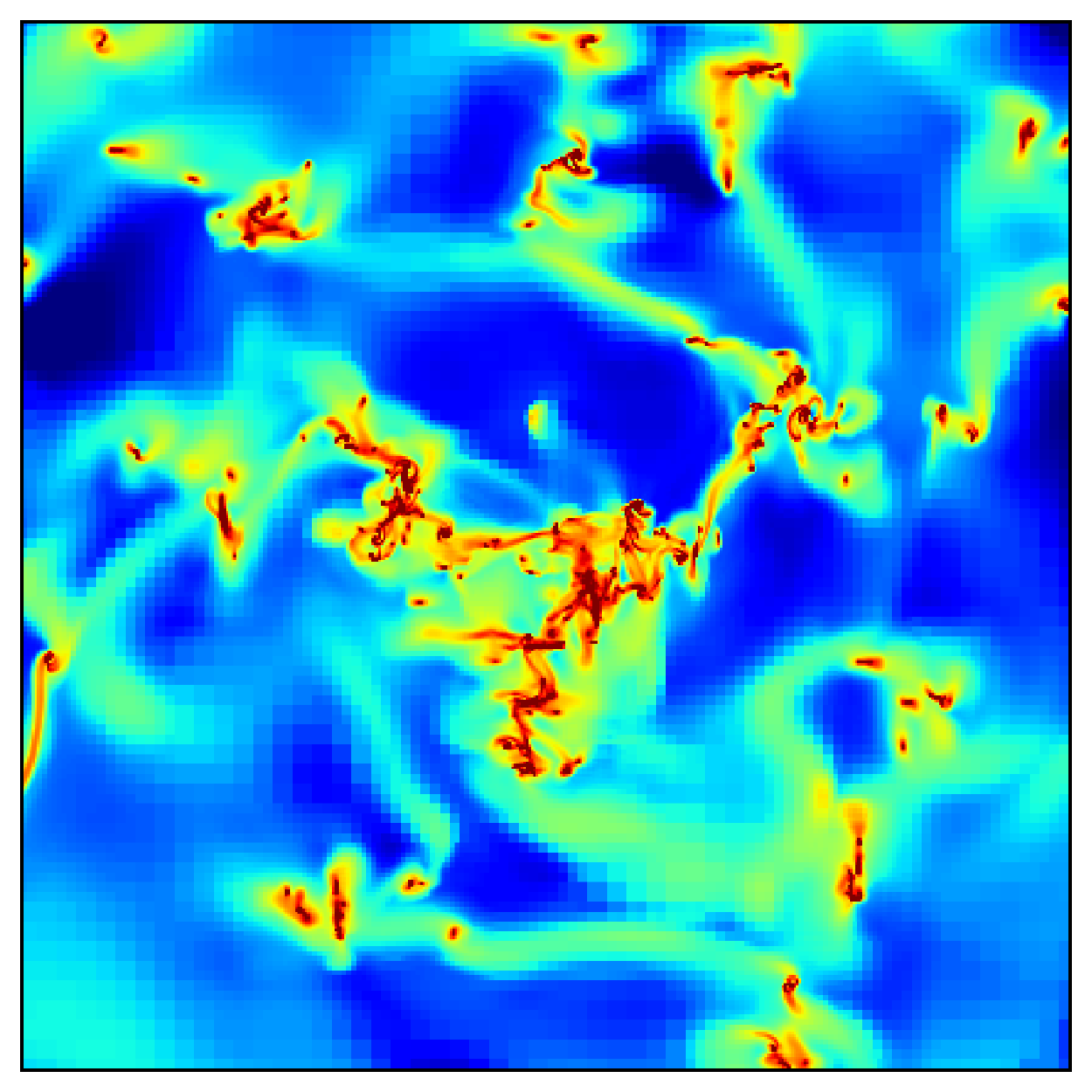}}
  {\includegraphics[width=0.24\linewidth]{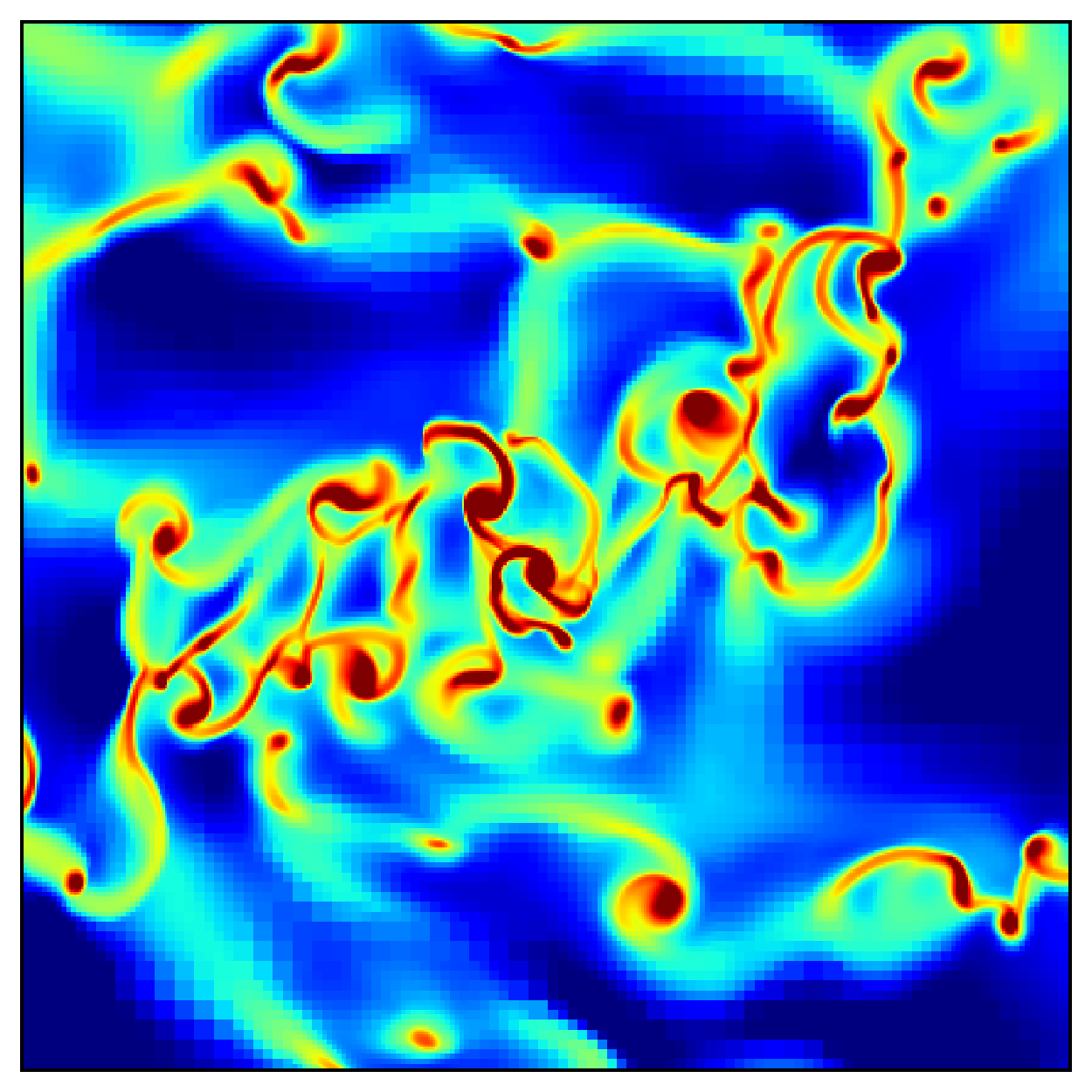}}
  {\includegraphics[width=0.24\linewidth]{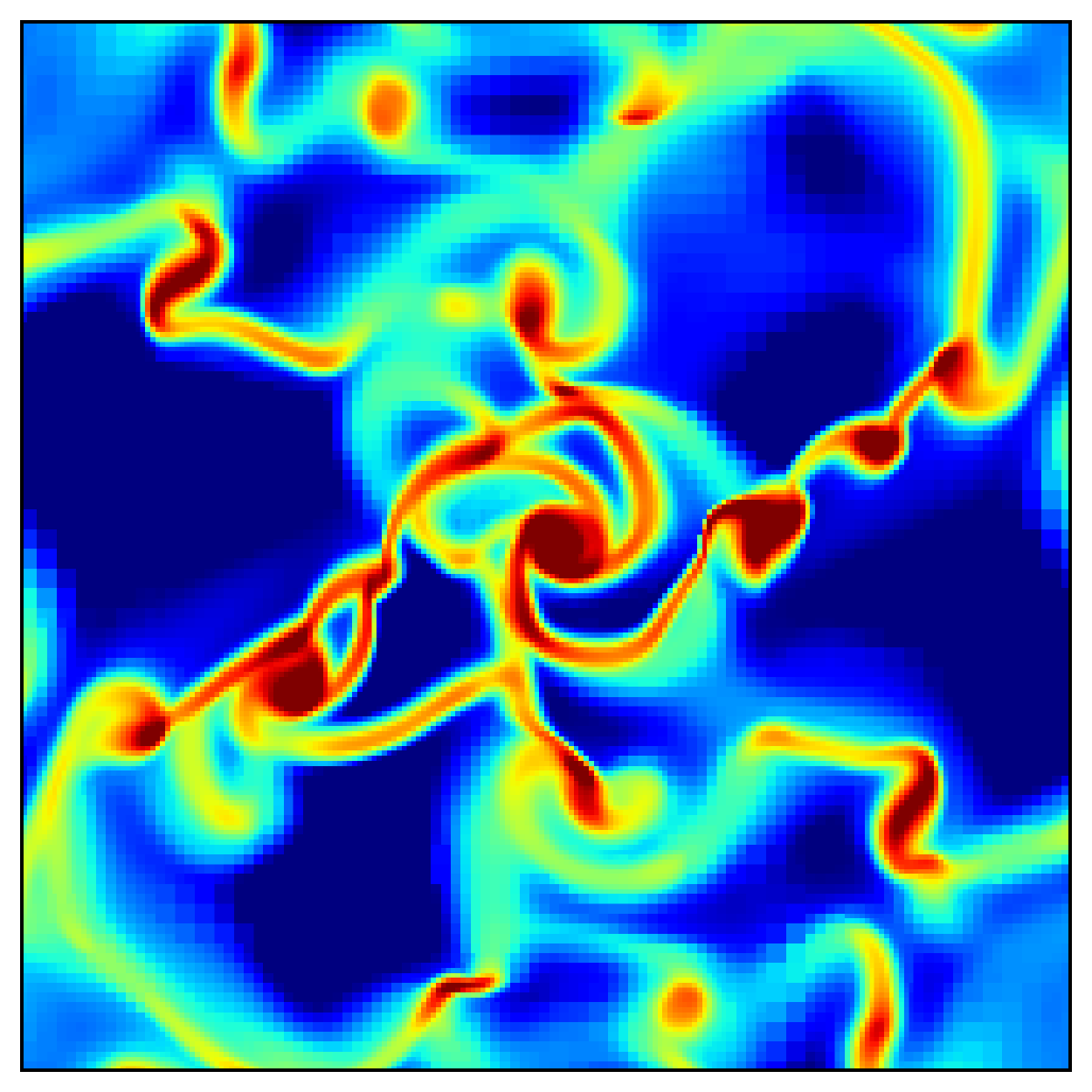}}\hfill

\includegraphics[width=.4 \linewidth]{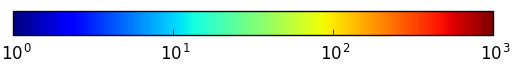} 

\caption{Surface density evolution of the discs of the different simulations. The images show the surface density $\mathrm{log10(\Sigma}) \ \mathrm{[M_{\odot} \; pc^{-2}}]$ face-on view of the gas disc at times $t = 100, 200, 400$ and 600 Myr for the four rows. Each panel shows a $20 \times 20 \ \mathrm{kpc^2}$ map centered on the disc. The different columns represent the four different simulations. The colors are limited to the range $1 - 10^3 \ \mathrm{M_{\odot} \ pc^{-2}}$ while densities of $10^6 \ \mathrm{M_{\odot} \ pc^{-2}}$ are reached in the centre of some clumps ($MS$). The snapshot at $655$ Myr of $MS$ can be seen in Figure \ref{fig:LOS_FWHM_set}.  \label{fig:surface_density_face_on}}

\end{figure*}

  \begin{figure*} 
\centering
\subfloat[ \label{fig:disk_sd_440}]
  {\includegraphics[width=.44\linewidth]{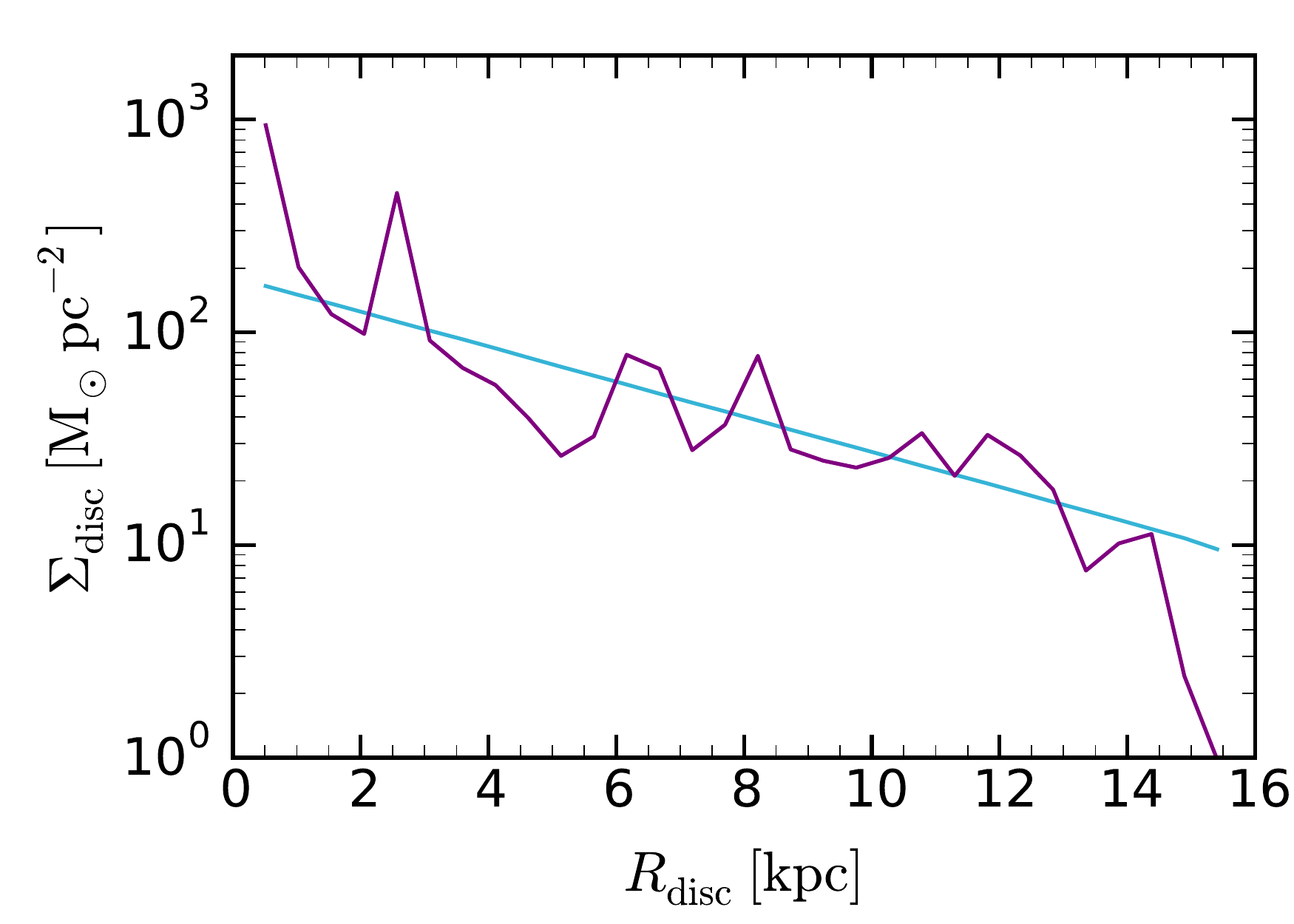}}
\subfloat[ \label{fig:disk_vrot}]
  {\includegraphics[width=.44\linewidth]{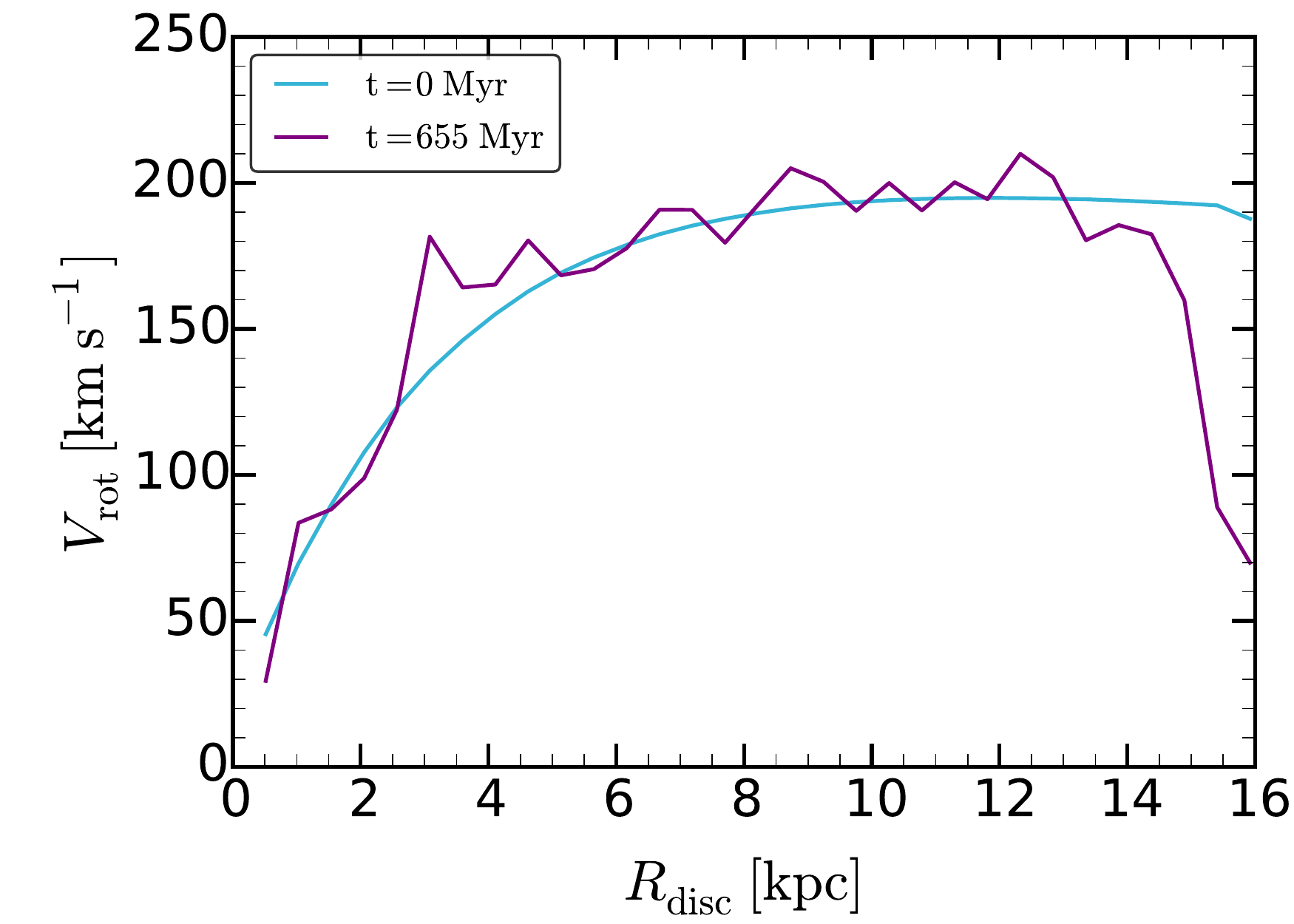}} 

\caption{Profiles of run $MS$ (500 pc bins): (a) Average gas surface density profile of the initial disc (pale blue) and at t = 655 Myr (purple).  (b) Mass-weighted average rotation velocity of the gas.
\label{fig:disk_profiles_sd}}
\end{figure*}

\subsection{Numerical setup}
\label{subsec:Simulation code}
\subsubsection{The high-resolution simulations} 
\label{subsubsec:The high-resolution simulations}
The simulations (Tab. \ref{tab:Main differences of the simulations}) are performed with the hydrodynamical AMR (adaptive mesh refinement) code RAMSES \citep{2002A&A...385..337T}. The gas is self-gravitating and the thermodynamics are regulated by an isothermal EoS. The hydrodynamical equations are being solved by using the HLL Riemann solver \citep{Harten_83} and making use of the MinMod slope limiter  \citep{1986AnRFM..18..337R}. The galaxy is centered in an isolated 48 kpc simulation box. The dark matter is added as a static external density field to the source term of the Poisson solver. To avoid artificial fragmentation in a gravitationally collapsing gas the Jeans length is typically required to be resolved by at least four cells $N_{\mathrm{J}} = 4$ \citep{1997ApJ...489L.179T} for all refinement levels. This refinement strategy sets and limits the initial resolution which is decreasing with radius (Figure \ref{fig:amr_levels_indisk}), due to the density gradient and the isothermal EoS of the galaxy. For the main simulation $MS$ with $N_{\mathrm{J}} = 19$ we apply most of the analysis and compare it with runs using $N_{\mathrm{J}} = 4$ which are reaching a lower initial resolution ($SR, LR, ULR$). The sizes of the first forming structures (rings) can be characterised by  the Toomre length \citep{2015MNRAS.448.1007B} and are therefore initially resolved in all runs (Table \ref{tab:Main differences of the simulations}) with several resolution elements. In the fragmented discs the simulations reach in the densest regions (clumps) a maximum resolution of $\Delta x_{\mathrm{min}} = 2.9 \ \mathrm{pc}$ for the main simulation ($MS$) (two times higher than in the previous work \citealp{2015MNRAS.448.1007B}). For the run $SR$, we reach $\Delta x_{\mathrm{min}} = 5.9 \ \mathrm{pc}$.  
 At the maximum refinement level higher densities could collapse further when resolved with less than 4 cells, as pressure gradients remain unresolved. This could lead to numerical fragmentation. It is common practice to ensure at least 4 cells \citep{1997ApJ...489L.179T} 
\begin{equation}
  N_{\mathrm{P}} = \frac{ \lambda_{\mathrm{J}} }{ \Delta x_{\mathrm{min}} } \geq 4,
\end{equation}
by introducing an artificial pressure floor (APF) which results in a constant minimum jeans length 
\begin{equation}
\label{eq:Jeans length}
 \lambda_{\mathrm{J}} = \sqrt{ \frac{ \pi c_{\mathrm{s,J}}^2 }{ G \rho} } = \mathrm{constant} ,
\end{equation}
that can be resolved in the simulations (see Section \ref{sec:Introduction}). The additional pressure can be expressed by the sound speed
\begin{equation} 
  c_{\mathrm{s,J}}^2 = \frac{G}{\pi} N_{\mathrm{P}}^2 \ \Delta x^2_{\mathrm{min}} \ \rho, \ \ \ 
  \mathrm{for} \ c_{\mathrm{s,J}} \gtrapprox 10 \ \mathrm{km \ s^{-1}}.
\end{equation}
The equation is dependent on the number of cells per Jeans length $N_{\mathrm{P}}$, the simulation's maximum resolution $\Delta x_{\mathrm{min}}$, the density $\rho$ in the computational cell and the gravitational constant $G$ (see Figure \ref{fig:pressure_floor}). Therefore, we obtain with $N_{\mathrm{P}}=7$ in run $MS$ the minimum resolved length of $L_{\mathrm{MinJeans}}=20.5 \ \mathrm{pc}$. Typically, $N_{\mathrm{P}}=4$ is used in simulations, comparable with run $SR$ that has $L_{\mathrm{MinJeans}}=23.4 \ \mathrm{pc}$. We call the minimum resolved length in the simulations \textbf{the effective resolution}.
 The higher value for $N_{\mathrm{P}}$ in the main simulation ensures a very similar minimum resolved length as in run $SR$ which is reached in the collapsed regions, inside the clumps. This allows us to identify possible resolution effects on the initial fragmentation due to their different Jeans number $N_{J}$.

  \begin{table}
 \caption{Definition of the surface density regimes.}
 \label{tab:Surface density regions}
 \begin{tabular}{lcl}
  \hline

  Notation & $\Sigma$				& Regime$^a$ \\
           & [$\mathrm{M_{\odot} \ pc^{-2}}$]\\
  \hline
  \textcolor{Plum}{$\blacksquare$} total       &  $1 - 10^{6}$			& total considered density range\\
  \textcolor{black}{$\blacksquare$} $H$       &  $> 10^{3}$			& \textit{higher} densities within the clumps\\  
  \textcolor{yellow}{$\blacksquare$} $C$ 	  &  $> 10^{2}$		    & typical gas densities for the \textit{clumps}\\ 
  \textcolor{ForestGreen}{$\blacksquare$}  $A$         &  $10 - 10^{2}$ 	& elongated, \textit{arm}-like features\\ 
  \textcolor{blue}{$\blacksquare$} $L$ 		  &  $< 10$		    & \textit{low} density, inter-clump gas\\ 						
  \hline
  
 \end{tabular}
\footnotesize{$^a$ The notation corresponds to the evolved disc.}

\end{table}

\begin{figure}
\includegraphics[width=84mm]{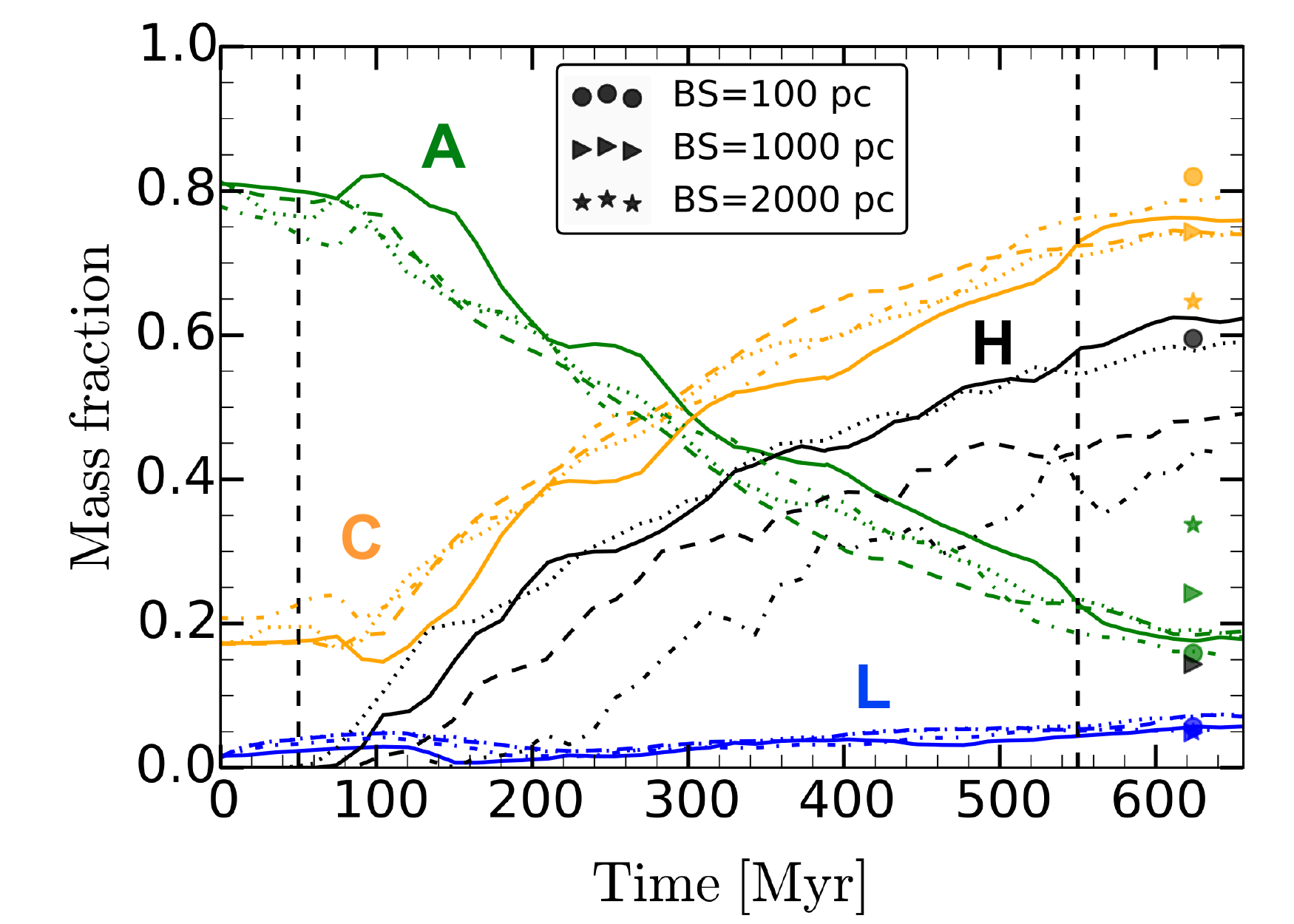} 
\caption{The evolution of the disc mass fractions for the different surface density regimes (for colors see Table \ref{tab:Surface density regions}) is very similar for the runs $MS$ (solid lines) and $SR$ (dotted lines), only the higher densities (black lines) are not reached by the runs $LR$ and $ULR$ (dashed and dashed-dotted black lines). All runs show a similar mass fraction for the clump areas (orange). The vertical dashed black lines indicate the beginning and the end of the fragmentation process in the disc. The symbols illustrate the effects of different beam-smearing at later times (see Section \ref{sec:The effects of beam smearing on the main clump properties} ).  \label{fig:mass_factions}} 
\end{figure}

\begin{figure*}
\centering
\includegraphics[width=.99\linewidth]{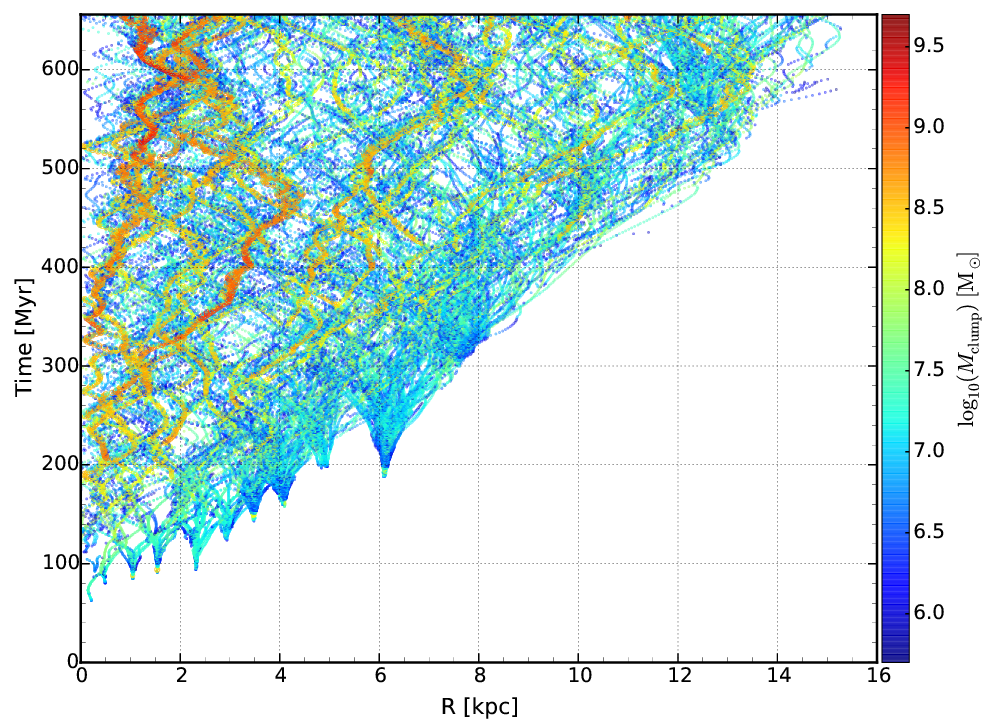} 
\caption{For run $MS$: Overview of the clump mass evolution ($\Delta t \sim 1.5$ Myr) and their radial positions within the face-on disc. The origin of the $\sim 10^7 \ \mathrm{M_{\odot}}$ clumps in the ring-like structures as well as their fast merging to $\sim 10^8 \ \mathrm{M_{\odot}}$ (within 50 Myr) is very well visible. Larger masses are more prominent at small radii. The mass range shown is limited to the minimum of $5 \times 10^5 \ \mathrm{M_{\odot}}$. Clumps with masses above $> 10^8 \ \mathrm{M_{\odot}}$ are indicated by larger symbol sizes.
\label{fig:clump_in_disk_evolution}} 
\end{figure*}

\subsubsection{The low-resolution simulations} 
\label{subsubsec:The low-resolution simulations}
The initial resolution in run $LR$ and $ULR$ are already reaching the maximum possible limit in the central part of the galaxies  (Figure \ref{fig:amr_levels_indisk}). To ensure a resolved Jeans length also there, the artificial pressure support is already in the beginning  effecting the setup in run $LR$ for radii R=0-0.25 kpc and more strongly in run $ULR$ between R=0-5 kpc (see in Figure \ref{fig:pressure_floor}  the enhanced sound speed at the low densities for the setup). In the fragmented disc a maximum resolution of $\Delta x_{\mathrm{min}} = 46.9 \ \mathrm{pc}$ for $LR$ and for the run $ULR$ only $\Delta x_{\mathrm{min}} = 93.8 \ \mathrm{pc}$ can be reached. But, due to the APF, the minimum structures that can be resolved are in run $LR$ limited to $L_{\mathrm{MinJeans}}=187.5 \ \mathrm{pc}$ and for run $ULR$ to $L_{\mathrm{MinJeans}}=375 \ \mathrm{pc}$.\\

\begin{figure*} 
\centering
\subfloat[ \label{fig:mass_tot_per_massbin}]
  {\includegraphics[width=.33\linewidth]{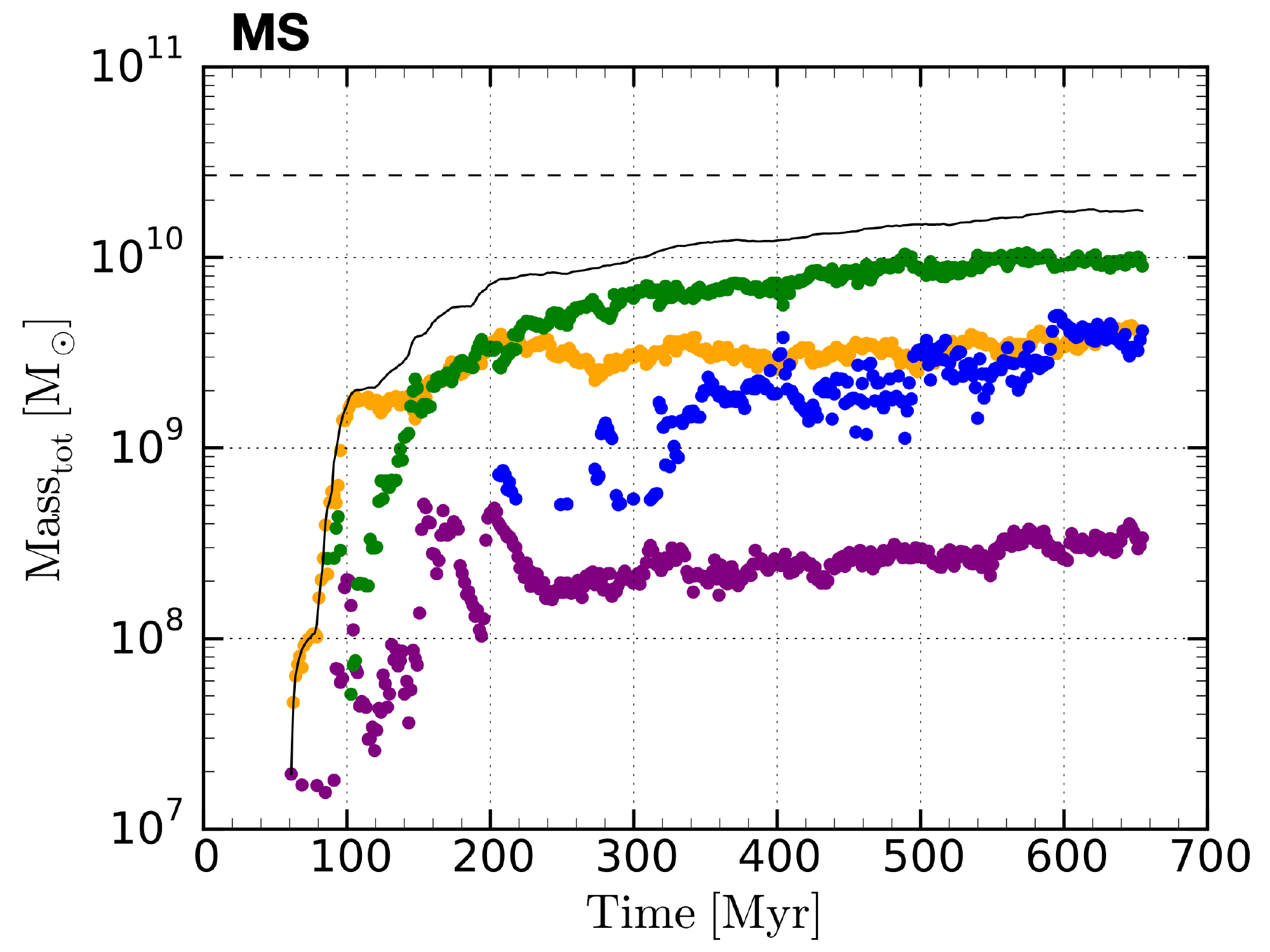}}\hfill
\subfloat[  \label{fig:size_average_evolution}]
  {\includegraphics[width=.33\linewidth]{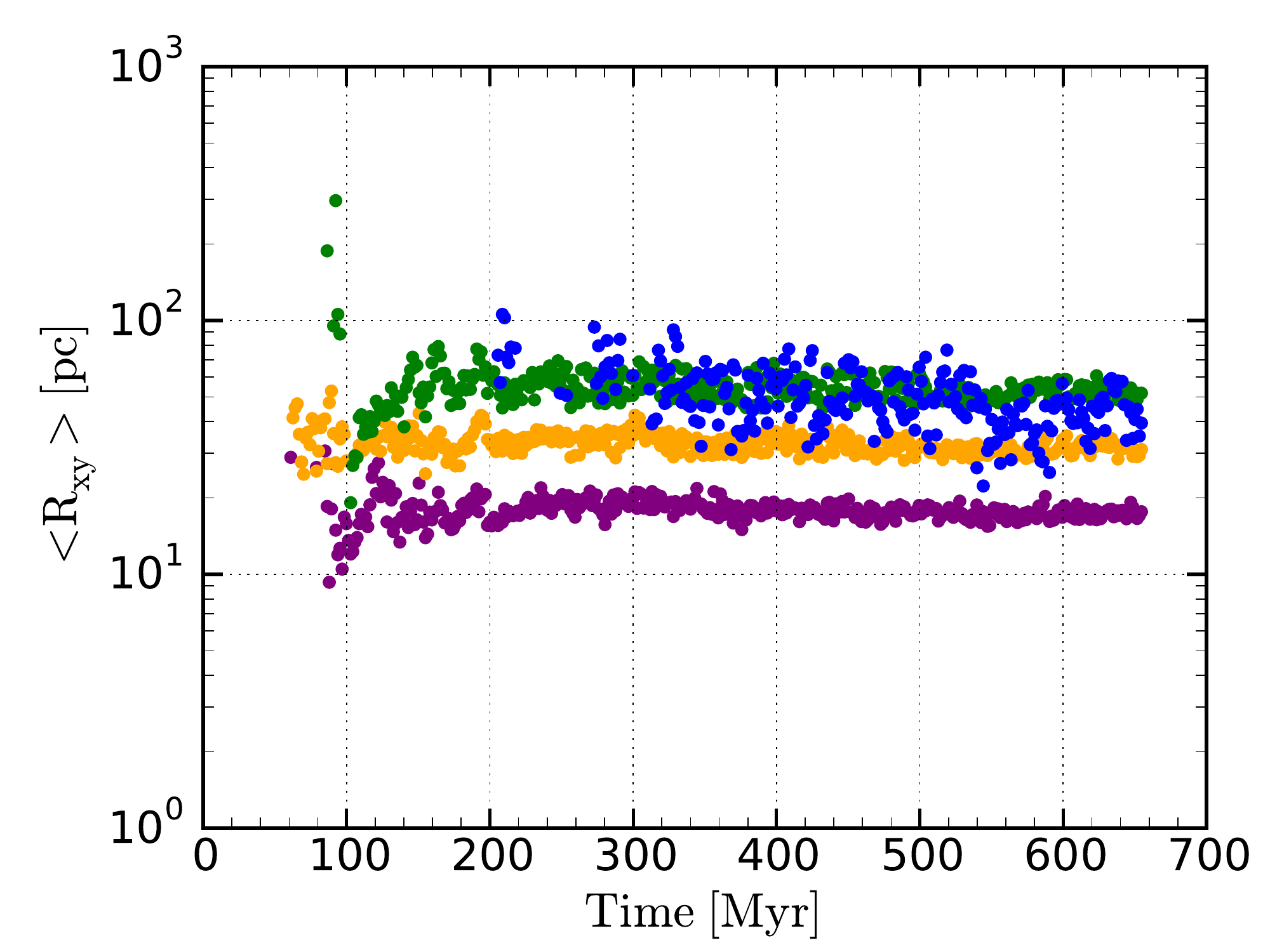}}
\subfloat[  \label{fig:hight_average_evolution}]
  {\includegraphics[width=.33\linewidth]{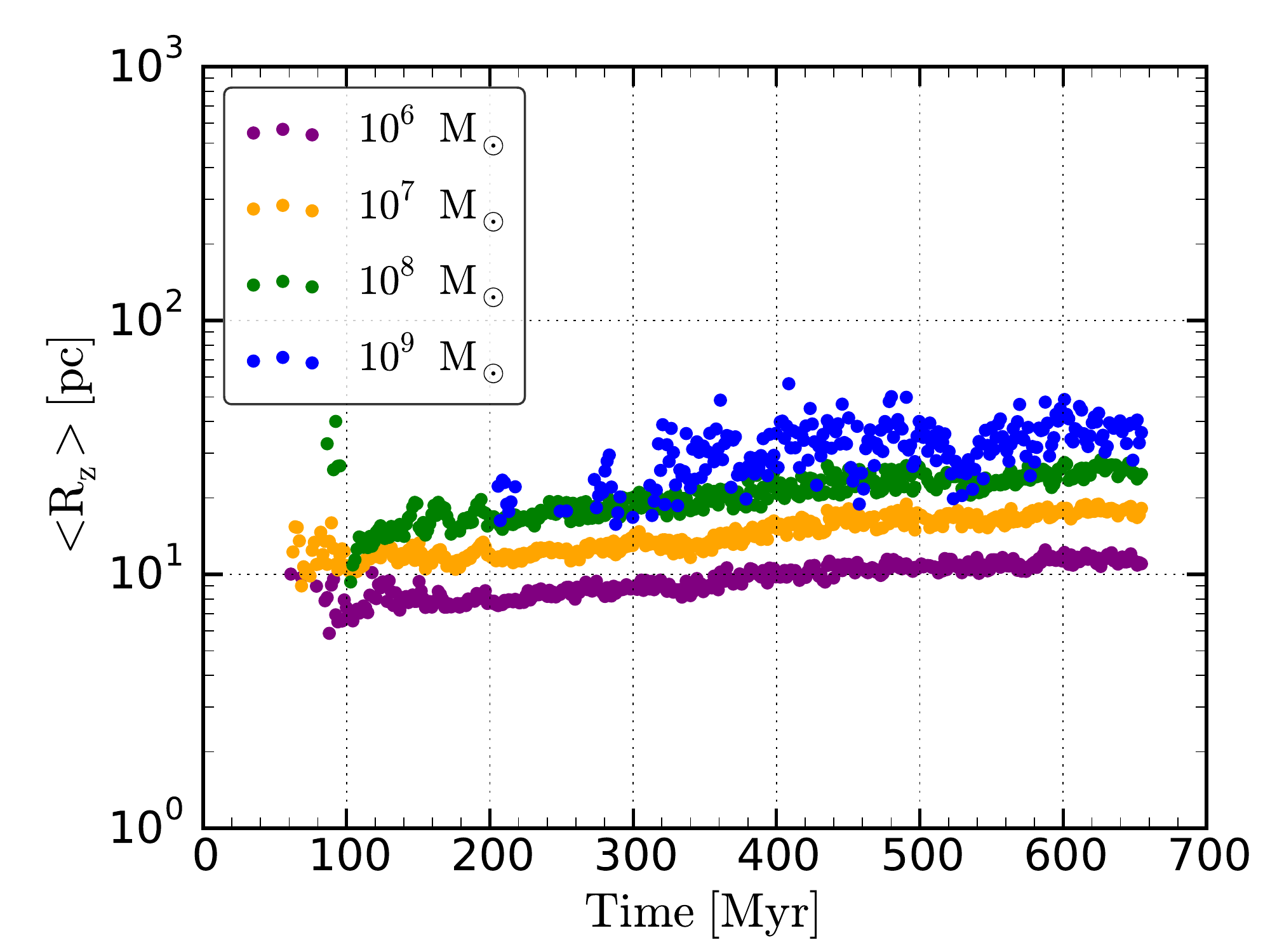}}

\subfloat[ \label{fig:N4_mass_tot_per_massbin}]
  {\includegraphics[width=.33\linewidth]{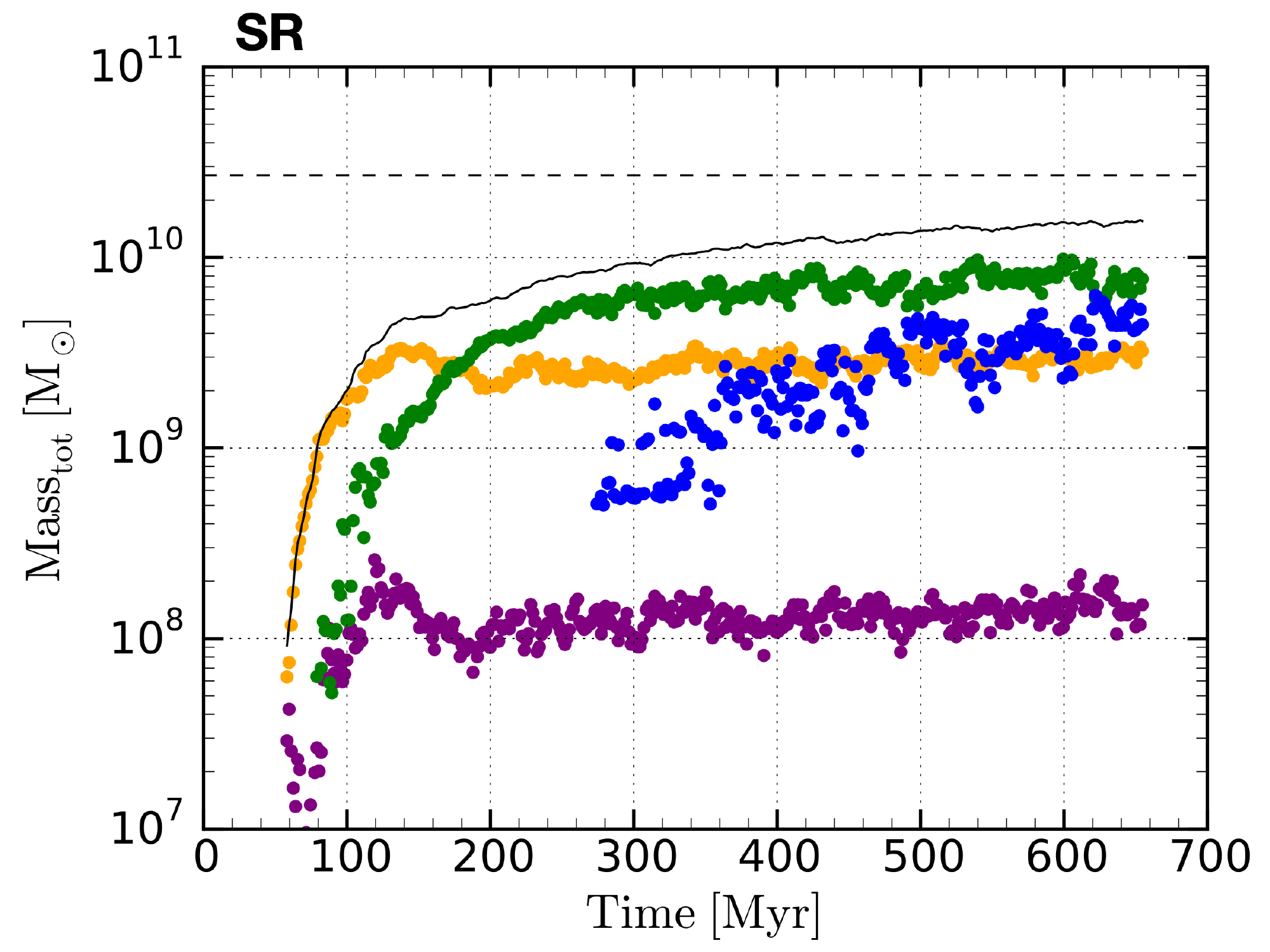}}\hfill
\subfloat[  \label{fig:N4_size_average_evolution}]
  {\includegraphics[width=.33\linewidth]{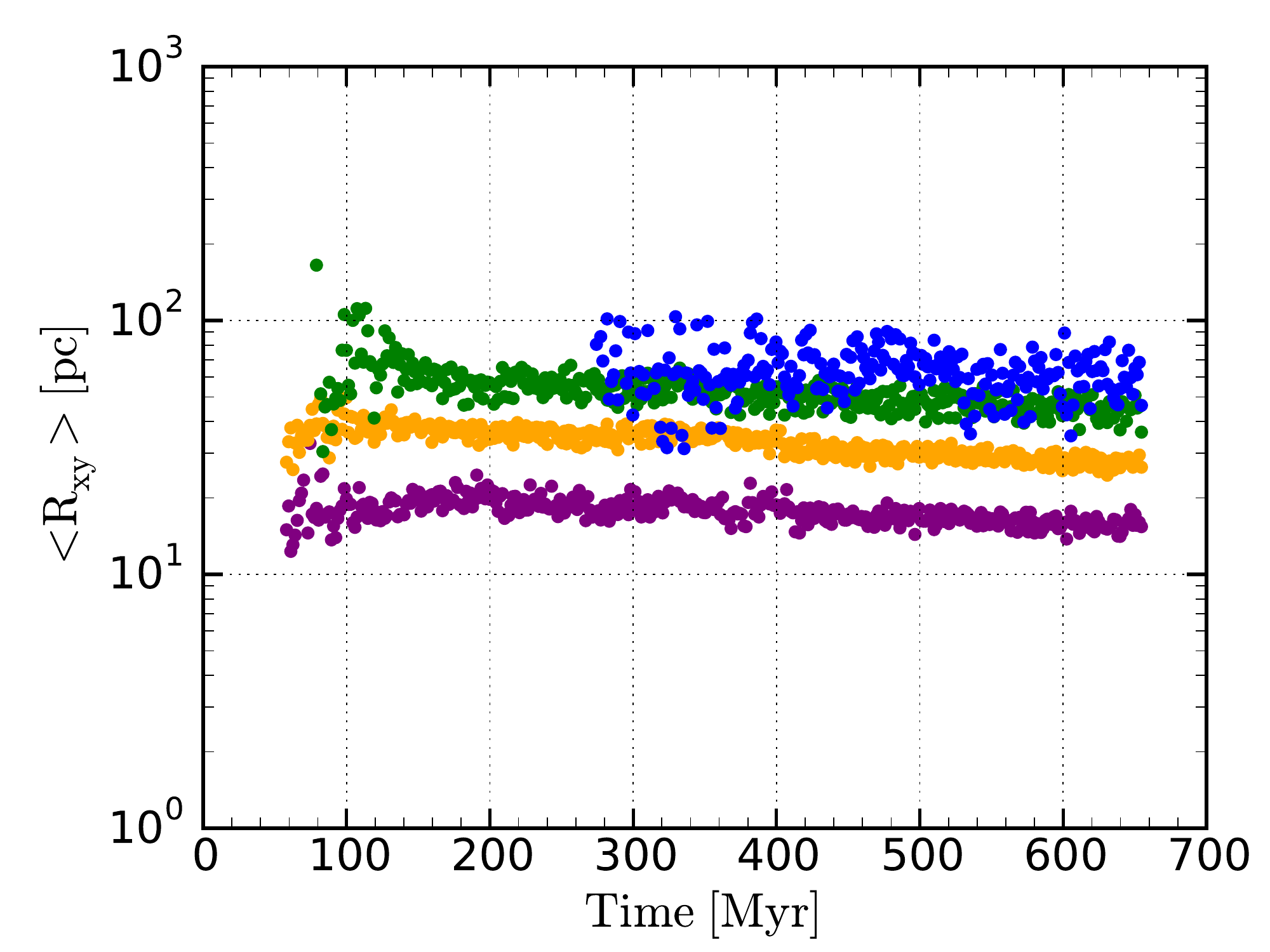}} 
\subfloat[  \label{fig:N4_hight_average_evolution}]
  {\includegraphics[width=.33\linewidth]{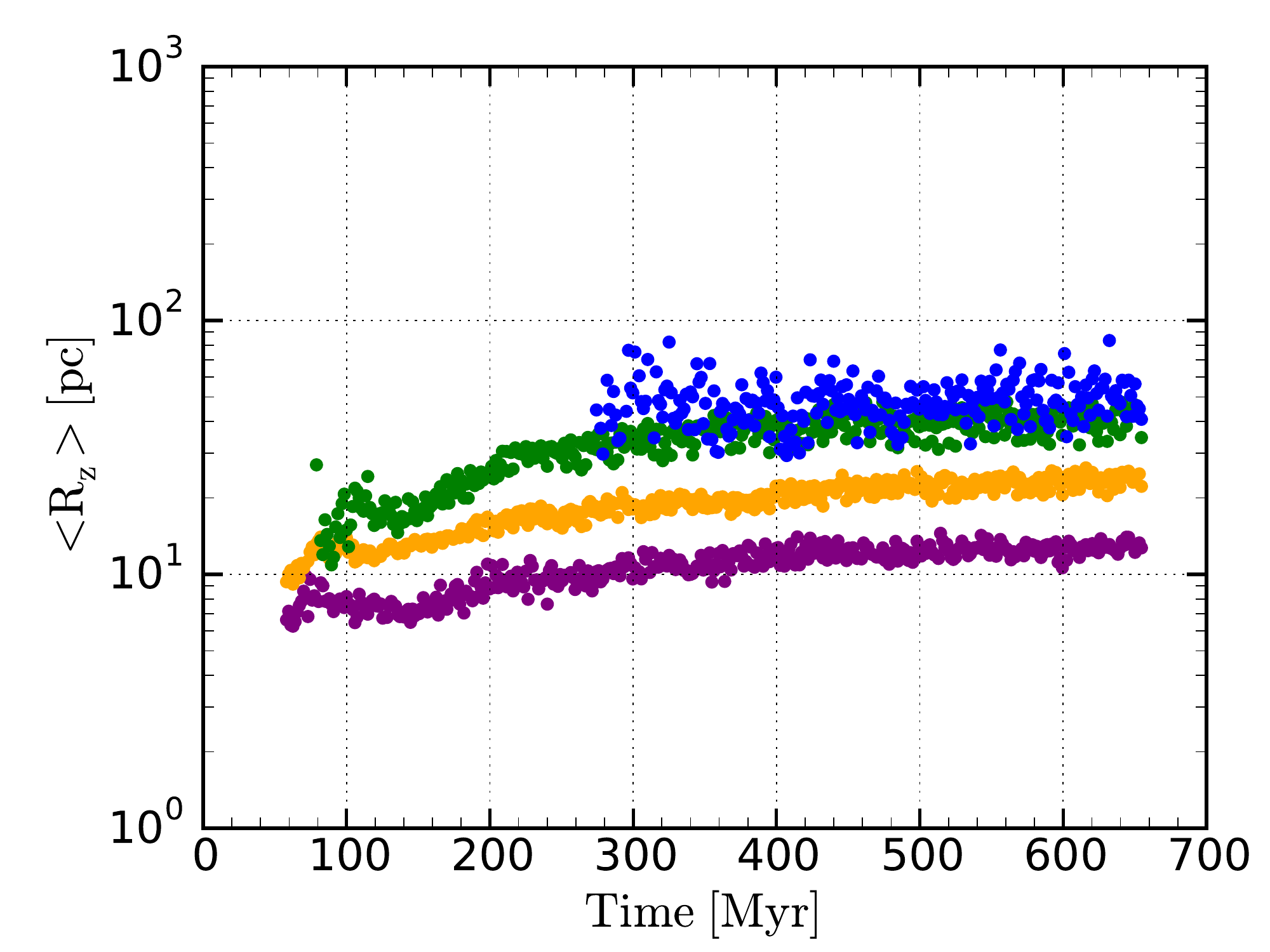}}   
   
\caption{Evolution of the identified clumps and their properties over time ($\Delta t \sim 1.5$ Myr). The first row corresponds to the main simulation $MS$ and the second row to the run $SR$. (a), (d): The total mass within all clumps in black and the colored symbols represent the total mass within each mass bin ]$5 \times 10^{i-1} - 5 \times 10^{i}$] ($\ \mathrm{M_{\odot}}$). The total disc mass is illustrated by the dashed black line. (b), (e): Evolution of the clump radii in the plane $R_{\mathrm{xy}}$ and for (c), (f) in vertical direction  $R_{\mathrm{z}}$. 
\label{fig:number_mass_clumps}}
\end{figure*}

\begin{figure*} 
\centering
\subfloat[ \label{fig:lsr100_mass_tot_per_massbin}]
  {\includegraphics[width=.33\linewidth]{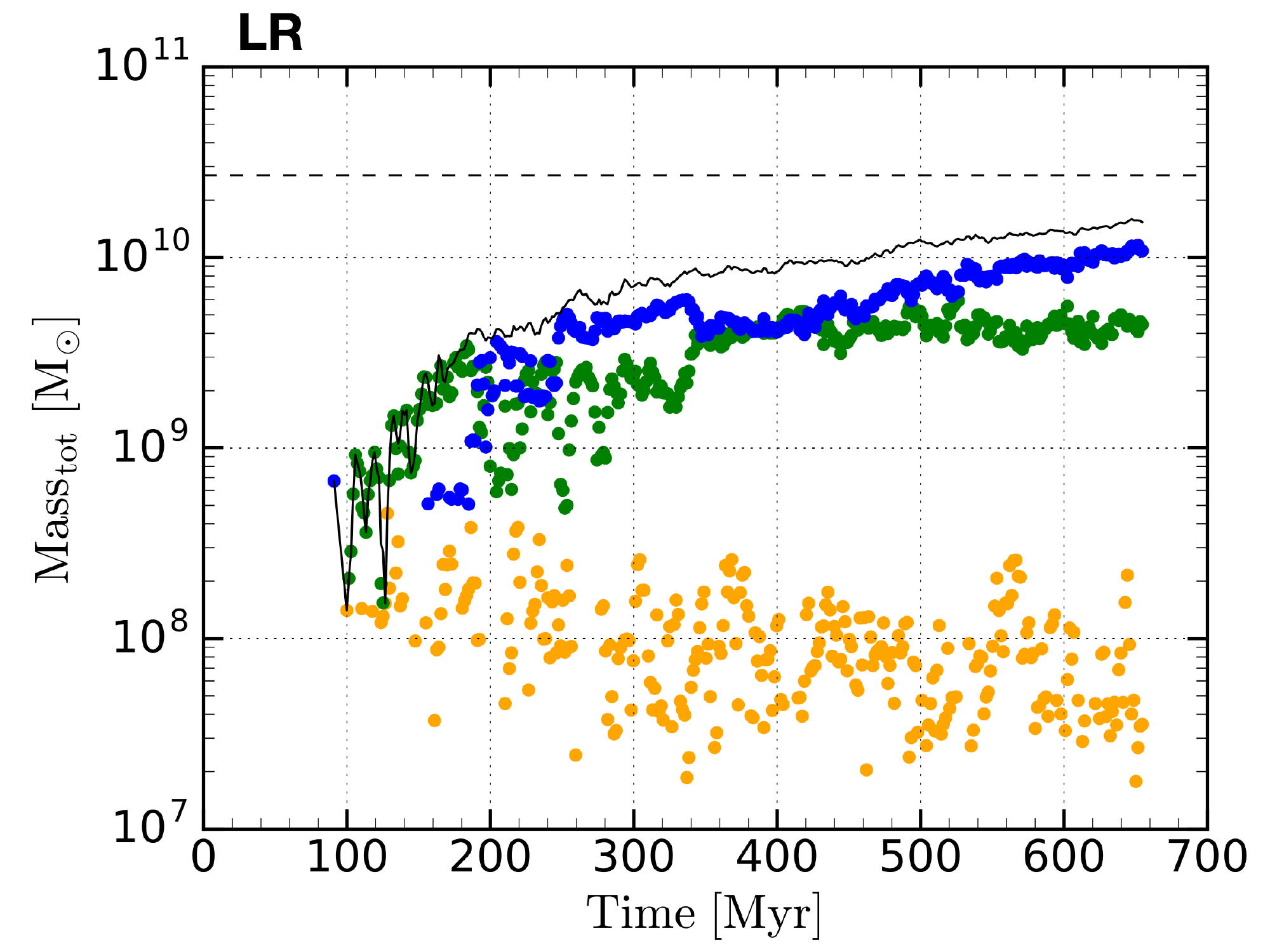}}\hfill
\subfloat[  \label{fig:lsr100_size_average_evolution}]
  {\includegraphics[width=.33\linewidth]{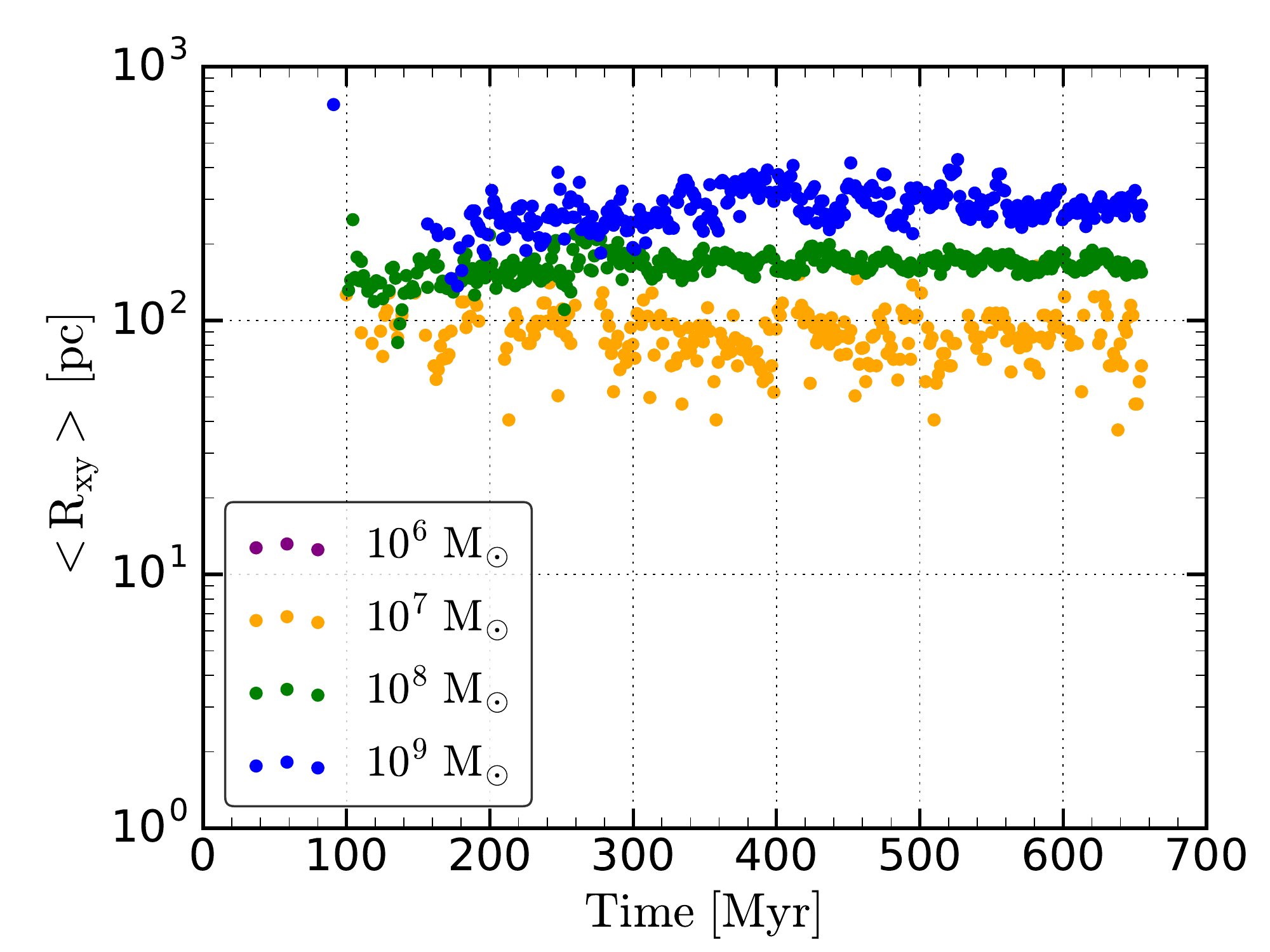}}
\subfloat[  \label{fig:lsr100_size_average_evolution}]
  {\includegraphics[width=.33\linewidth]{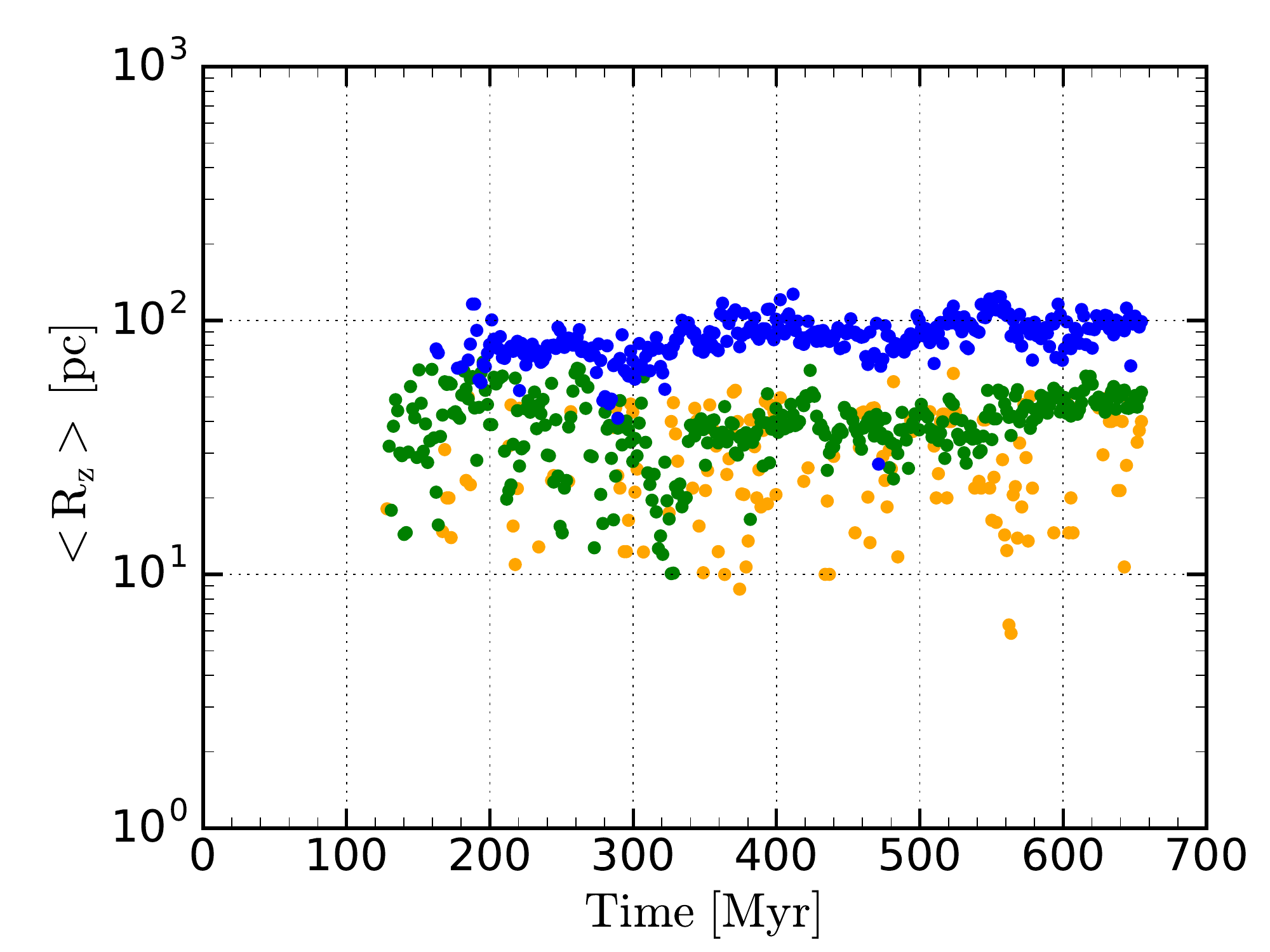}}

\subfloat[ \label{fig:ulres_mass_tot_per_massbin}]
  {\includegraphics[width=.33\linewidth]{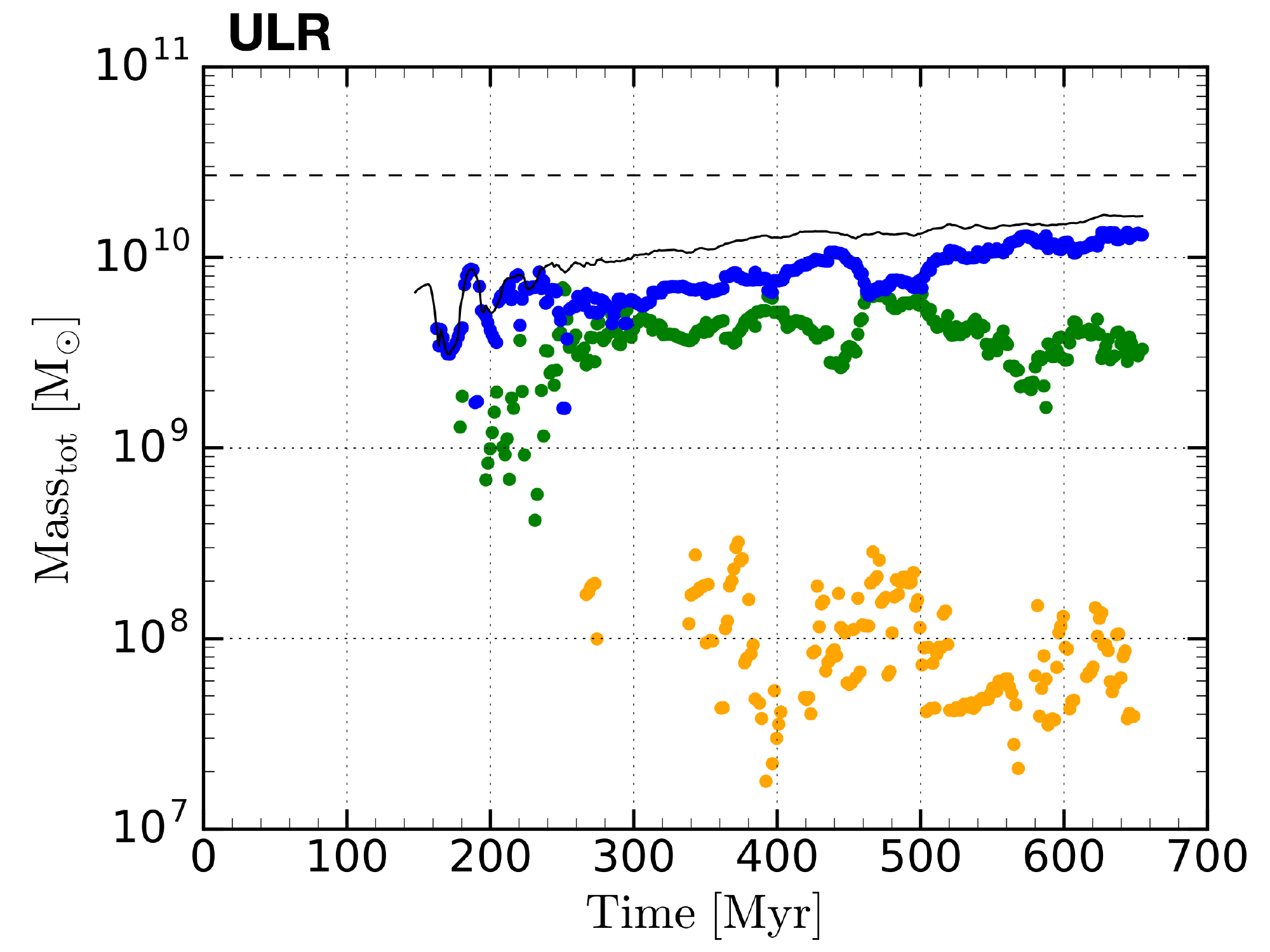}}\hfill
\subfloat[  \label{fig:ulres_size_average_evolution}]
  {\includegraphics[width=.33\linewidth]{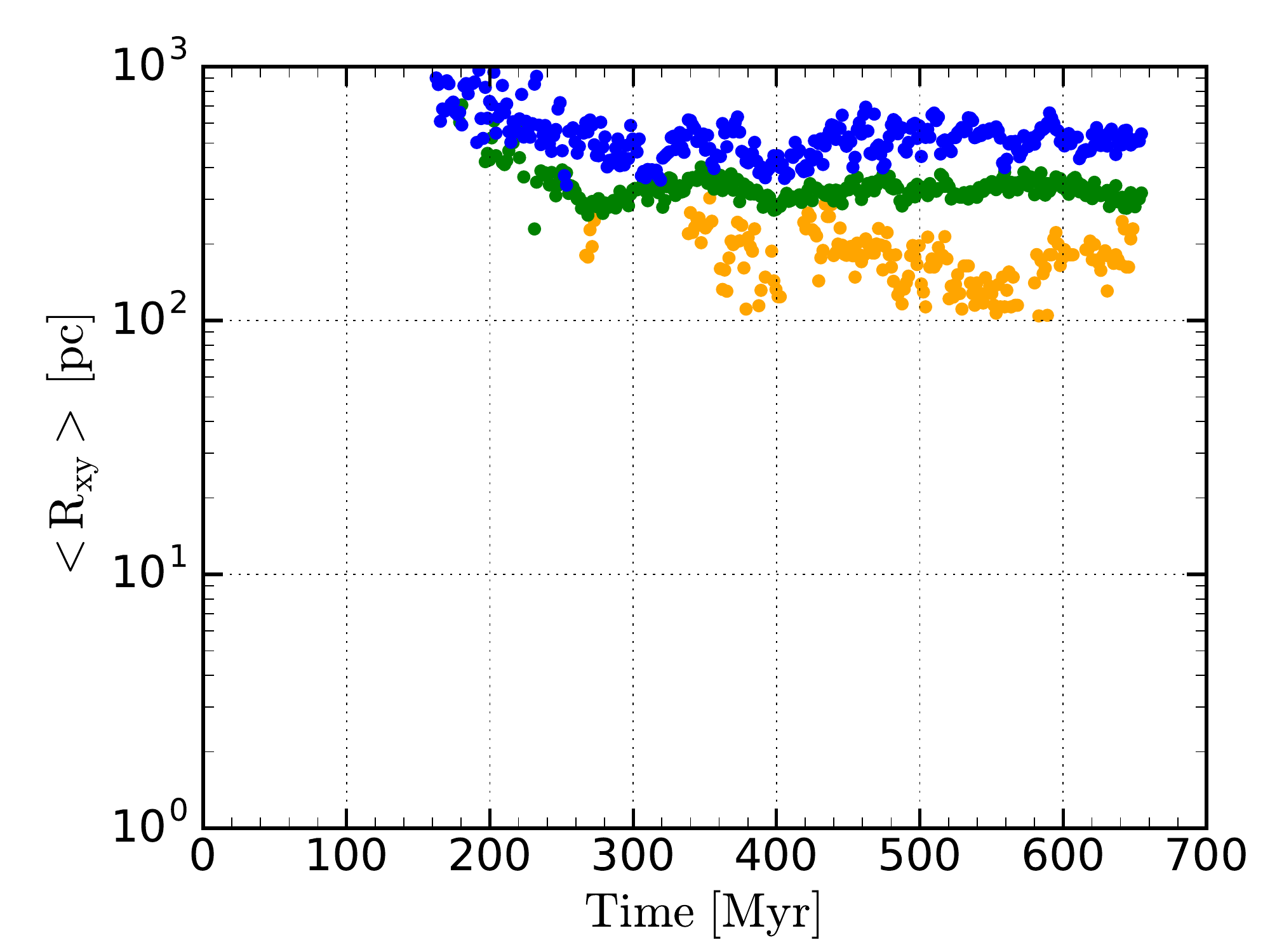}} 
\subfloat[  \label{fig:ulsr_hight_average_evolution}]
  {\includegraphics[width=.33\linewidth]{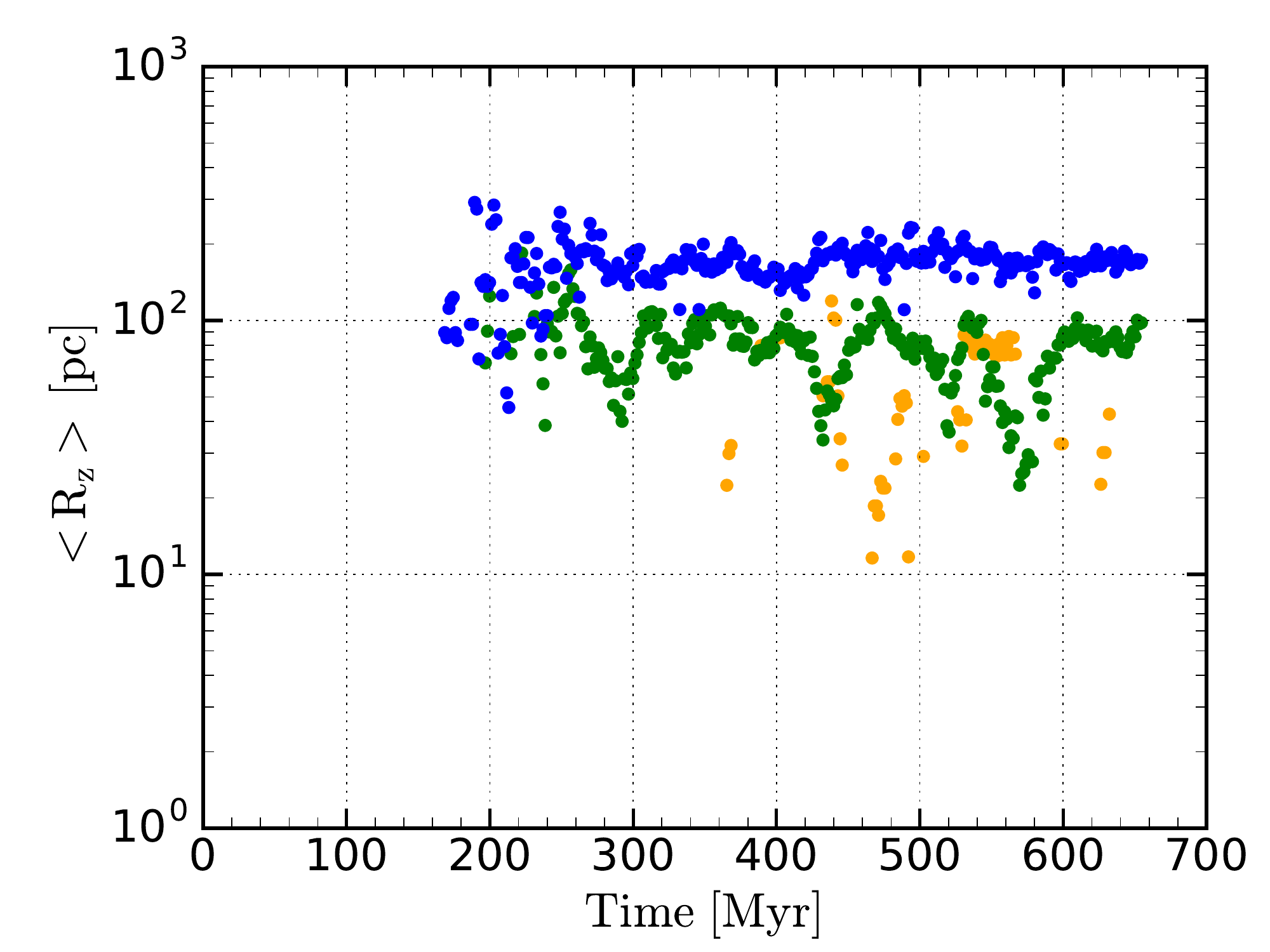}}

\caption{Evolution of the identified clumps and their properties over time ($\Delta t \sim 1.5$ Myr). The first row corresponds to the simulation $LR$ and the second row to the run $ULR$. (a), (d): The total mass within all clumps in black and the colored symbols represent the total mass within each mass bin ]$5 \times 10^{i-1} - 5 \times 10^{i}$] ($\ \mathrm{M_{\odot}}$). The total disc mass is illustrated by the dashed black line. (b), (e): Evolution of the clump radii in the plane $R_{\mathrm{xy}}$ and for (c), (f) in vertical direction  $R_{\mathrm{z}}$.  
\label{fig:LR_number_mass_clumps}}
\end{figure*}

\subsection{Clump definitions}
\label{subsec:Clump definition}
We use two possibilities to measure the clump properties. First, with the 3D density approach we compare the clump characteristics between the different simulations. Second, the surface density images of the galaxy allow to 
    characterise the clump clustering and to compare the identified objects with observations.

\subsubsection{3D density approach}
\label{subsubsec:3D density approach}
The clumps are identified in the simulation on the fly \citep{2014MNRAS.445.4015B, 2015ComAC...2....5B} by considering gas densities above the threshold $n_{\mathrm{H}} \geq 100 \ \mathrm{cm^{-3}}$ for the runs $MS$ and $SR$ which are typical for molecular gas and also potential sites for star formation (at much higher densities). The clumps in the simulations with less resolution not only reach much lower maximum densities, but also experience in general a shift towards lower densities which makes a redefinition of the density threshold unavoidable (see Section 
\ref{subsubsec:Comparison between the different runs}). The clumps are well represented by densities $n_{\mathrm{H}} \geq 10 \ \mathrm{cm^{-3}}$ in run $LR$ and for run $ULR$ for densities $n_{\mathrm{H}} \geq 1 \ \mathrm{cm^{-3}}$ and lead to a similar total mass. Similarly, the density thresholds for the star formation are adapted in cosmological simulations in dependence on resolution. At first the clump finder algorithm divides the density field by a watershed segmentation into regions (sub-structure) associated to a certain density peak. Peaks with a low density contrast to the background are merged to a neighbour through their densest saddle point (noise removal). Finally, the left sub-structure is recursively merged to form larger associated objects based on a threshold.
 We define the clump radius in the xy-plane as
\begin{equation}
\label{def:radius_xy}
	R_{\mathrm{xy}} = \sqrt{ \dfrac{A_{\mathrm{xy}}}{\upi} } = \sqrt{ \dfrac{\Delta x \Delta y}{4} },	
\end{equation}
where $\Delta x /2$ and $\Delta y / 2$ are the maximal semi-axes in the plane of the area
\begin{equation}
\label{def:A_xy}
A_{\mathrm{xy}} = \dfrac{\upi \ \Delta x \Delta y}{4} .
\end{equation}
For the vertical size definition we take the average of the maximal inferred radii in the xz- and yz-plane with
\begin{multline}
\label{def:radius_z}
R_{\mathrm{z}} = \dfrac{R_{\mathrm{xz}} + R_{\mathrm{yz}} }{2} = \dfrac{1}{2} \left( \sqrt{ \dfrac{A_{\mathrm{xz}}}{\upi} } + \sqrt{ \dfrac{A_{\mathrm{yz}}}{\upi} } \right) \\
= \dfrac{1}{4} \left( \sqrt{ \Delta x \Delta z } + \sqrt{ \Delta y \Delta z } \right). 	
\end{multline}

\subsubsection{2D images}
\label{subsubsec:2D images}
2D images are created by observing the surface density line-of-sight and are spatially convolved by a Gaussian kernel of different full width at half maximum (FWHHM = 0.1-2kpc). For the identification of the clumps and clump clusters we apply the blob detection algorithm of the \texttt{scikit-image} library \citep{scikit-image} with the Laplacian of Gaussian approach on the beam smeared images in linear scale (Section \ref{sec:The effects of beam smearing on the main clump properties}). It computes the Laplacian of Gaussian with successively increasing the standard deviation and stacks them up in a cube. The local maxima in the cube are the positions of the objects from where we estimate their FWHM sizes $D_{\mathrm{FWHM}}$ (geometrical mean) and masses $M_{\mathrm{FWHM}}$. After every calculation step of a CC, its region is cut out from the image to avoid double inspection of the mass which would be rarely occurring.  We continuously change the detection threshold, for small convolutions we find a surface density of $\Sigma = 70 \ \mathrm{M_{\odot} \ pc^{-2}}$ to avoid spurious detections and for the largest beam smearing (BS) $\Sigma = 25 \ \mathrm{M_{\odot} \ pc^{-2}}$ to avoid under-detections due to a decreased and flatter surface density.

\section{Simulation results}
\label{sec:Results}
In the following (Section \ref{subsec:General properties}), we compare the general properties of the different simulations and characterize in particular the main simulation $MS$ (Table \ref{tab:Main differences of the simulations}). The detailed clump properties are analysed separately in Section \ref{subsec:Clump properties related to the initial resolution}.

\subsection{General properties}
\label{subsec:General properties}

\subsubsection{Disc evolution}
\label{subsec:General disc evolution}
The gas disc in the main simulation $MS$ (Figure \ref{fig:surface_density_face_on}, left column) fragments into rings ($> 50 $ Myr) from inside-out in agreement with the local fastest growing wavelength, often referred to as Toomre length (see \citealp{2015MNRAS.448.1007B}). Later on, the individual ring-like structures collapse under their self-gravity from kpc scales to dense circular filaments where they subsequently break-up into many bound clumps with diameters of around $d_{\mathrm{clump}} \sim 100$ pc for higher densities (see Section \ref{subsec:Clump properties related to the initial resolution}). Several clumps begin to interact with each other or merge. The disc naturally develops a clumpy and irregular morphology and after $\sim 550$ Myr the total disc has been fragmented. The disc in the simulation with the lower initial resolution $SR$ develops a similar clumpiness while not all the rings fully emerge compared to $MS$. In the low resolution simulations $LR$ and $ULR$ we find marginally developing rings which are not collapsing and the clumps are directly forming with larger sizes. The APF operates for run $ULR$ already initially in the center of the galaxy and changes the initial setup after the first timestep. The consequence is a marginally stable condition which allows for non-axisymmetric instabilities, and spiral-arm like features grow (see also \citealp{Wang:2010wr}). The runs $MS$ and $SR$ have a lot of clumps which are merging or grouping to clusters on several hundred pc to kpc scales \citep{2015MNRAS.448.1007B}. The cluster properties and their evolution will be described in Section \ref{sec:The effects of beam smearing on the main clump properties}. After $t = 655$ Myr for $MS$, the surface density (Figure \ref{fig:disk_sd_440}) still reflects on average the initial exponential profile with local variations. Strong migration to the centre of the galaxy is not apparent, but local accumulation of mass. Furthermore, the rotation curve (Figure  \ref{fig:disk_vrot}) keeps its initial shape on average (mass weighted). 

 \begin{figure*} 
\centering
\subfloat[ \label{fig:mass_shift}]
  {\includegraphics[width=.33\linewidth]{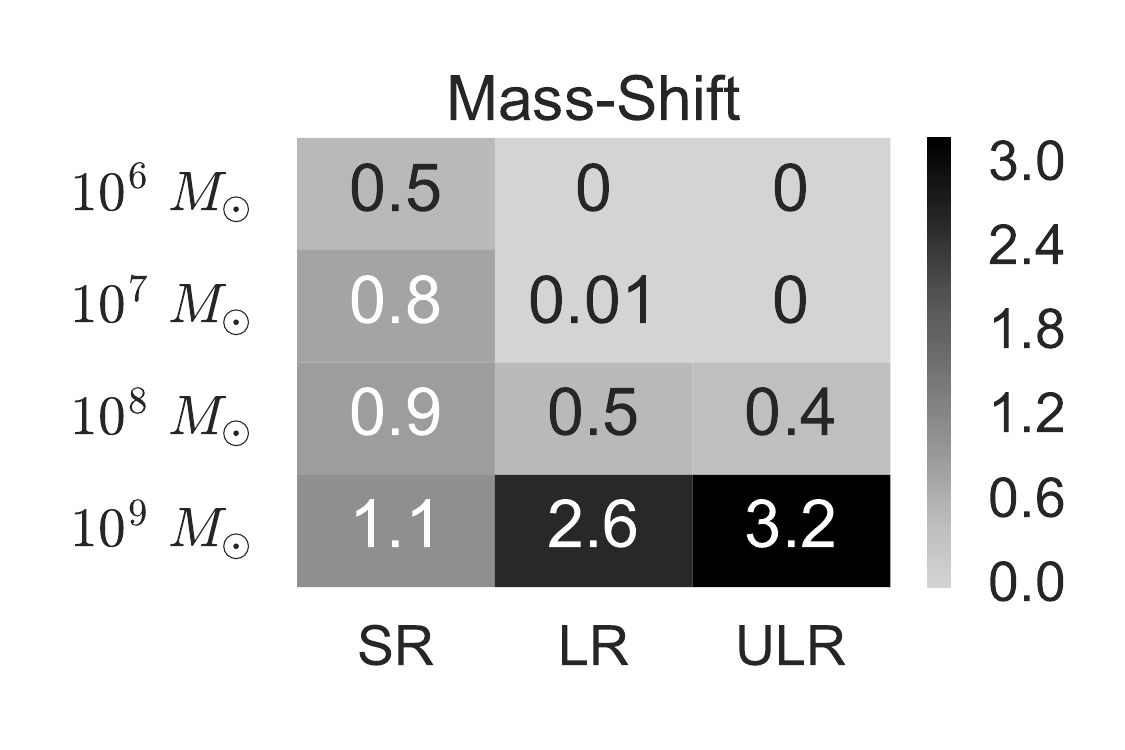}}\hfill
\subfloat[  \label{fig:size_shift}]
  {\includegraphics[width=.33\linewidth]{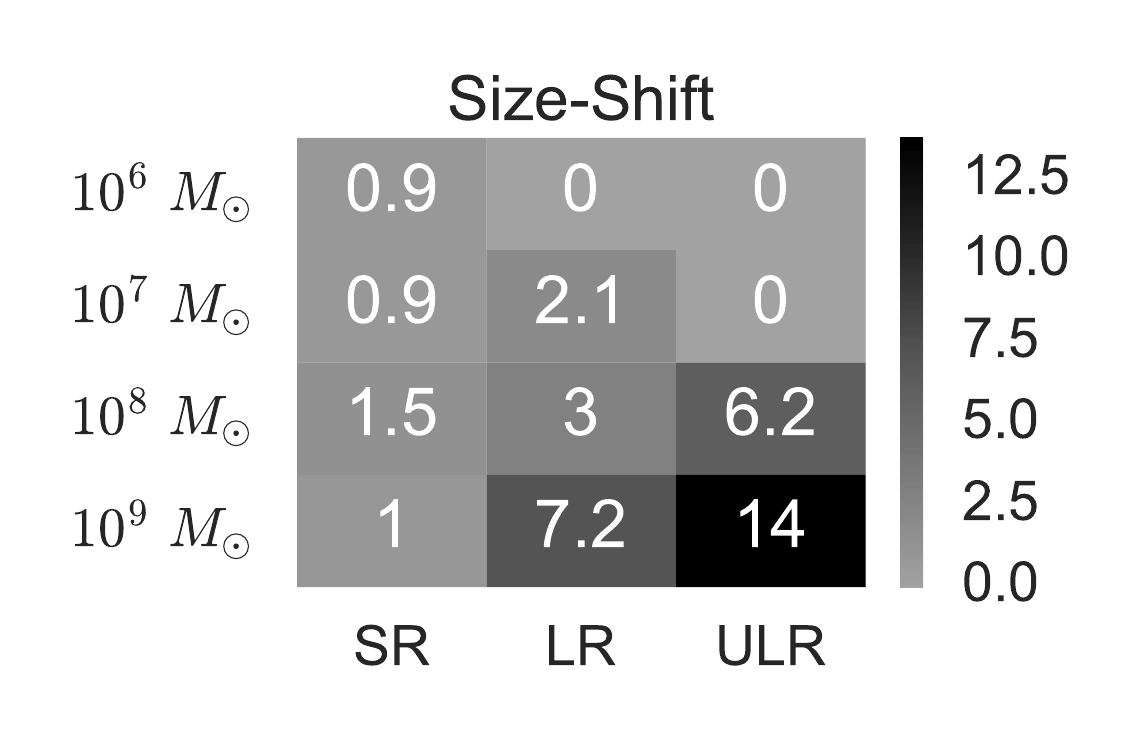}}
\subfloat[  \label{fig:hight_shift}]
  {\includegraphics[width=.33\linewidth]{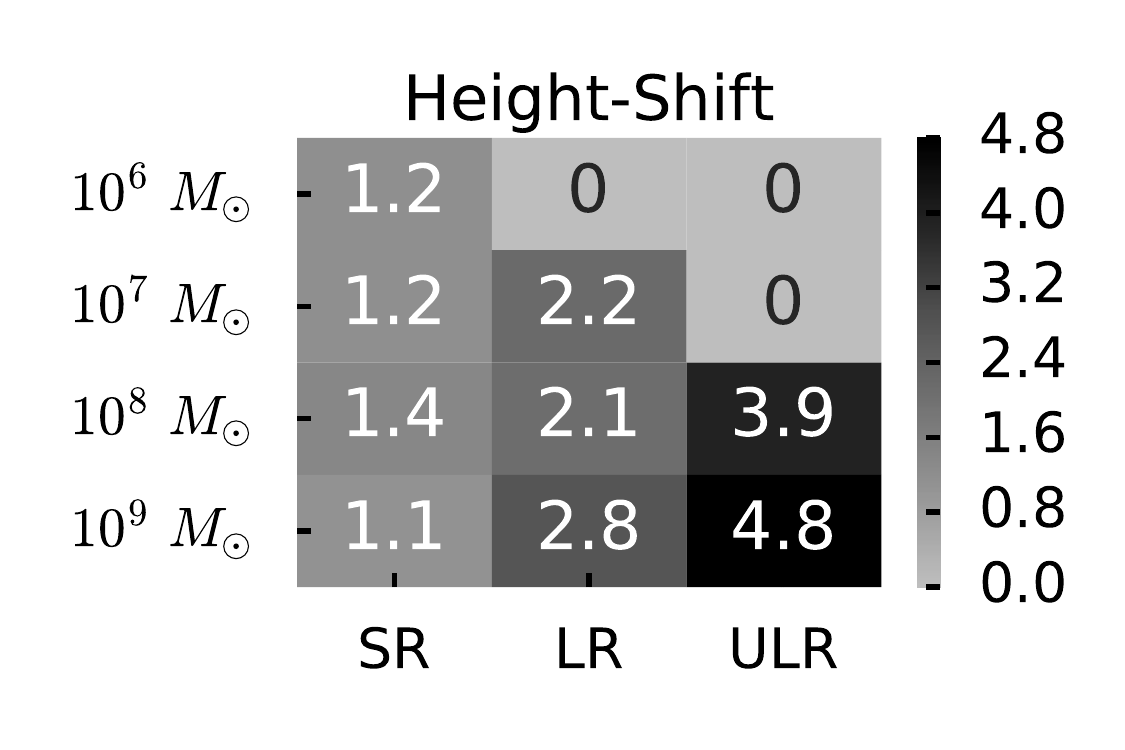}}
  
\caption{ The clump properties of the different runs relative to the main simulation $MS$ for each mass bin at t=655 Myr. Several trends from high to low resolution are present: (a) The mass is shifted towards more massive clumps and the low-mass clumps are suppressed more and more. (b) The clump size is dramatically increasing for most clump masses, (c) the vertical height is getting larger too. 
\label{fig:shifts}}
\end{figure*}

\begin{figure} 
\centering
\includegraphics[width=84mm]{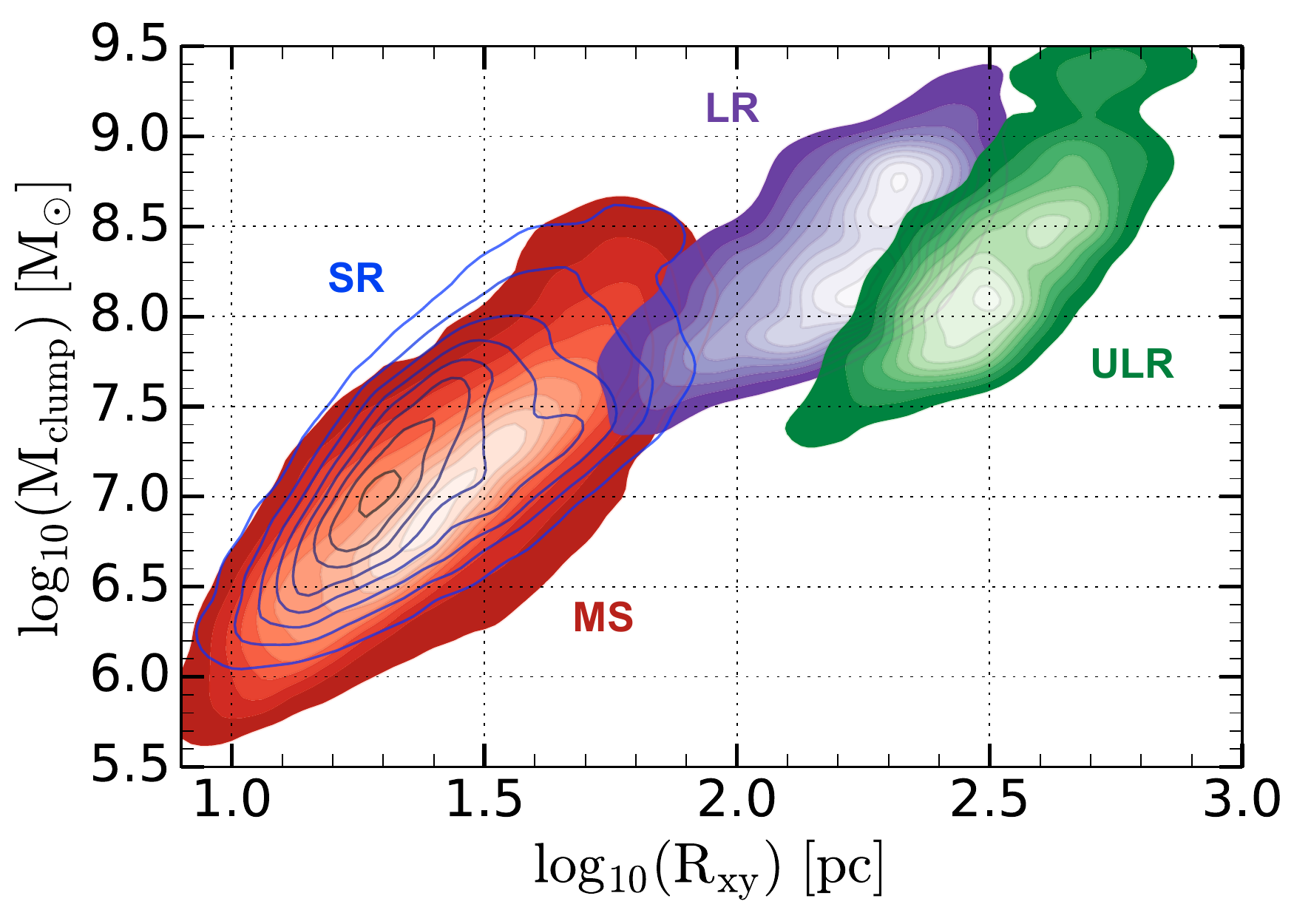}
\caption{Clump mass-radius relation for the different runs averaged over the time 450-655 Myr. The data is represented by a Kernel density estimation and illustrates the abundance of clumps over the time range. The white color in each color-map corresponds to the most frequent clumps and the red shading to run $MS$, the purple to $LR$ and the green to run $ULR$. Run $SR$ is represented by blue closed contour lines while a smaller circumference corresponds to a higher abundance. \label{fig:M-R_relation}}
\end{figure}

\begin{figure} 
\centering
\includegraphics[width=84mm]{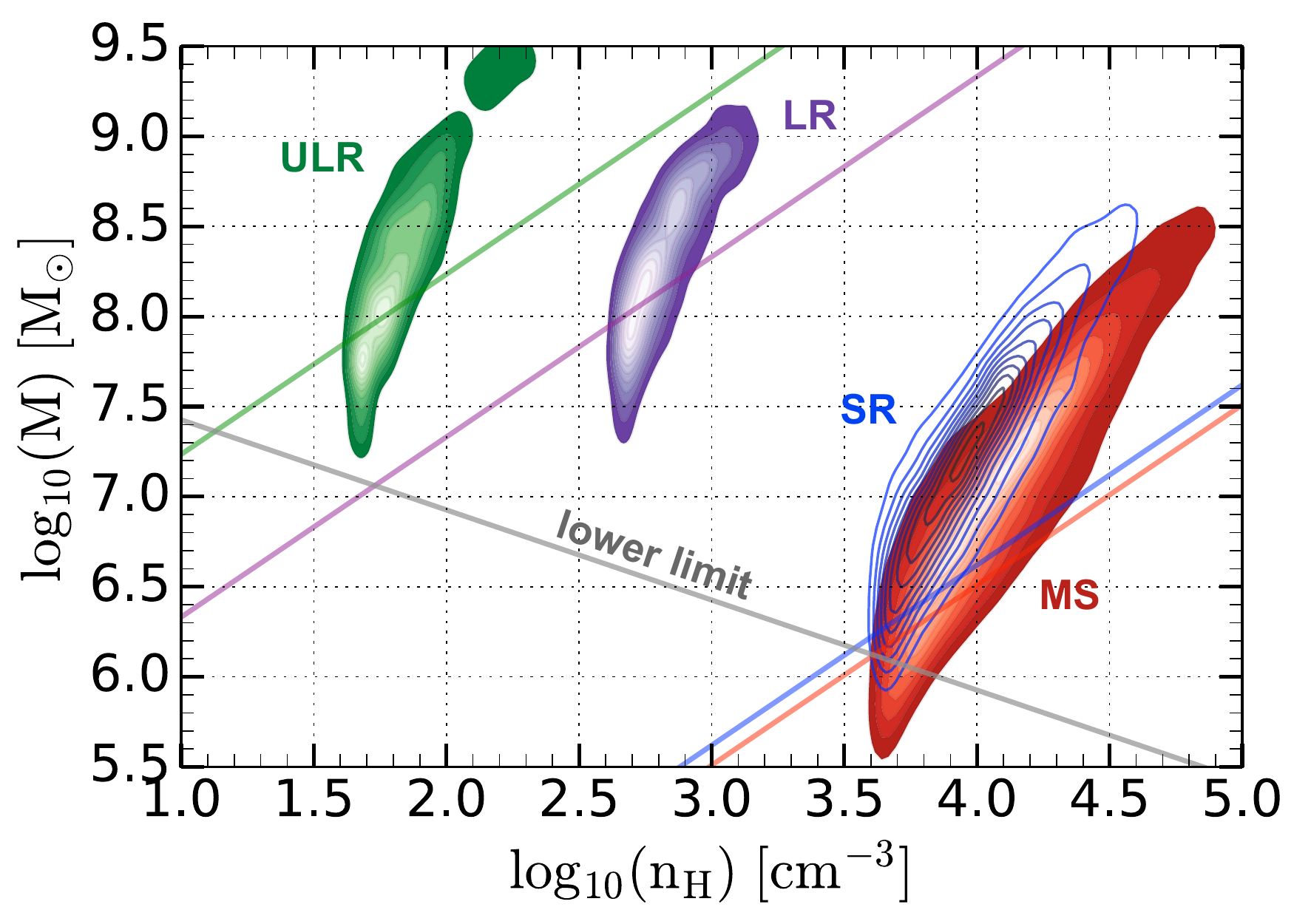}
\caption{The average density and mass of the clumps between 450-655 Myr of the different runs are represented by a kernel density estimation (colored shadings and contour lines as in Figure \ref{fig:M-R_relation}. The Jeans mass and their corresponding average densities expected from the artificial pressure floor of the different simulations are represented by the colored lines. The areas above and on the left side of the individual lines are possible clump properties. The grey line gives the lower limit for Jeans masses given by the minimum sound speed $c_{\mathrm{s}} \sim 10 \ \mathrm{km \ s^{-1}}$ and is only reached by the smallest clumps in run $MS$.
\label{fig:JeansMass_Density}}
\end{figure}

\subsubsection{Mass redistribution}
\label{subsec:Mass redistribution}
Despite the different evolution of the fragments at different resolution, we find similarities for the total mass fraction of the different structures in the galaxy. For the following analysis we distinguish between four surface density regimes within the disc (see Table \ref{tab:Surface density regions}). The initially limited surface density range is redistributed to values over several orders of magnitudes $\Sigma \sim 10^0 - 10^6 \ \mathrm{M_{\odot} \ pc^{-2}}$ at $t = 655$ Myr. \textcolor{orange}{\textbf{Regime $C$}} defines densities which roughly describe the clump areas with $\Sigma > 10^{2} \ \mathrm{M_{\odot} \ pc^{-2}}$ and represents the majority of the mass ($76 \%$ for $MS$) for the evolved disc (Figure \ref{fig:mass_factions}) and in the beginning already for the central disc region. At first the disc consists of lower densities in the range  $\Sigma = 8 \times 10^{0}- 2 \times 10^{2} \ \mathrm{M_{\odot} \ pc^{-2}}$ which in the evolved disc characterises the gas surrounding the clumps in the elongated, \textit{arm}-like features (\textcolor{ForestGreen}{\textbf{regime $A$}}) with a mass fraction of $18 \%$, since the main mass is with time transferred to  higher densities. The \textit{low} density gas (\textcolor{blue}{\textbf{regime $L$}}) with $\Sigma < 10 \ \mathrm{M_{\odot} \ pc^{-2}}$ represents the smallest fraction of the mass ($2-6 \%$) and does not change significantly over time. In the beginning it resides at the outer edge of the galaxy and in the evolved disc it is also found between the main gravitationally bound structures, distributed over the total disc, but less common in the central region of the galaxy. After the disc is fully fragmented (after $\sim 550$ Myr), the mass fractions of the different regimes are quasi-steady for the rest of the simulation (Figure \ref{fig:mass_factions}). Within the resolution study, we find significantly lower mass fractions for the low resolution runs $LR$ and $ULR$ for the highest densities within the clumps $\Sigma > 10^{3} \ \mathrm{M_{\odot} \ pc^{-2}}$ (\textcolor{black}{\textbf{Regime $H$}}). All other regimes show similar mass fractions for the full resolution study.

\begin{figure*}
\centering
\subfloat[  \label{fig:gascontour_L10}]
  {\includegraphics[width=.49\linewidth]{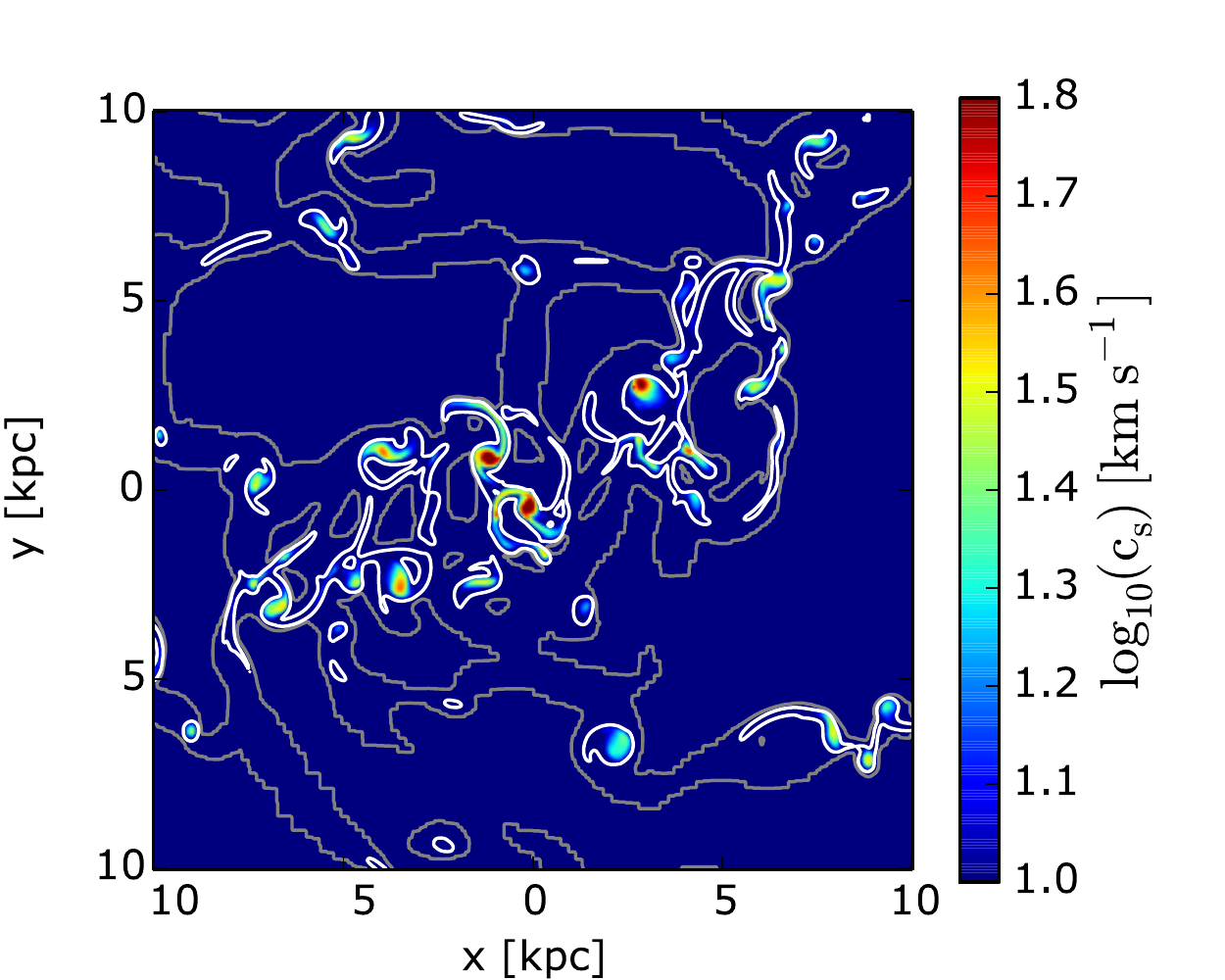}}
\subfloat[  \label{fig:gascontour_L9}]
  {\includegraphics[width=.49\linewidth]{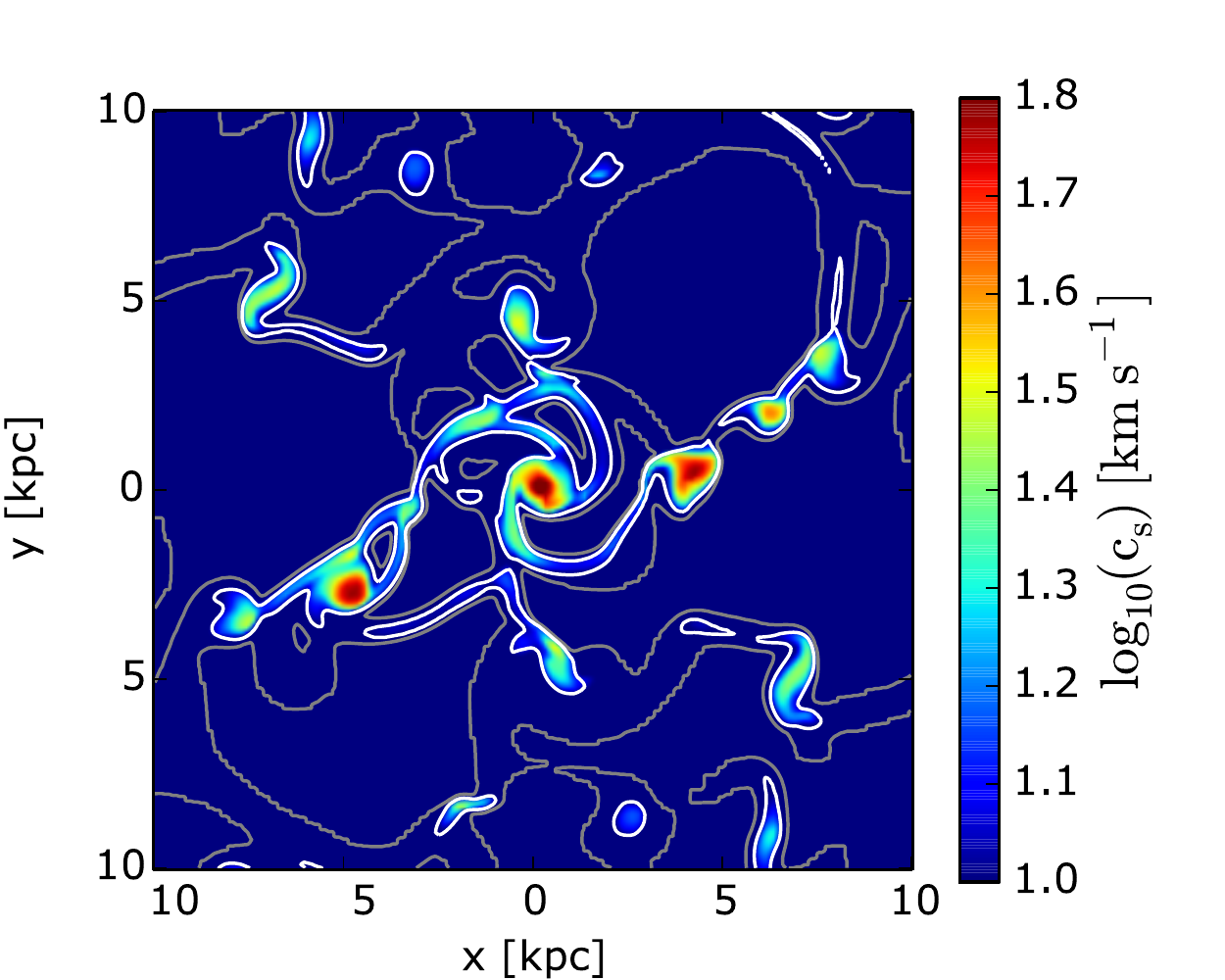}} \hfill
\caption{The average mass weighted sound speed of the face-on discs for the run $LR$ (left) and the run $ULR$ (right) at 600 Myr (see Figure \ref{fig:surface_density_face_on} for the corresponding surface density plot). Very high sound speeds are present within the clump regimes, indicated by the white contours at 100 $\mathrm{M_{\odot} \ pc^{-2}}$. The grey contours correspond to 10 $\mathrm{M_{\odot} \ pc^{-2}}$ and show the arm like features.
\label{fig:cs_lowres} }
\end{figure*}

\subsection{Clump statistics}
\label{subsec:Clump properties related to the initial resolution}

\subsubsection{General clump evolution of $MS$}
\label{subsec:General clump definition}
The overall clump evolution of $MS$ is shown in Figure \ref{fig:clump_in_disk_evolution} with their radial positions in the disc and their individual masses for timesteps $\Delta t \sim 1.5 \ \mathrm{Myr}$. A large number of clumps fragment in the dense circular filament-like structures (11 rings, the outermost at $\sim 8$ kpc). Their typical initial masses are between $\sim 10^6 - 5 \times 10^7  \ \mathrm{M_{\odot}}$. Very early they gravitationally interact with structures formed further in and further out. They are locally attracted by their next neighbours which leads for half of the clumps within one ring to a movement towards the galaxy centre and the other half towards larger radii, by conserving the angular momentum. This oscillating behaviour continuous and the clumps move steadily inwards and outwards, which induces a strong mixing. Many merge with each other, to e.g. $10^8 \ \mathrm{M_{\odot}}$ within 50 Myr and only a few above $10^9 \ \mathrm{M_{\odot}}$ are visible on the map (mainly at small disc radii). When following the path of the most massive clumps over time, it can be seen that they are temporarily present and some are losing material when crossing the path of fast encounters (e.g. at $R_{\mathrm{disc}} \sim 4 \ \mathrm{kpc}$, $\sim 450 \ \mathrm{Myr}$) but then again accumulating mass. 
\subsubsection{Detailed clump properties of $MS$}
\label{subsec:Detailed clump properties}
For the following analysis we consider the mass bins:
\begin{center}
$\sim 10^6 \ \mathrm{M_{\odot}} \equiv \mathrm{]5 \times \ 10^5 : 5 \times 10^6] \ M_{\odot} }$,\\
$\sim 10^7 \ \mathrm{M_{\odot}} \equiv \mathrm{]5 \times \ 10^6 : 5 \times 10^7] \ M_{\odot} }$,\\
 $\sim 10^8 \ \mathrm{M_{\odot}} \equiv \mathrm{]5 \times \ 10^7 : 5 \times 10^8] \ M_{\odot} }$, \\
 $\sim 10^9 \ \mathrm{M_{\odot}} \equiv \mathrm{]5 \times \ 10^8 : 5 \times 10^9] \ M_{\odot} }$.\\
\end{center}
We recall (Figure \ref{fig:clump_in_disk_evolution}) that the initial clumps which form within the rings have $\sim 10^6-10^7 \mathrm{M_{\odot}}$. The relatively steady amount of clumps in the mass bin of $10^7 \ \mathrm{M_{\odot}}$ is in balance between newly forming  $\sim 10^7 \ \mathrm{M_{\odot}}$ clumps and merging to $\sim 10^8 \ \mathrm{M_{\odot}}$ clumps. The more massive clumps form later since it needs, e.g. 100 clumps with $10^7 \ \mathrm{M_{\odot}}$ to merge to $10^9 \ \mathrm{M_{\odot}}$ but are more frequent in the central disc region due to a larger volume filling-factor. All mass bins stay quasi-steady after the disc is fully fragmented ($>$ 550 Myr) and most of the mass resides in clumps of $\sim 10^8 \ \mathrm{M_{\odot}}$ with a total of $\sim  10^{10} \ \mathrm{M_{\odot}}$ and are distributed over the whole disc (Figure \ref{fig:clump_in_disk_evolution}). The clumps in the mass bin $10^6 \ \mathrm{M_{\odot}}$ play with a total mass $\sim 3 \times 10^8 \ \mathrm{M_{\odot}}$ only a minor role compared to the total mass of clumps above $> 10^7 \ \mathrm{M_{\odot}}$. The clump masses are related to different sizes, ranging between $R_{\mathrm{xy}} \simeq 20-60$ pc (Figure \ref{fig:size_average_evolution}). The clumps have a disc-like shape and lie initially in the xy-plane and are well represented by the radius definition in Section \ref{subsec:Clump definition}. Interactions between the clumps, especially in the central disc region, lead to tilting with respect to the  equatorial plane which explains the slow decrease of the mean radius $R_{\mathrm{xy}}$ over time and the small increase of the measured mean vertical sizes $R_{\mathrm{z}}$ (Figure.\ref{fig:hight_average_evolution}). This is especially the case for the massive merger-products, e.g. the tilted $10^9 \ \mathrm{M_{\odot}}$ clumps, where the size definition does not work anymore (measured in xy-plane of the galaxy) and they appear with similar xy-sizes as the $10^8 \ M_{\odot}$ clumps. The minimum thickness in the very center of the clumps is given by the artificial pressure floor with $R_{\mathrm{z}} \sim 10 \ \mathrm{pc}$ which is for the diameter equal to the given minimum Jeans length (see Section \ref{sec:Simulation Methods}, \ref{results_subsec:Intrinsic clump profiles}). This can be especially seen in the beginning of the statistics for $R_{\mathrm{z}}$. The diameters of the clumps with $10^6 \ \mathrm{M_{\odot}} $ are close to the minimum Jeans length (factor $\sim 1.7$ difference), while the clumps with $10^7 \ \mathrm{M_{\odot}}$ are larger by a factor of $\sim 3$ and the mergers with $10^8 \ \mathrm{M_{\odot}}$ are even $\sim 5$ times larger and therefore better resolved.

\begin{figure*}
\centering
\subfloat[ \label{fig:440_tstep_rho_cl101}]
  {\includegraphics[width=0.33\linewidth]{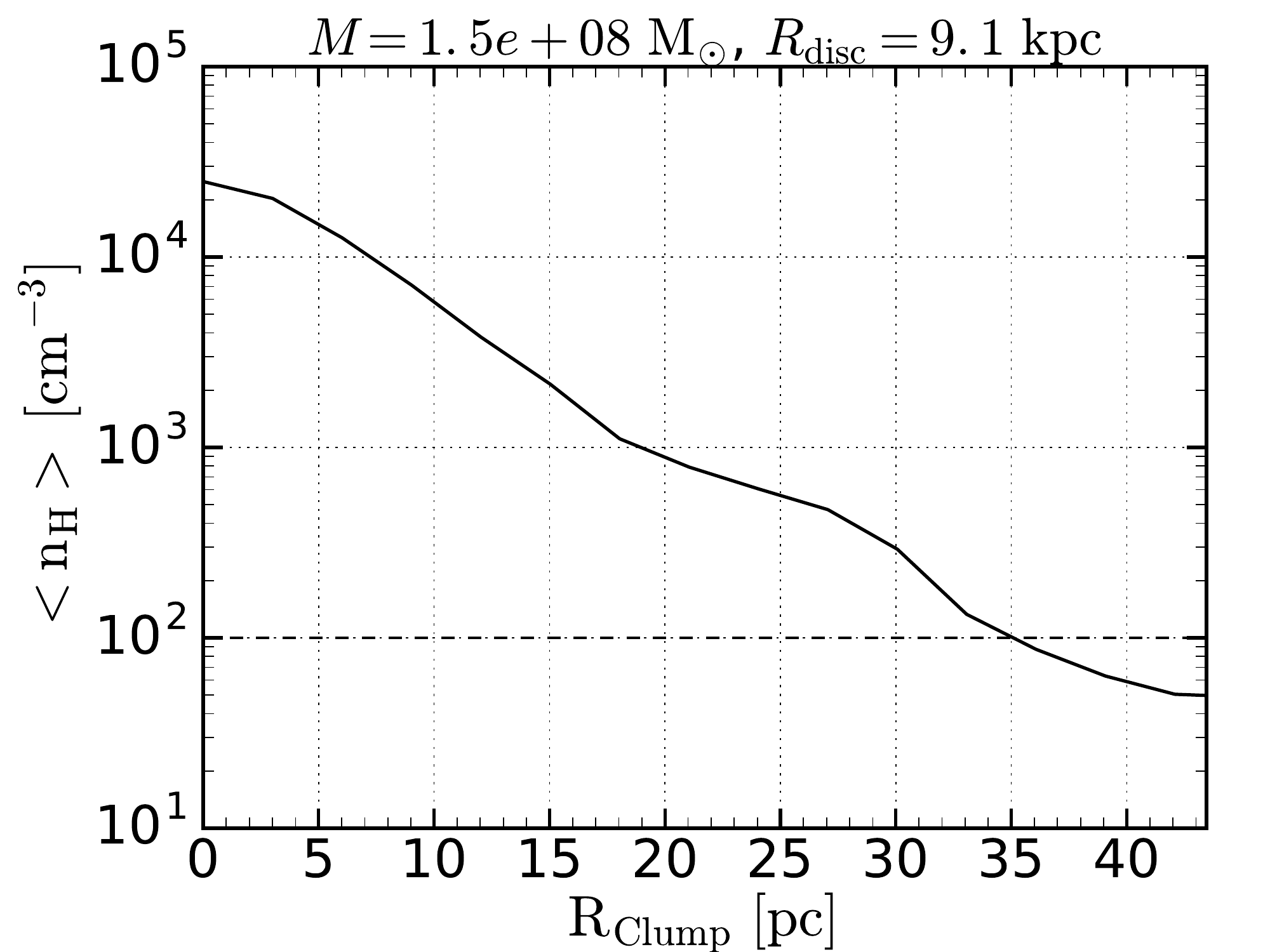}}
\subfloat[ \label{fig:440_tstep_vrot_cl101}]
  {\includegraphics[width=0.33\linewidth]{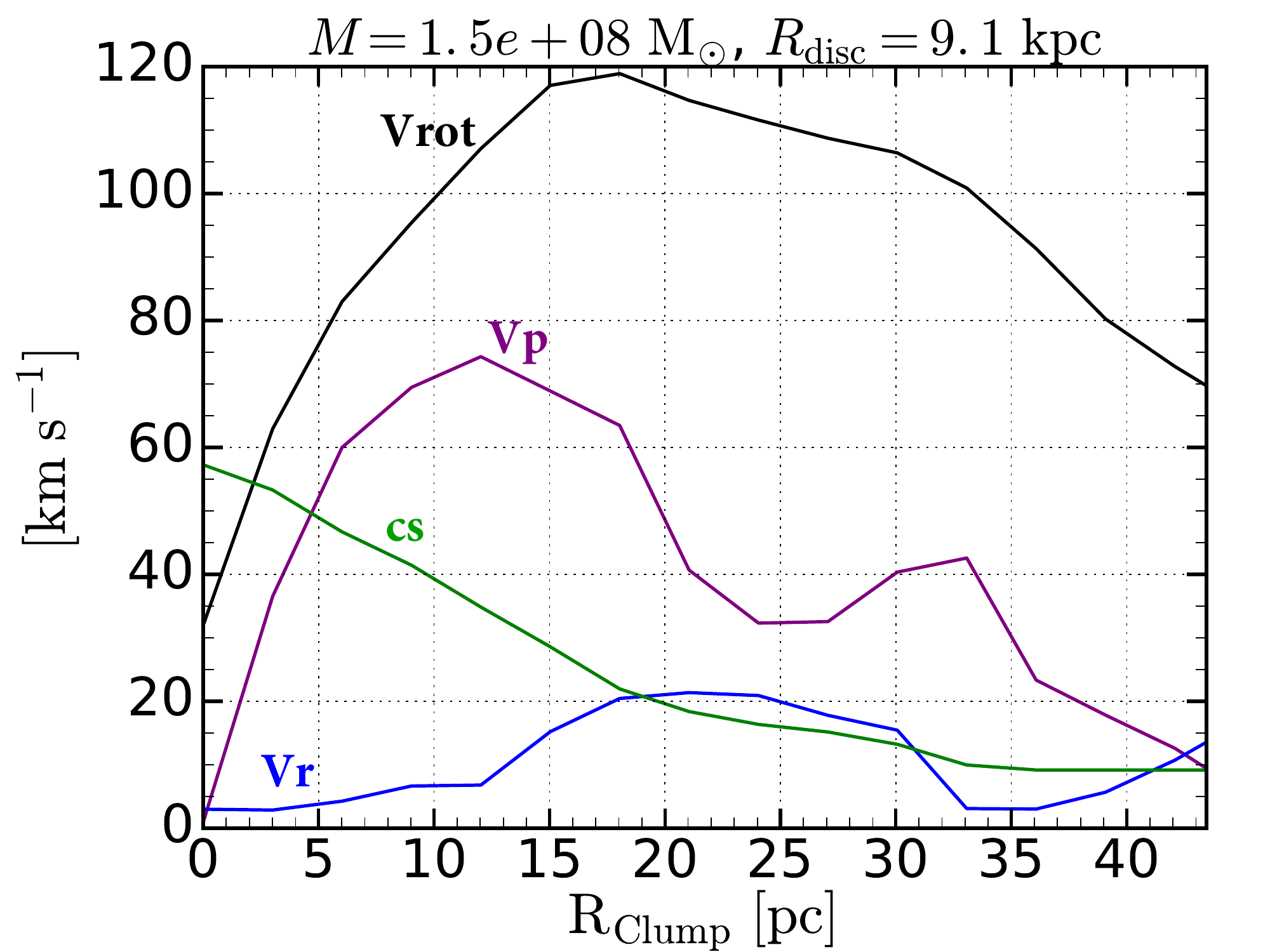}}
\subfloat[ \label{fig:440_tstep_ratio_vrot_cs_cl101}]
  {\includegraphics[width=0.33\linewidth]{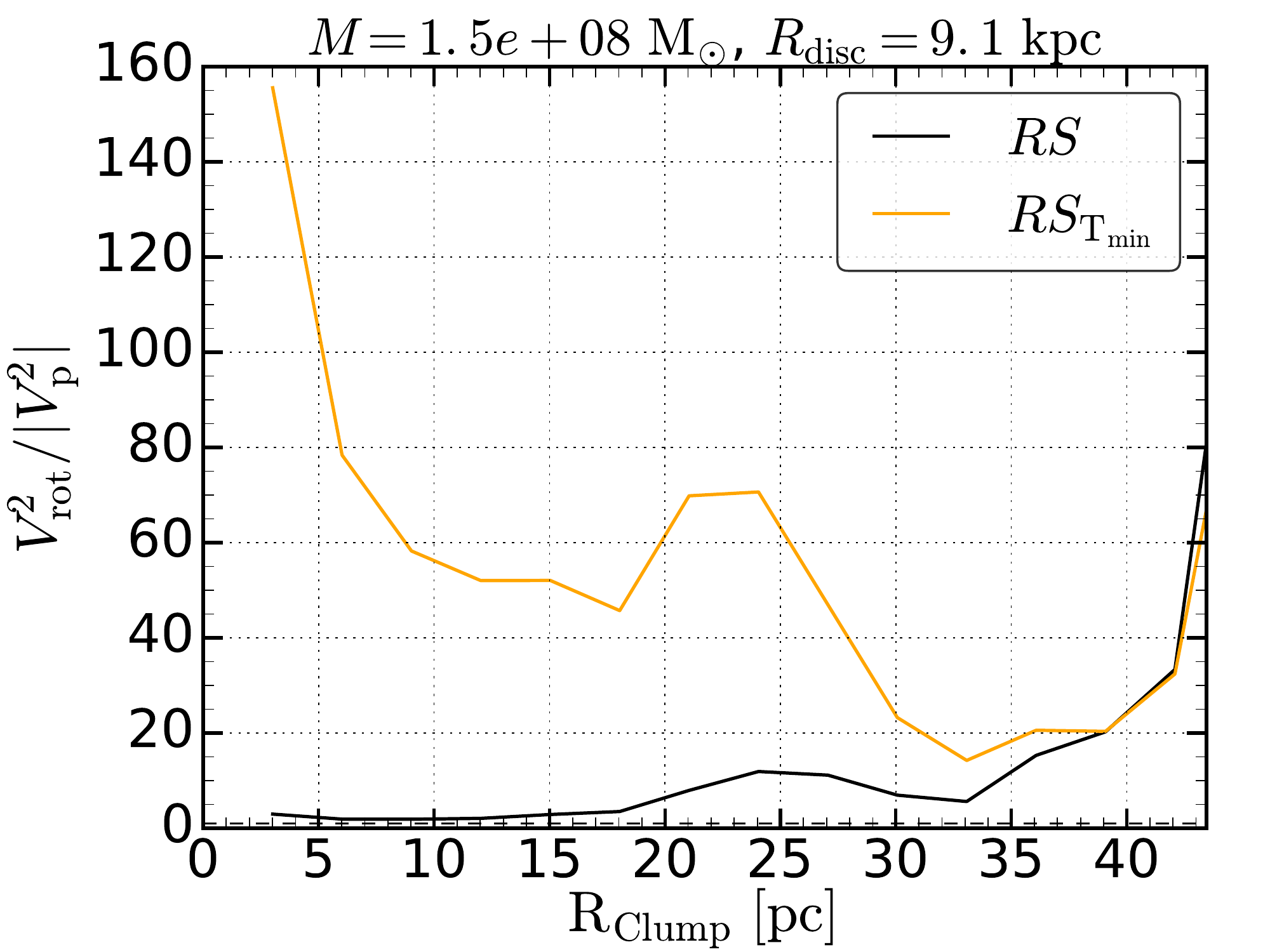}} \hfill

\subfloat[ \label{fig:440_tstep_rho_cl132}]
  {\includegraphics[width=0.33\linewidth]{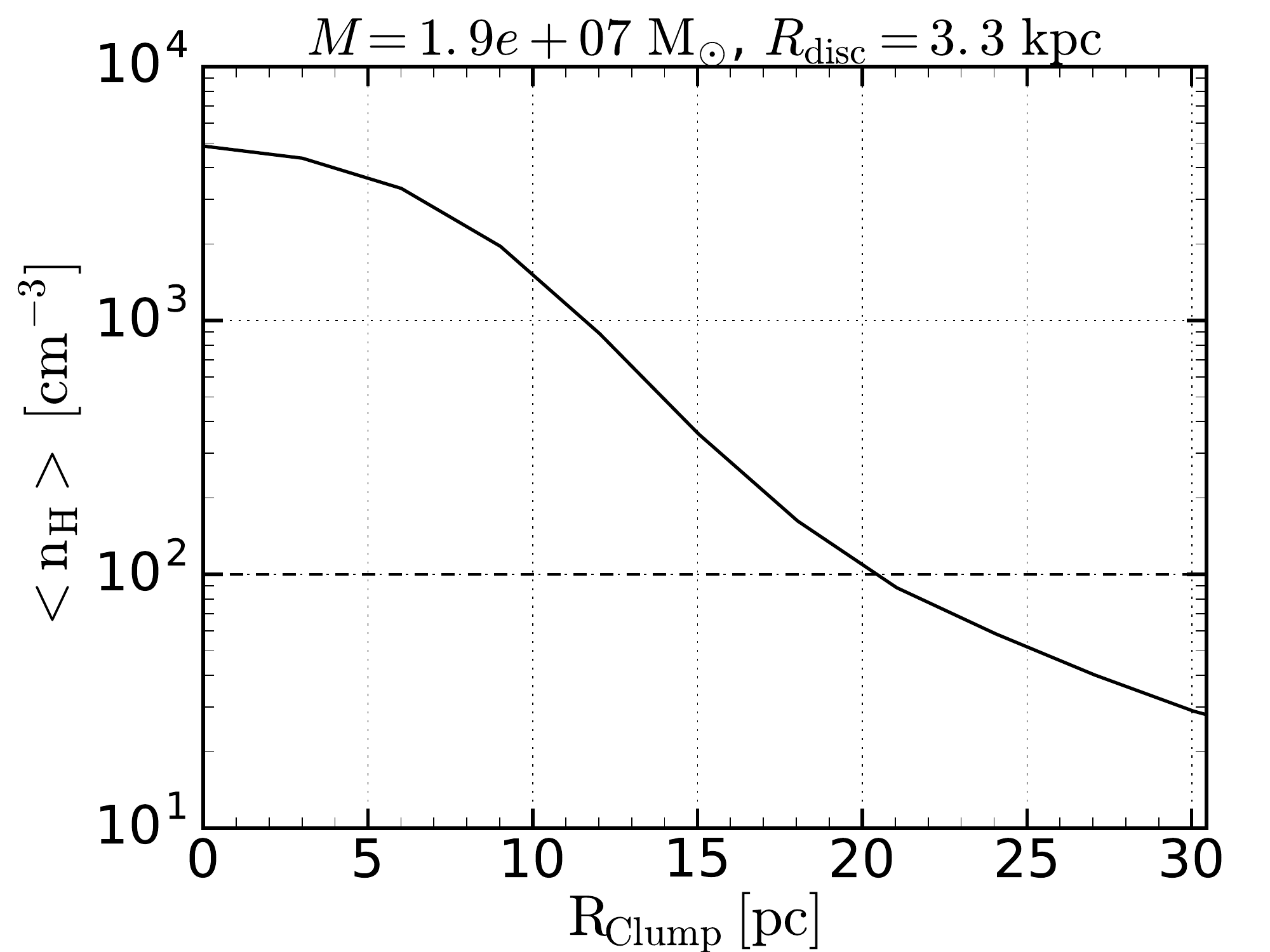}}
\subfloat[ \label{fig:440_tstep_vrot_cl132}]
  {\includegraphics[width=0.33\linewidth]{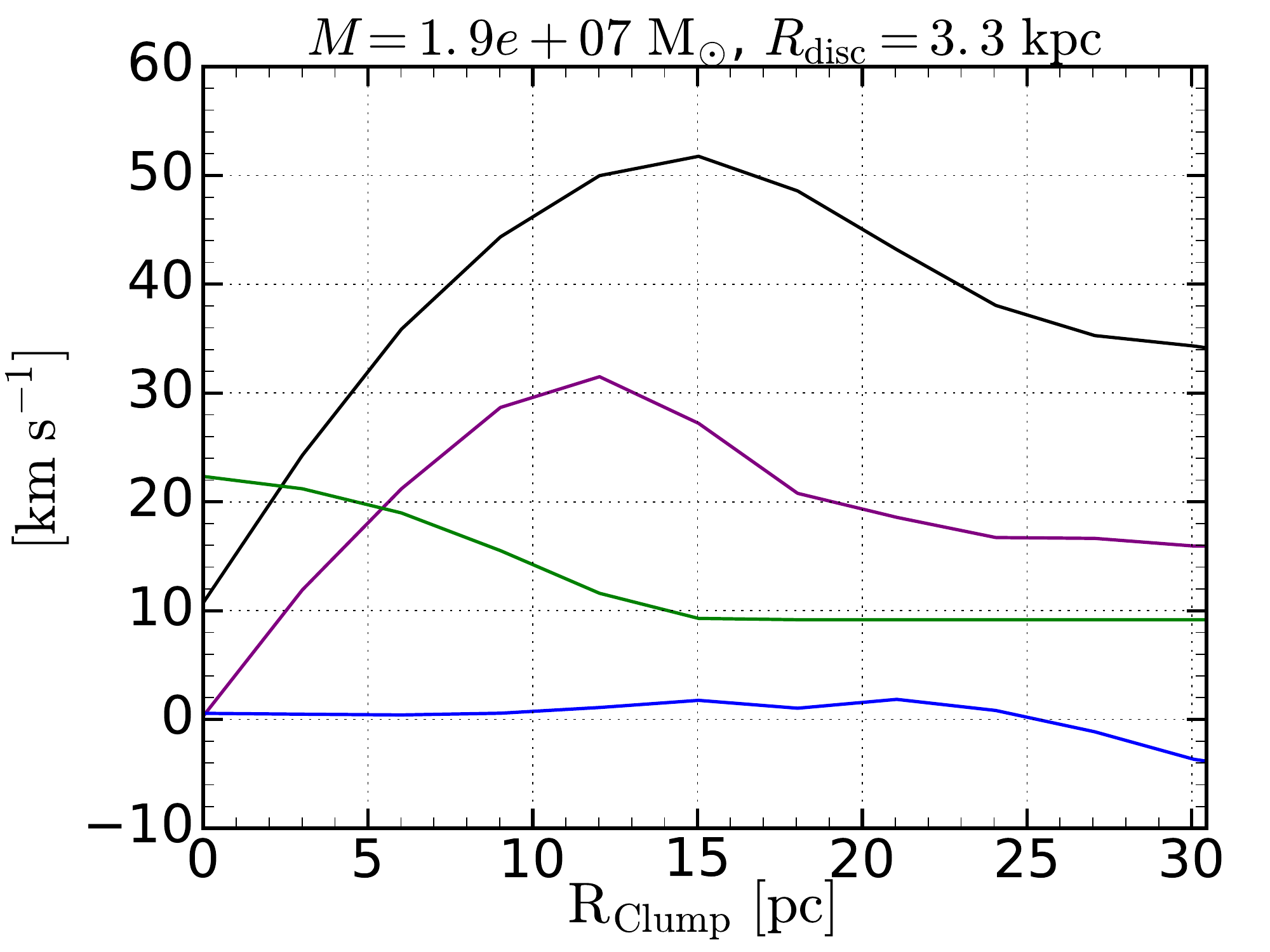}}
\subfloat[ \label{fig:440_tstep_ratio_vrot_cs_cl132}]
  {\includegraphics[width=0.33\linewidth]{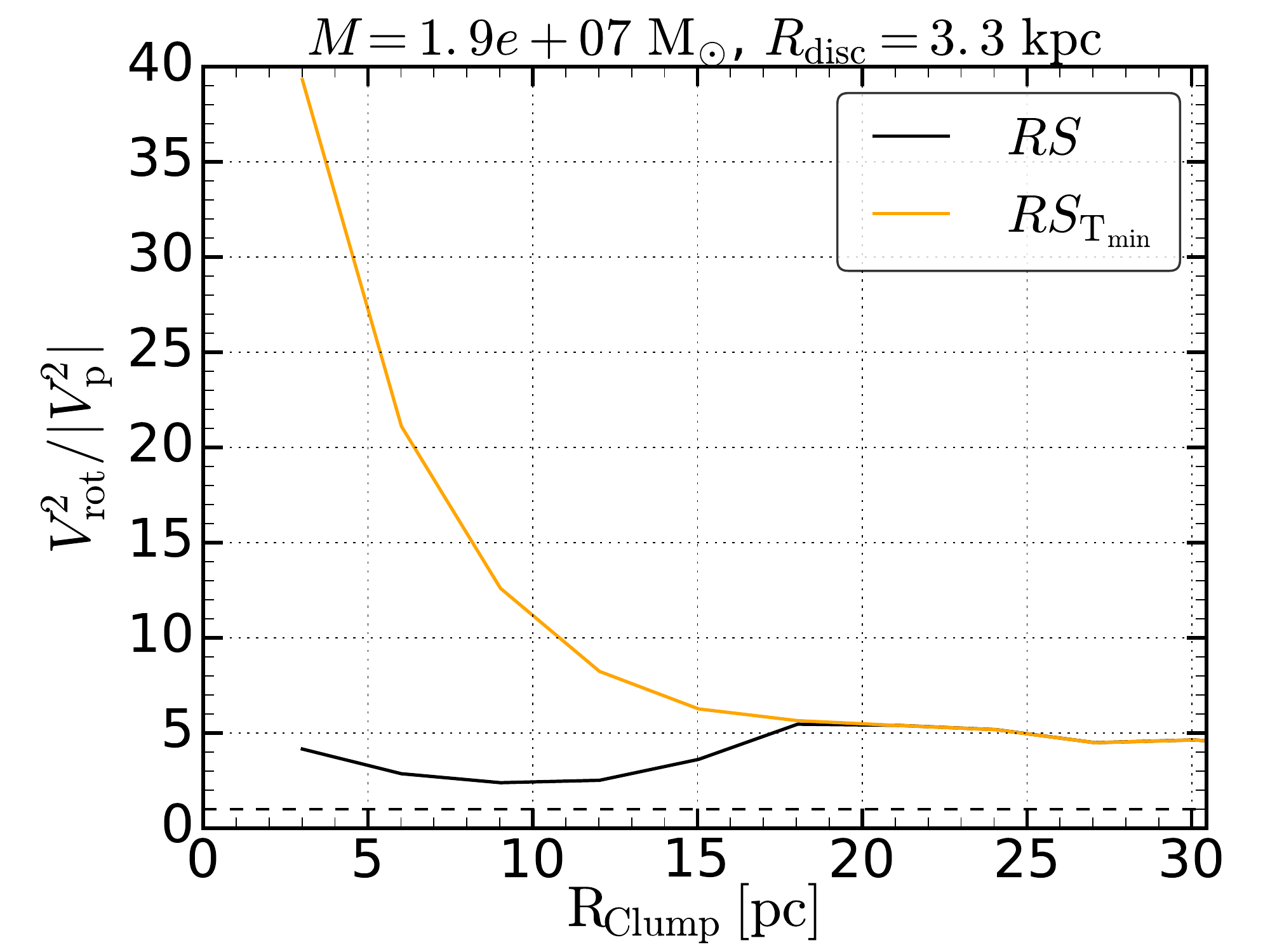}} \hfill  

\subfloat[ \label{fig:440_tstep_rho_cl38}]
  {\includegraphics[width=0.33\linewidth]{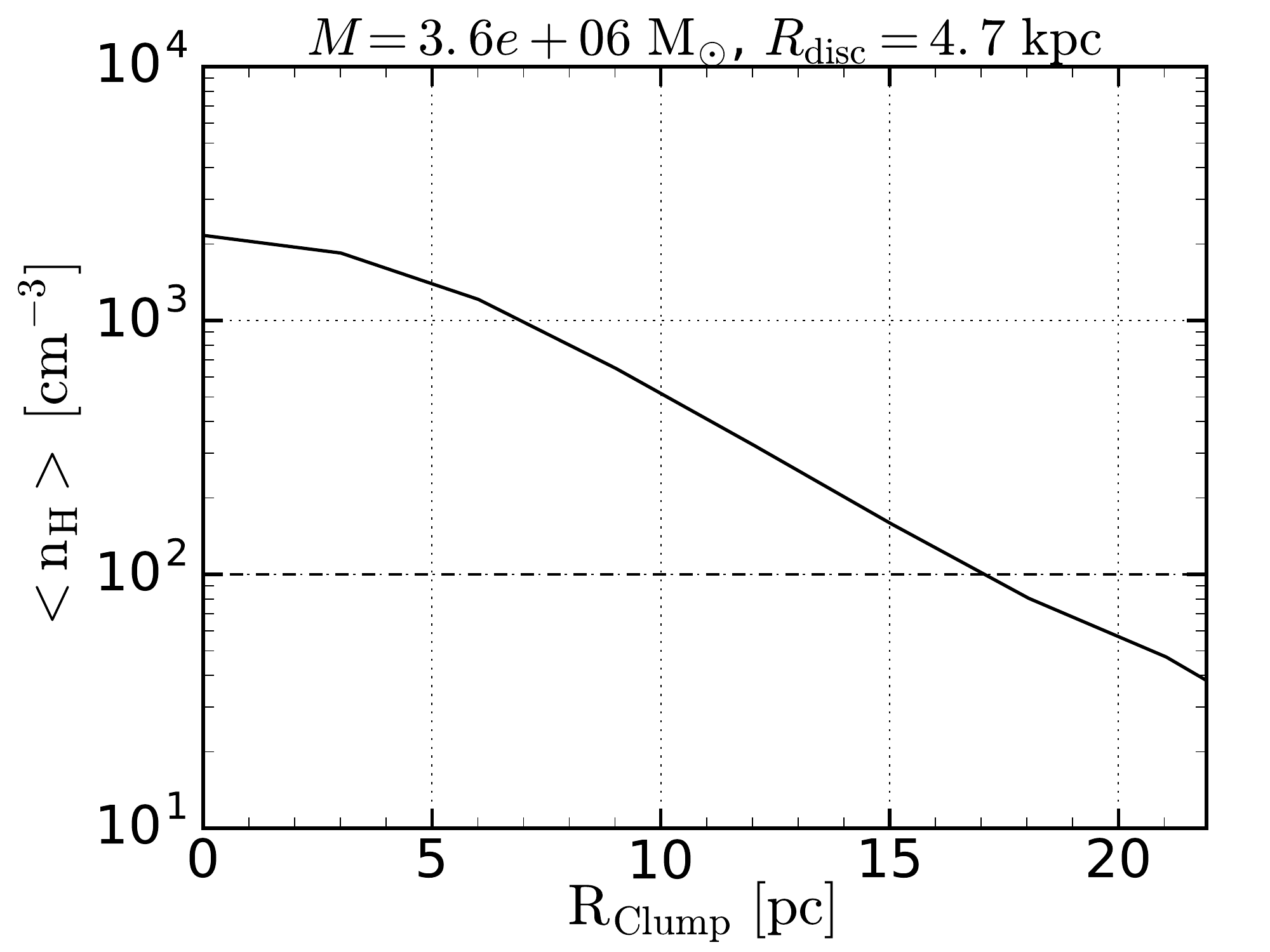}}
\subfloat[ \label{fig:440_tstep_vrot_cl38}]
  {\includegraphics[width=0.33\linewidth]{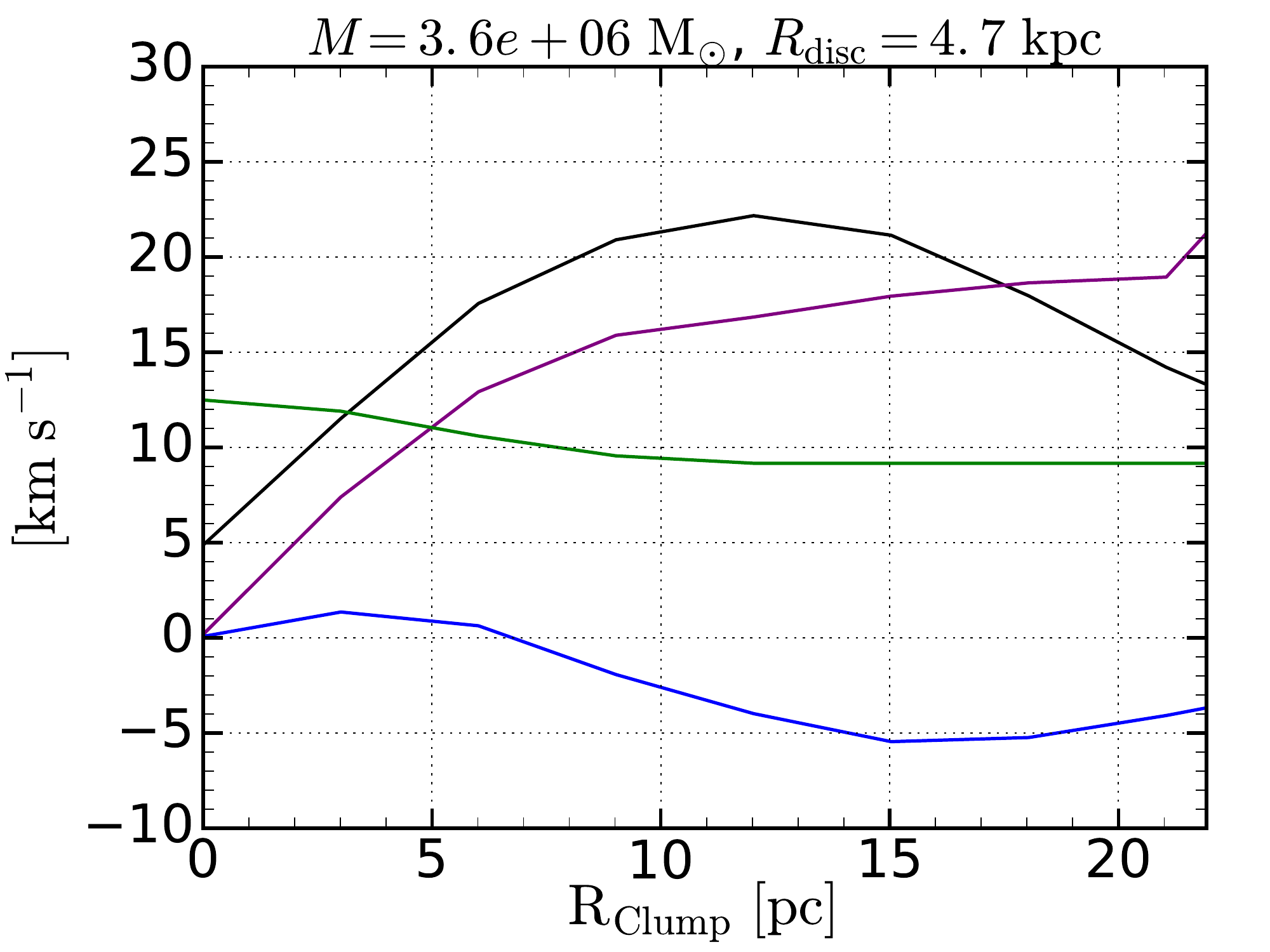}}
\subfloat[ \label{fig:440_tstep_ratio_vrot_cs_cl38}]
  {\includegraphics[width=0.33\linewidth]{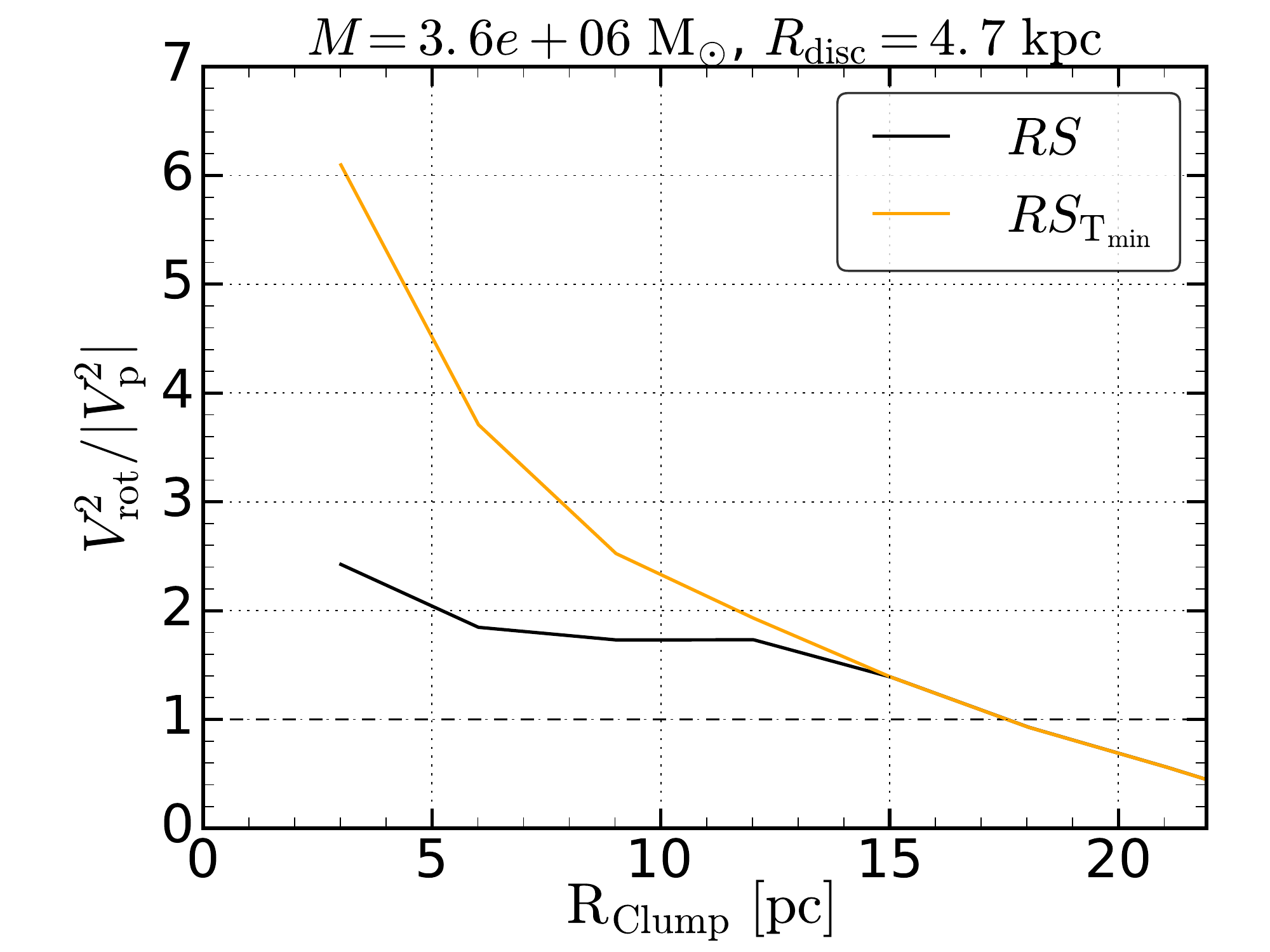}} \hfill

\caption{Examples of radial profiles of three typical clumps (t = 655 Myr) in the main simulation $MS$. The first row corresponds to the clump $V1$ with a mass of $\sim 2 \times 10^8 \ \mathrm{M_{\odot}}$, the second row to $V2$ with $\sim 2 \times 10^7 \ \mathrm{M_{\odot}}$ and the last row to clump $V3$ with $\sim 4 \times 10^6 \ \mathrm{M_{\odot}}$. (a), (d), (g) show the density profile; (b), (e), (h) give the kinematic profiles of the rotational velocity $v_{\mathrm{rot}}$, the gradient of the thermal pressure $v_{\mathrm{p}}$ (Equation \ref{eq:Jeans equation}) , radial velocity $v_{\mathrm{r}}$ and the sound speed $c_{\mathrm{s}}$; (c), (f) (i) show the ratio between the measured rotation and the effective velocity dispersion (black line) and the orange line the ratio between the measured rotation and the minimum sound speed of $c_{\mathrm{s}} \sim 10 \ \mathrm{km \ s^{-1}}$. \label{fig:clump_profile_01}}
\end{figure*}

\begin{figure*} 
\centering
\includegraphics[width=84mm]{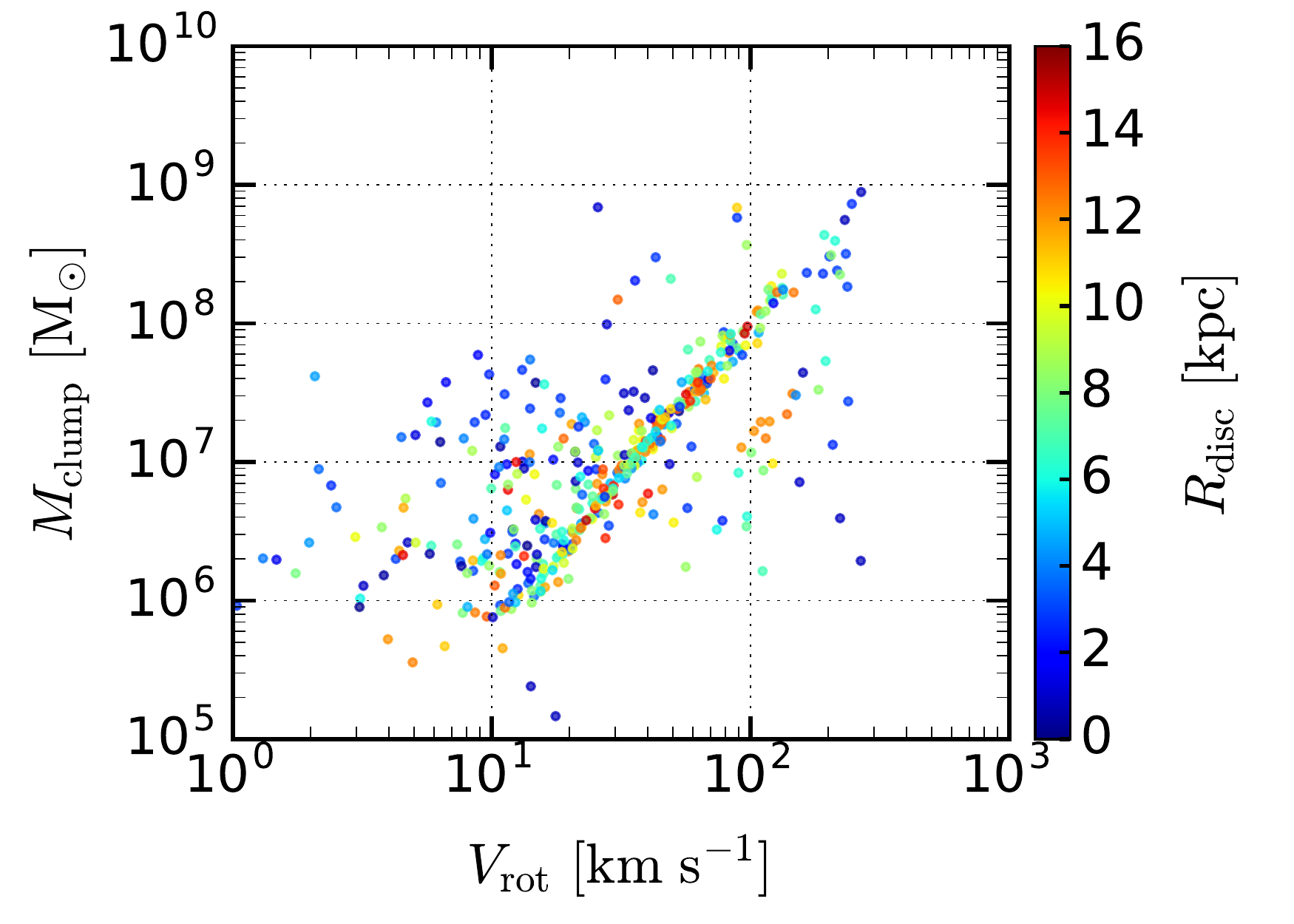}
\caption{The comparison between the clump masses and their maximum rotation velocities at t=655 Myr. Their radial positions within the disc are indicated by the color. \label{fig:mass_radius_color_coded}}
\end{figure*}

\begin{figure*}
\centering
\subfloat[ \label{fig:rot_support}]
  {\includegraphics[width=.33\linewidth]{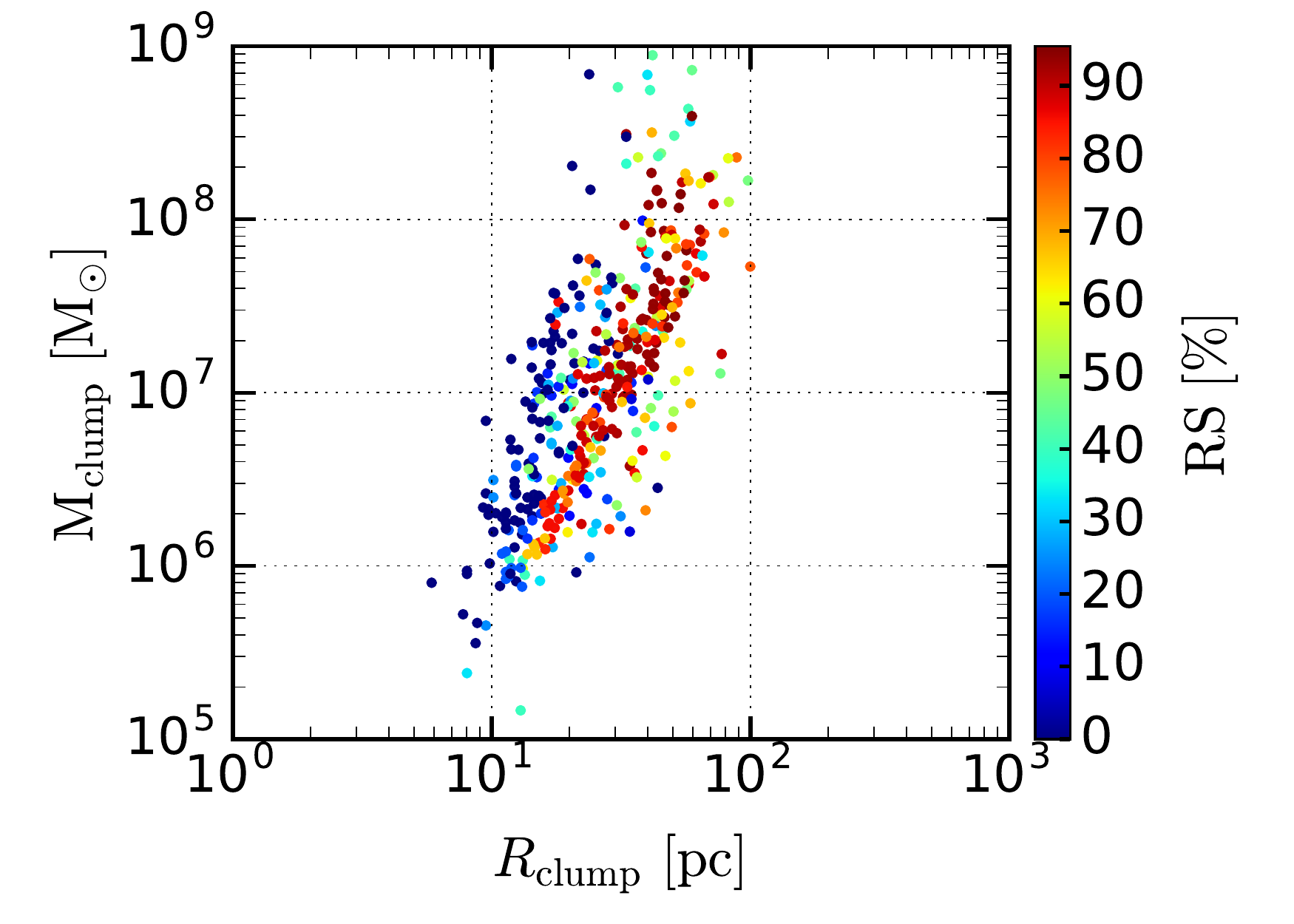}}
\subfloat[ \label{fig:rot_support50}]
  {\includegraphics[width=.33\linewidth]{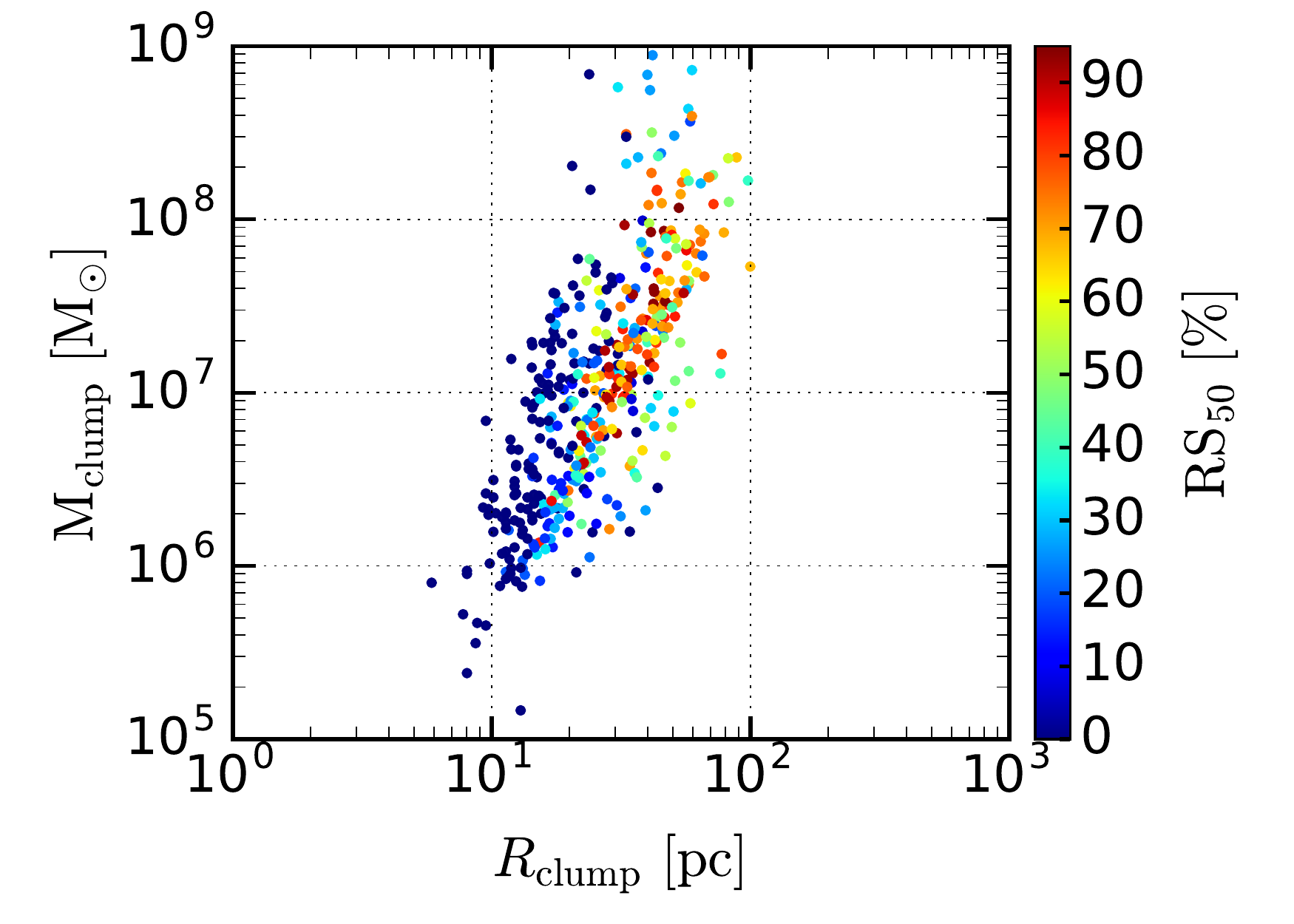}}
\subfloat[ \label{fig:cs_support}]
  {\includegraphics[width=.33\linewidth]{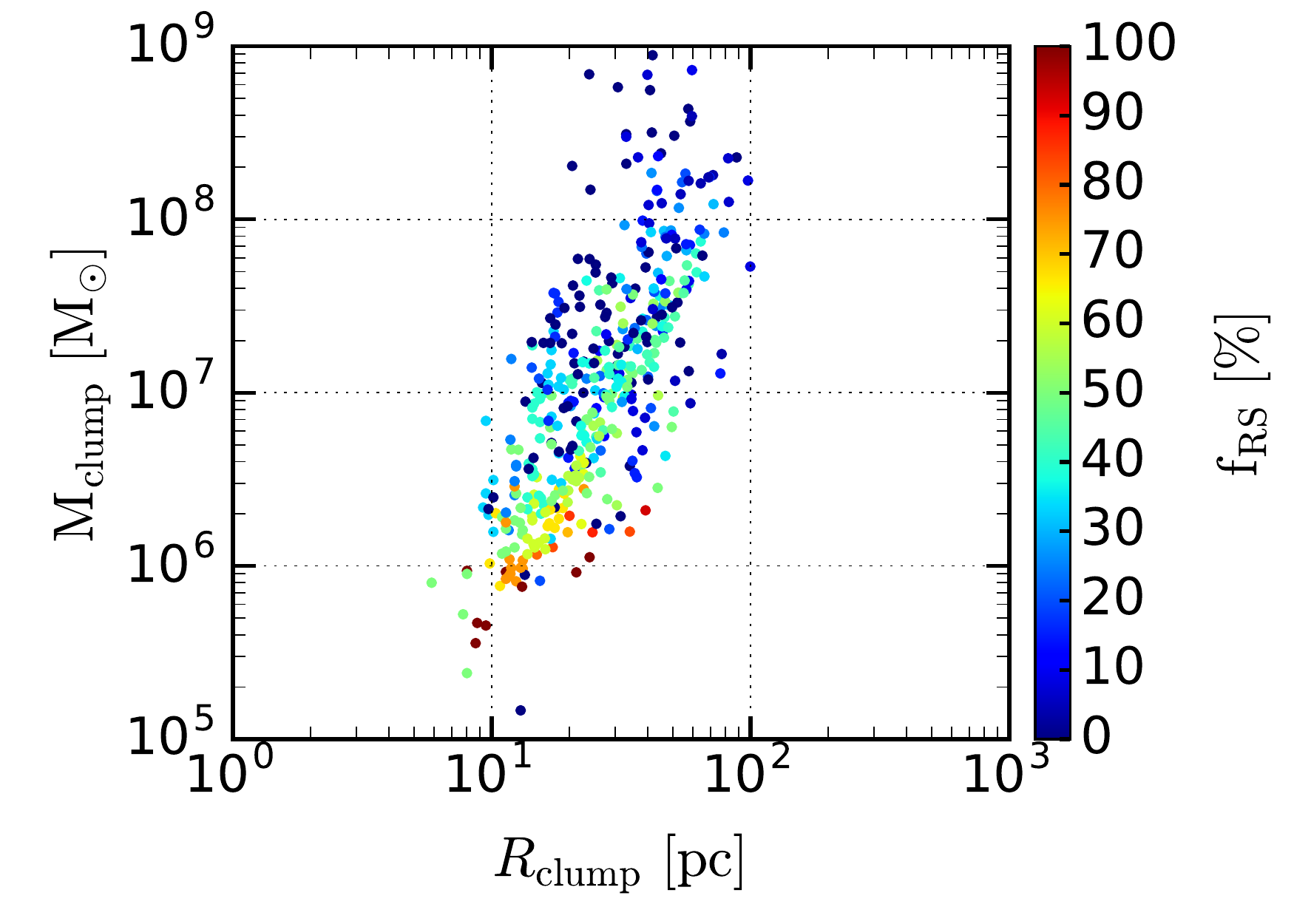}}
 
\caption{(a) Fraction of each clump that fulfills the rotational support criterium $RS > 1$ (Equation \ref{eq: rotational support}). (b) Fraction of each clump that is rotationally supported by at least $50$ percent, defined by the criterium $RS_{\mathrm{50 \%}} > 2$ (Equation \ref{eq: rotational support_50}). (c) Shows how much of the clump is dominated by the artificial pressure support over the pressure coming from the minimum temperature of the simulation (Equation \ref{eq: cs support}).\label{fig:rot_cs_support}} 
\end{figure*}

\subsubsection{Comparison between the different runs}
\label{subsubsec:Comparison between the different runs}

In this Section we compare the evolutionary clump properties between the different runs and the main simulation (Table \ref{tab:Main differences of the simulations}). In general, we find from high to low resolution a strong shift towards lower densities, higher masses and larger sizes (Figure \ref{fig:pressure_floor}, see clump-mass symbols). The clump properties change systematically with the associated artificial pressure floor. Firstly, this is reflected in the maximum densities that are reached on average in each clump. For example, the typical maximum density in the $10^8 \ \mathrm{M_{\odot}}$ clumps is for $MS$ more than three orders of magnitude higher than for $ULR$. Secondly, the density threshold for the clump definition does not capture all clumps anymore for the lower resolution runs $LR$ and $ULR$. Therefore, we chose the minimum density by visual inspection of the dense clump regions. With this definition the total mass of the clumps are in all runs very similar and only by 6-13 percent lower than in $MS$. This is in line with the mass estimation in Section \ref{subsec:Mass redistribution}, see Figure \ref{fig:mass_factions}. Overall, the clump statistics in this section describe the  "high density" part of the clump regions. They are closely surrounded by more gas with lower densities which is  $\sim 30$ percent more in mass quantified  by the surface density approach (see Figure \ref{fig:mass_factions}). For the run $SR$ (Figure \ref{fig:number_mass_clumps}, lower panel) we find in general a very similar clump-mass distribution from beginning on and similar sizes despite that  lower densities are reached and the same density threshold for the clump-finder as in run $MS$ is used. The measured maximum heights are similar too, as the minimum Jeans length due to the pressure floor is not deviating much. We find a very small trend towards higher values since the clumps tend to tilt on average more. The smaller the clump masses the more they are suppressed, while for the $10^9 \ \mathrm{M_{\odot}}$ mass bin a very small trend to more mass is apparent (see also Figure \ref{fig:mass_shift}). When lowering the resolution ($LR$ and $ULR$, Figure \ref{fig:LR_number_mass_clumps}) the amount of clumps below $10^9 \ \mathrm{M_{\odot}}$ are more and more suppressed and their mass contributes less (Figure \ref{fig:mass_shift}). A significant trend towards larger sizes and heights per mass bin is apparent (Figure \ref{fig:size_shift}, \ref{fig:hight_shift}). The maximum heights are roughly increasing with the imposed minimum Jeans length (Figure \ref{fig:hight_shift}) and the clumps tend to be tilted less than in the higher resolution cases. 
 Almost all clumps are larger than the minimum Jeans length imposed by the pressure floor. Only the $10^7 \ \mathrm{M_{\odot}}$ clumps for $LR$ are smaller than the minimum Jeans length but are also almost not present anymore. \\
In summary we find that most of the mass resides in clumps of $10^8 \ \mathrm{M_{\odot}}$ (mergers) in the higher resolution runs and a shift towards $10^9 \ \mathrm{M_{\odot}}$ (no-mergers) for lower resolutions. The size (in the xy-plane) of the clumps with $10^8 \ \mathrm{M_{\odot}}$ increases up to $\sim 6$ times and for $10^9 \ \mathrm{M_{\odot}}$ by a factor of $\sim 14$ for lower resolution.

\subsubsection{Mass-radius relation}
\label{results_subsec:Mass-Radius Relation}
The relations between the clump masses and their radii for all runs are shown in Figure \ref{fig:M-R_relation}  (average between $400-655$ Myr). The kernel density estimation gives us the abundance of clumps with given mass-radius properties (Figure \ref{fig:M-R_relation}) within the time range. For the higher resolution simulations $MS$ and $SR$ we find the most abundant clumps at around $10^7 \ \mathrm{M_{\odot}}$ and we recall that most of the mass resides in clumps with around $10^8 \ \mathrm{M_{\odot}}$. Furthermore, mostly masses of $5 \times 10^8 \ \mathrm{M_{\odot}}$ represent the $10^9 \ \mathrm{M_{\odot}}$ mass bin. The relation for simulation $SR$ is slightly shifted towards smaller sizes at the same mass and a larger density threshold for the clump finder could compensate the effect. We recall that not as high densities are reached than in run $MS$. For the lower resolution runs $LR$ and $ULR$ the density shift is more visible and the clump finding has been adapted to a smaller density threshold. The smaller masses are completely absent and the relation is for both runs shifted in parallel to each other. It is apparent that the $10^8 \ \mathrm{M_{\odot}}$ clumps are different when comparing the low and high resolution simulations. Clumps for run $LR$ are 3 times larger (Figure \ref{fig:size_shift}) than in $MS$ and a factor 6 larger for $ULR$ and are therefore correspondingly much less dense.

\subsubsection{Clump's relation to the artificial pressure floor}
\label{results_subsec:Mass-Radius Relation}
In the previous sections we explored in general the influence of the resolution on the clump properties. Here we investigate the impact of the APF more directly on the clump mass and density. If the Jeans mass corresponding to the APF is similar to the clump's mass, we can speak of a clump with artificially given properties. In the simulation code the Jeans length (Equation \ref{eq:Jeans length}) is applied on quadratic cells. Therefore, we define the Jeans mass as 
\begin{equation}
  M_{\mathrm{J}} =   \lambda_{\mathrm{J}}^3 \ <\rho>,
\end{equation}
which is 1.9 times larger than for the spherical case. Since the Jeans length is given as a constant at the highest resolution within the clump, we can write
\begin{equation}
  M_{\mathrm{J}} =  \left( N_{\mathrm{P}} \Delta x_{\mathrm{min}} \right)^3 \ <\rho>. 
\end{equation}
The APF dominated clumps have then $M_{\mathrm{clump}} \approx  M_{\mathrm{J}}$ for the average density  $<\rho_{\mathrm{clump}} > \approx  <\rho_{\mathrm{J}}>$, while $M_{\mathrm{clump}} <  M_{\mathrm{J}}$ is suppressed. We take all identified clumps at every timestep between 400-655 Myr (well fragmented disc) and calculate their average density and total mass. The kernel density estimation gives us the abundance of the clump's mass-density properties (Figure \ref{fig:JeansMass_Density}) within the time range. The bulk of the clumps (initial clumps in a ring) in simulation $LR$ and $ULR$ are well described by the minimum Jeans mass-density relation. The initial clumps in the main simulation $MS$ and in run $SR$ do clearly not lie on the APF induced relationship. Clumps above the lines are less influenced by the APF and with distance to the relation are more and more rotationally supported, which will be shown in the next Sections. Another way to present the effect of the APF is the face-on projection of the disc with the mass weighted sound speed (Figure \ref{fig:cs_lowres}). The clumps of the runs $LR$ and $ULR$ show extremely high sound speeds between $c_{\mathrm{s,J}} \simeq 20-200 \ \mathrm{km \ s^{-1}}$, and their areas are larger for $ULR$ than for $LR$. 
 
\subsubsection{Intrinsic clump profiles of $MS$}
\label{results_subsec:Intrinsic clump profiles}
 To understand the nature of the clumps even better it is necessary to measure their intrinsic properties. We cut out a region around each clump, 25 percent larger than defined by the clump finder. The centre of mass defines the origin of the clumps coordinate system. The velocities in the clump rest frame are given by the local values in each cell subtracted by the mass weighted clumps linear velocity x,y,z components in the disc. For the radial profiles of different properties the values are mass weighted and averaged within 3 pc radial bins. \\
We selected three clumps as archetypes, corresponding to the masses of $\sim 2 \times 10^8 \ \mathrm{M_{\odot}}$ (clump V1) which represent most of the mass in the galaxy, $\sim 2 \times 10^7 \ \mathrm{M_{\odot}}$ (clump V2) which belongs to the group with the most frequent clumps and of $\sim 4 \times 10^6 \ \mathrm{M_{\odot}}$ (clump V3) which are mainly influenced by the APF. In Figure \ref{fig:clump_profile_01} we provide their intrinsic radial profiles. The volume densities show in general an exponential decline. The rotational velocities show a typical profile for a baryonic mass dominated rotationally supported exponential disc. The clumps maximum rotation is clearly related to the clump mass with $M_{\mathrm{clump}} \ \mathrm{[M_{\odot}]} \simeq 10^4 \times V_{\mathrm{rot,max}}^2 \ \mathrm{[km \ s^{-1}]}$ (Figure \ref{fig:mass_radius_color_coded} ) with ranges between $V_{\mathrm{max}} \sim 10-200 \ \mathrm{km \ s^{-1}}$. A similar relationship follows from the circular speed of the exponential disc at the radius for the peak velocity (see Equation 2.165 in \citep{2008gady.book.....B}, substituted by the mass within $R_{\mathrm{h}}$ with $M_{\mathrm{h}}=2 \pi \Sigma_{\mathrm{0}} R_{\mathrm{h}}^2$ and the ratio $y=\frac{2.15 R_{\mathrm{h}}}{2 R_{\mathrm{h}}}$ with the peak radius $2.15 \times R_{\mathrm{h}}$). The different clumps along the mass-radius relation (Figure \ref{fig:M-R_relation}) are distributed over all the disc (Figure \ref{fig:mass_radius_color_coded}) and significant deviations from the relation (Figure \ref{fig:mass_radius_color_coded}) correspond to those clumps in the central region of the galaxy and are caused by tilts to the disc plane due to more frequent interactions as well as our definition of the clump radius, measured in the xy-plane. 
The sound speed peaks in the centre of the clump at highest density, induced by the artificial pressure floor, and declines with radius. The radial velocity dispersion is in general not larger than the sound speed. Both are calculated locally and then azimuthally averaged (mass-weighted).

\subsubsection{Rotational supported clumps}
\label{results_subsec:Rotational supported clumps}	
In the following we clarify the contributions of the clumps' rotation, the minimum sound speed $c_{\mathrm{s}} \sim 10 \ \mathrm{km \ s^{-1}}$ and the artificial pressure floor to their internal dynamics by analysing properties at the time step 655 Myr. For the analysis we assume that the clumps are in dynamical equilibrium, since their radial velocities are  negligible compared to their total rotational velocities (Figure \ref{fig:clump_profile_01} b, e, h). Then, the circular velocity can be expressed by the rotational velocity and the gradient of the thermal pressure (e.g. \citet{2008gady.book.....B}, Equations 4.229, 4.230)
\begin{equation}
\label{eq:Jeans equation}
V^2_{\mathrm{circ}} = V^2_{\mathrm{rot}} - \dfrac{r}{\rho} \ \dfrac{\partial ( \rho \ c^2_{\mathrm{s}} )}{\partial r} = V^2_{\mathrm{rot}} - V^2_{\mathrm{p}}.
\end{equation} 
The clump is locally rotationally supported if the first term of Equation \ref{eq:Jeans equation} dominates the second term, which we refer to as $V^2_{\mathrm{p}}$, expressed by the ratio 
\begin{equation}
\label{eq: rotational support}
RS = \dfrac{V^2_{\mathrm{rot}}}{ V^2_{\mathrm{p}} } > 1,
\end{equation}
while the pressure term is calculated numerically. For completeness, we plot the radial profiles of $V_{\mathrm{p}}$ in Figures \ref{fig:440_tstep_vrot_cl101}, \ref{fig:440_tstep_vrot_cl132}, \ref{fig:440_tstep_vrot_cl38}. The three examples (black line) in Figures \ref{fig:440_tstep_ratio_vrot_cs_cl101}, \ref{fig:440_tstep_ratio_vrot_cs_cl132}, \ref{fig:440_tstep_ratio_vrot_cs_cl38} show the radial dependence of Equation \ref{eq: rotational support}. The clumps are mainly supported by rotation. This is also true for the rest of the clumps at this timestep (Figure \ref{fig:rot_support}) where we measure intrinsically what fraction of the clump radius is dominated by the rotation term. For the clumps lying on the tight M-R relation, $RS$ dominates by at least $70 \%$ in the inner region. While on the lower mass-end of the relation the pressure plays an important role which can also be seen in Figure \ref{fig:440_tstep_vrot_cl38} for clump V3. The left side of the relation lacks on rotational signal due to their tilts to our measurement in the xy-plane. With the above measurement we only identified if there is a rotational support. To understand better its importance  we measure how much of the clump is rotationally supported by more than $50 \%$ (Equation \ref{eq:Jeans equation}), expressed by the ratio
\begin{equation}
\label{eq: rotational support_50} 
RS_{\mathrm{50}} = \dfrac{V^2_{\mathrm{rot}}}{| V^2_{\mathrm{p}} |} > 2.
\end{equation}
Figure \ref{fig:rot_support50} clearly illustrates that rotation dominates the main clumps ($> 10^7 \ \mathrm{M_{\odot}}$) by more than $70 \%$. Now it is even more visible that for the smaller clump masses the pressure term plays a stronger role. Here, we have not distinguished between the contribution of the artificial pressure floor or the minimum temperature. To compare both contributions, we ignore the artificial pressure floor (Figures \ref{fig:440_tstep_ratio_vrot_cs_cl101}, \ref{fig:440_tstep_ratio_vrot_cs_cl132}, \ref{fig:440_tstep_ratio_vrot_cs_cl38}) and calculate Equation \ref{eq: rotational support} with only the contribution of the minimum sound speed $c_{\mathrm{s}} \sim 10 \ \mathrm{km \ s^{-1}}$ and call it $RS_{\mathrm{T_{min}}}$. The rotational support in the central part of the clump appears now much higher and is outside towards larger radii similar to the profile where the artificial pressure floor is taken into account. This means that the hotter inner part has a stronger contribution from the artificial pressure floor and outside only the minimum sound speed plays a role.
To quantify the dominance of the artificial pressure floor over the pressure due to the minimum temperature of the simulation, we take the ratio between both of the $RS$ parameter (Figure \ref{fig:cs_support})
\begin{equation}
\label{eq: cs support} 
f_{RS} = \dfrac{RS}{RS_{\mathrm{T_{min}}}} = \dfrac{| V^2_{\mathrm{p,T_{min}}} |}{| V^2_{\mathrm{p}} |} \geq 0.9
\end{equation}
The rotational term drops out and and only the two pressure terms are compared while we have $RS_{\mathrm{T_{min}}} \geq RS$ (see Figures \ref{fig:440_tstep_ratio_vrot_cs_cl101}, \ref{fig:440_tstep_ratio_vrot_cs_cl132}, \ref{fig:440_tstep_ratio_vrot_cs_cl38}). This means that the artificial pressure is dominating whenever $f_{RS} < 1$ and for $f_{RS} = 1$ the minimum temperature is more important. We allow for a small deviation of $10$ percent and whenever $f_{RS}$ has values between $0.9 -1$ the clump is locally supported by the pressure induced by the minimum temperature of the simulation. This is mainly the case for smaller clump masses (Figure \ref{fig:cs_support}) below  $< 10^7 \ \mathrm{M_{\odot}}$. With increasing mass more and more of the clumps pressure gradient is dominated by the artificial pressure floor, but plays a smaller role than their rotational support (Figure \ref{fig:rot_support50}).

 \begin{figure*}
\centering
  {\includegraphics[width=0.49\linewidth]{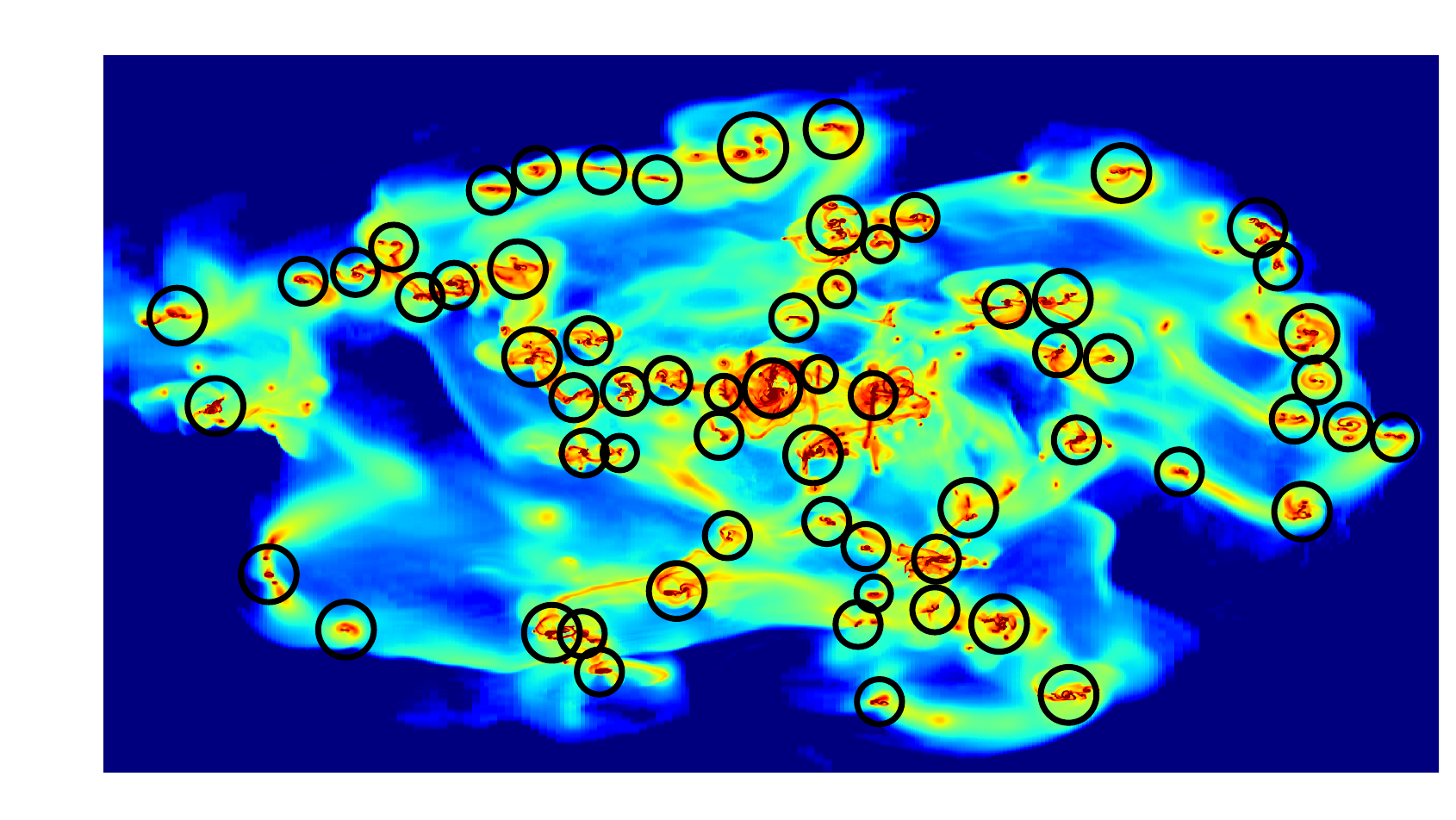}
	\includegraphics[width=0.49\linewidth]{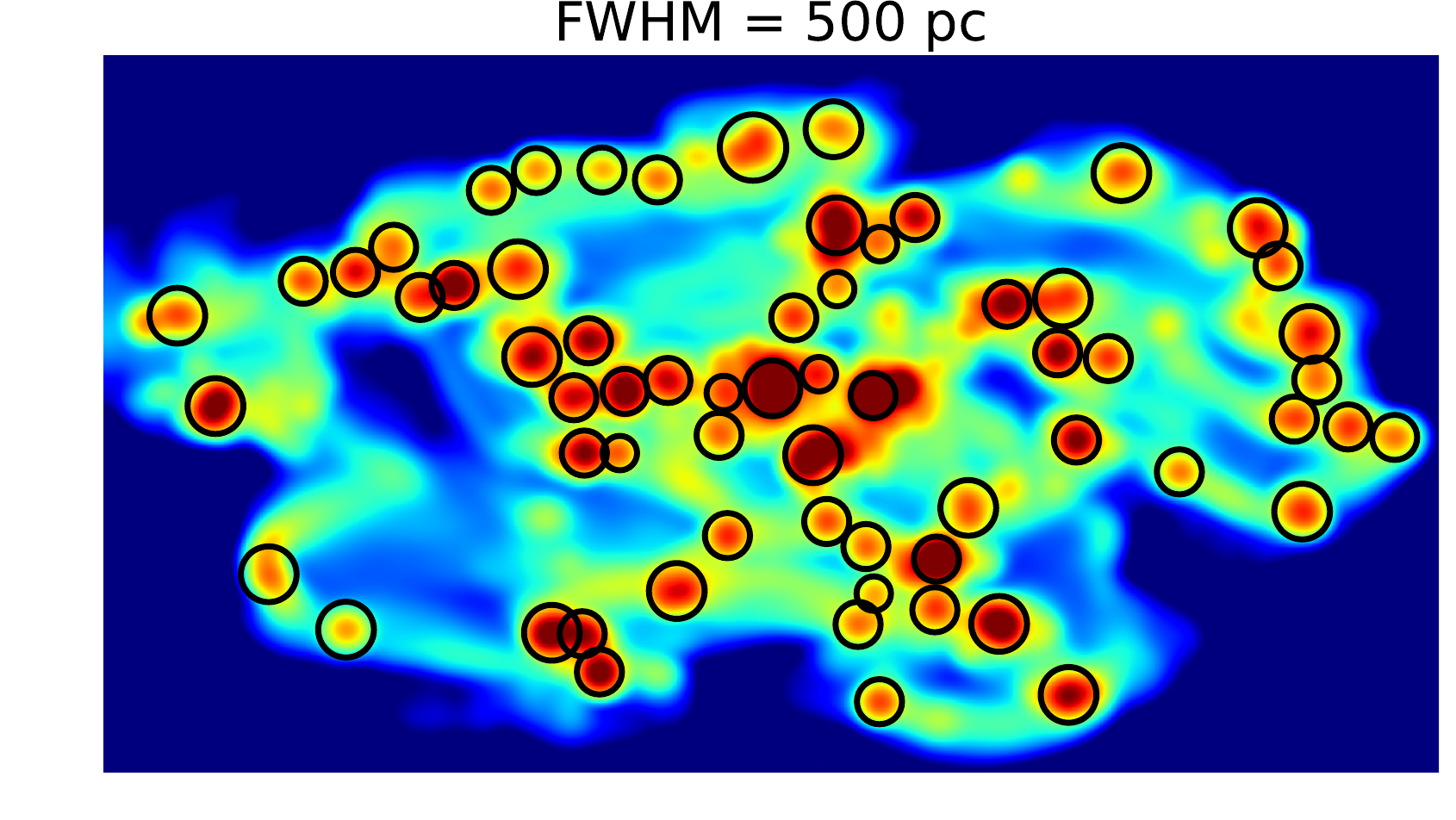}}\hfill

  {\includegraphics[width=0.49\linewidth]{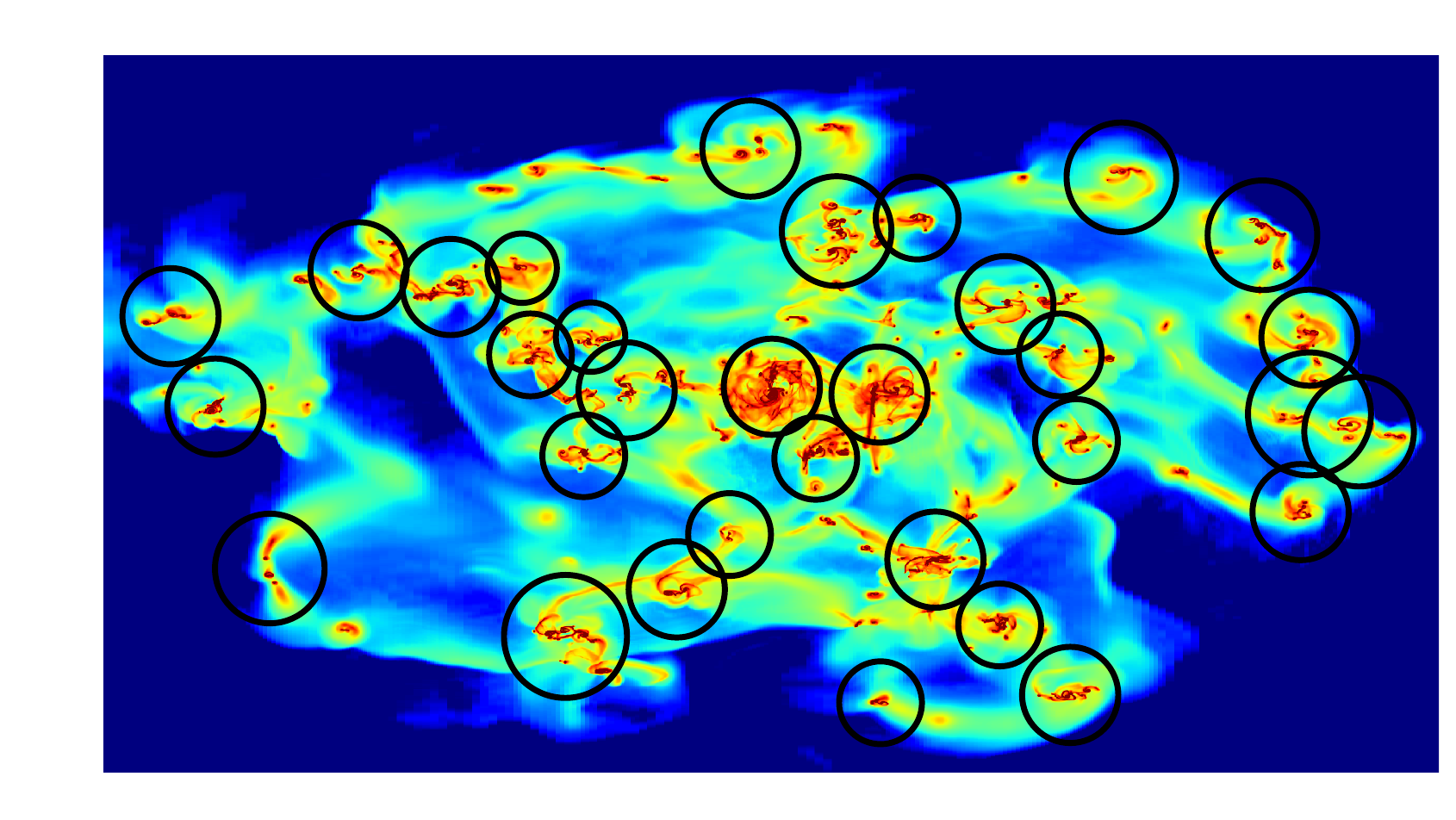}
  \includegraphics[width=0.49\linewidth]{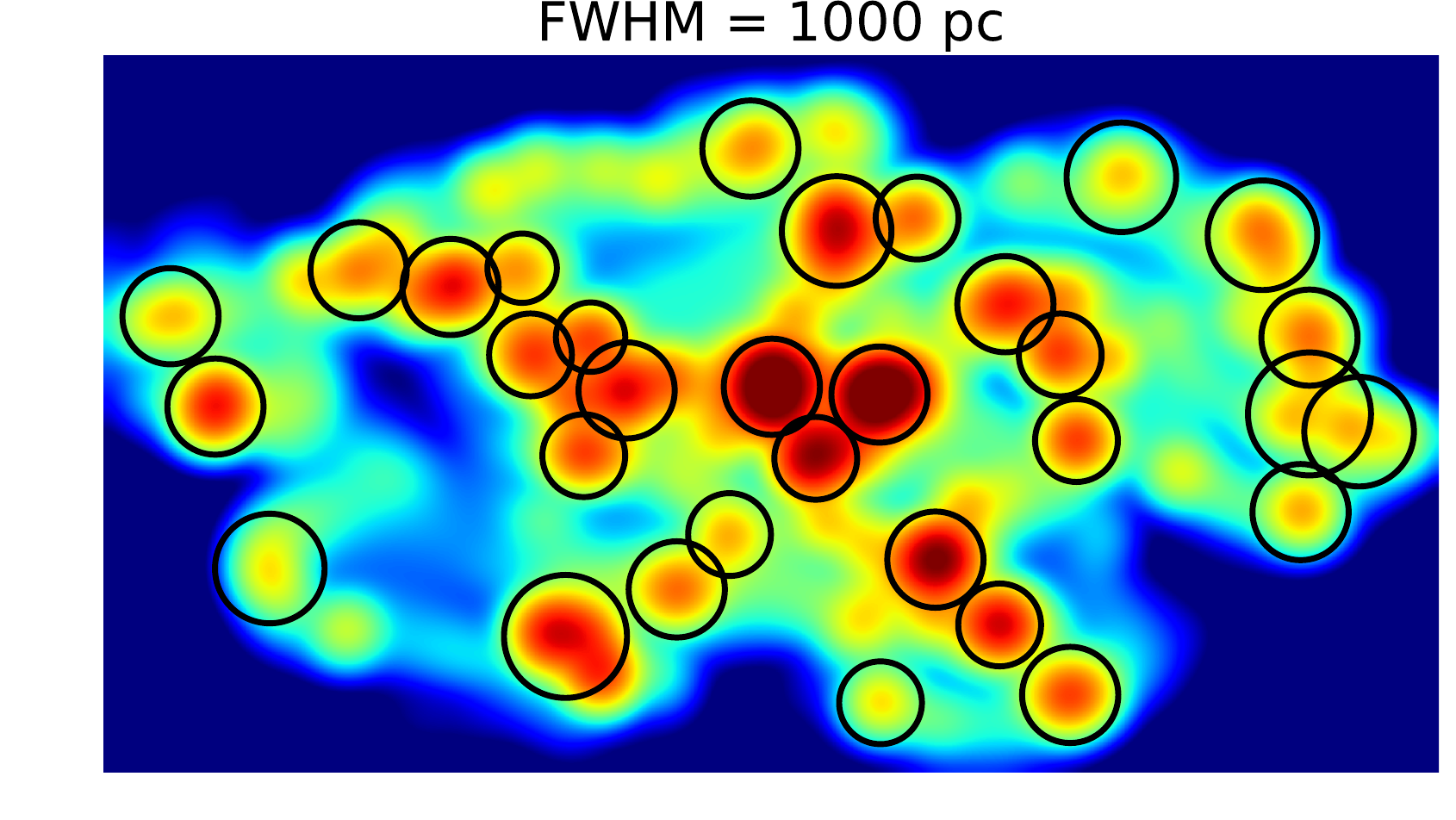}}\hfill

  {\includegraphics[width=0.49\linewidth]{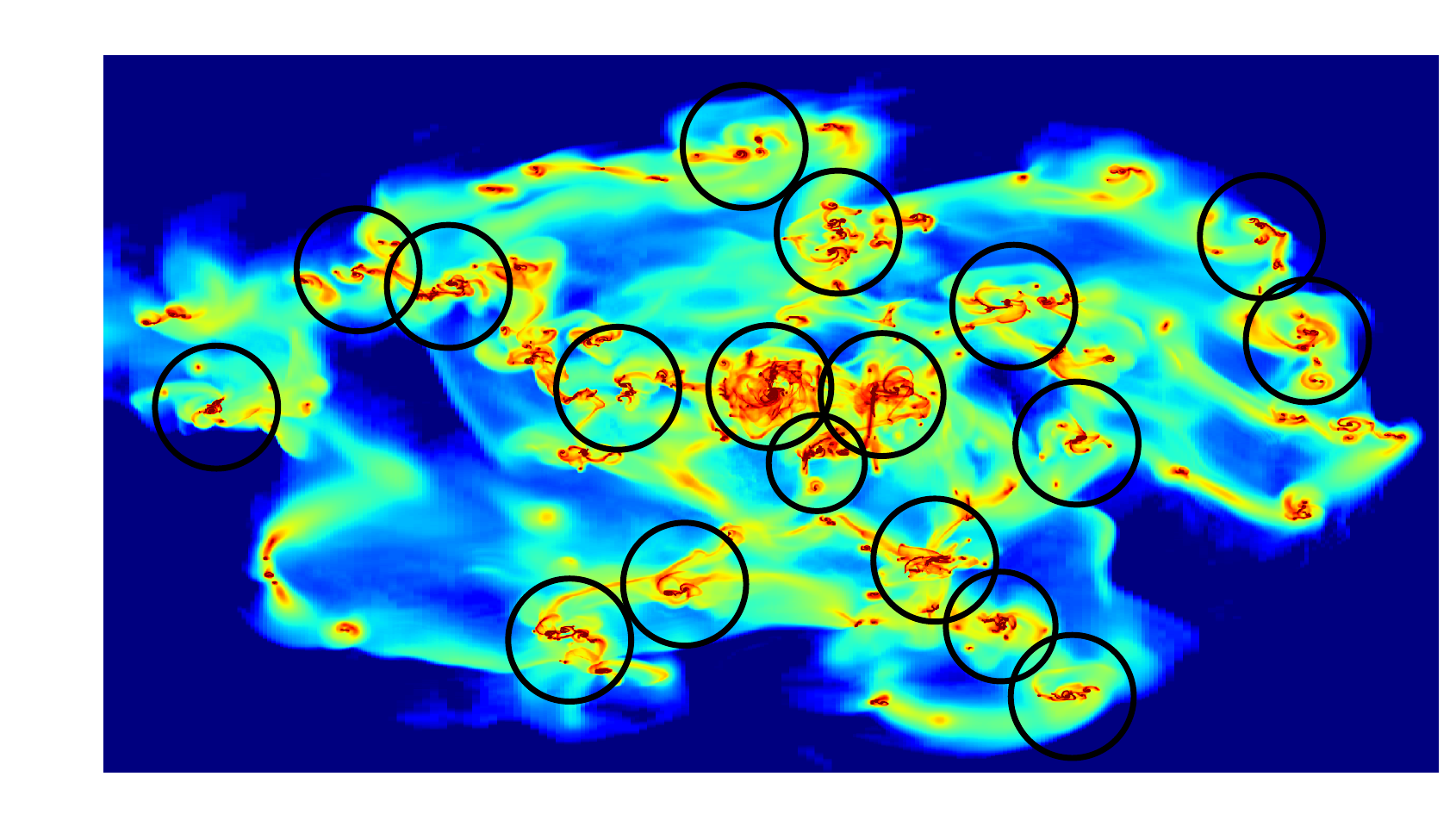}
  \includegraphics[width=0.49\linewidth]{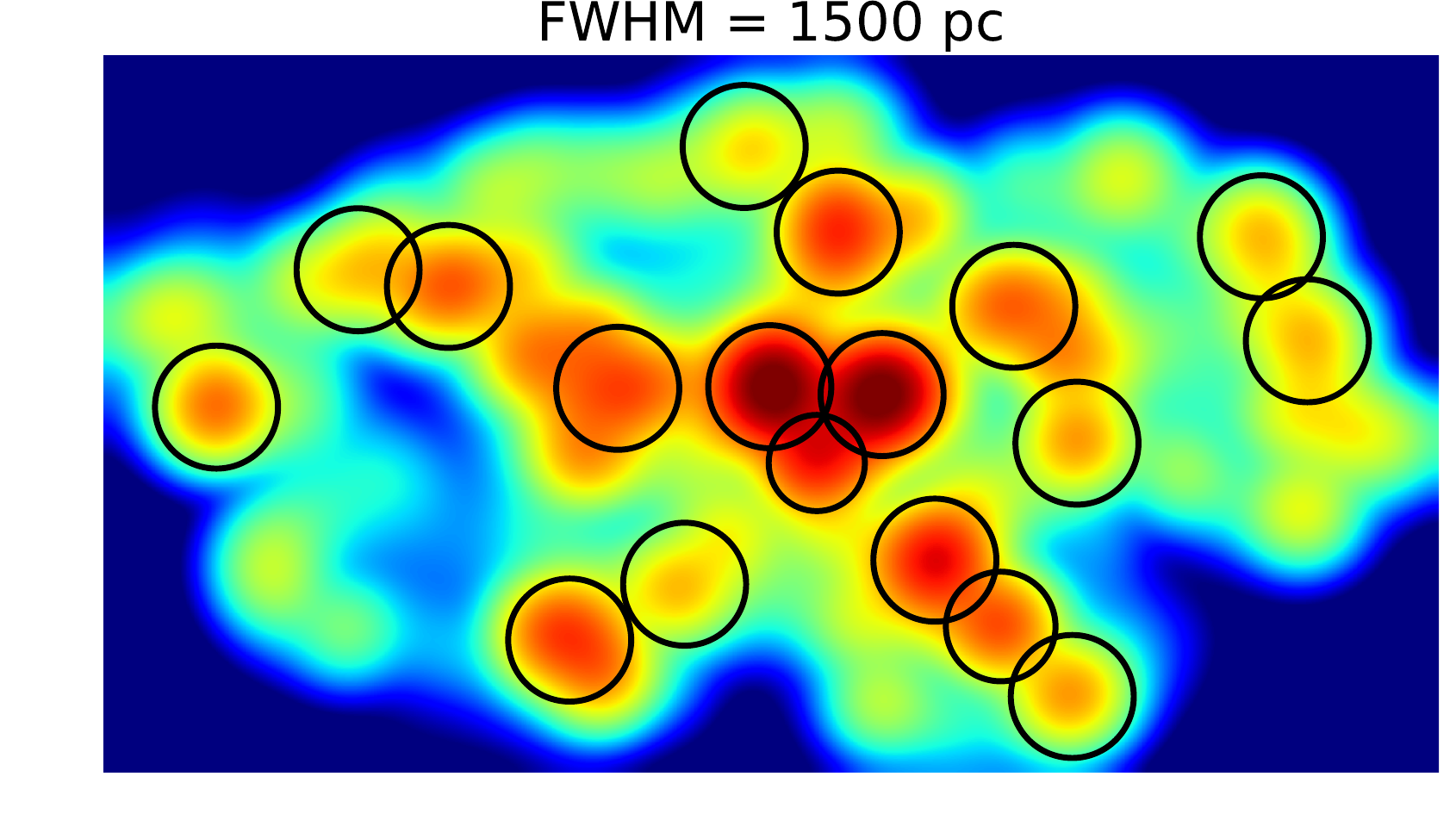}}\hfill

  {\includegraphics[width=0.49\linewidth]{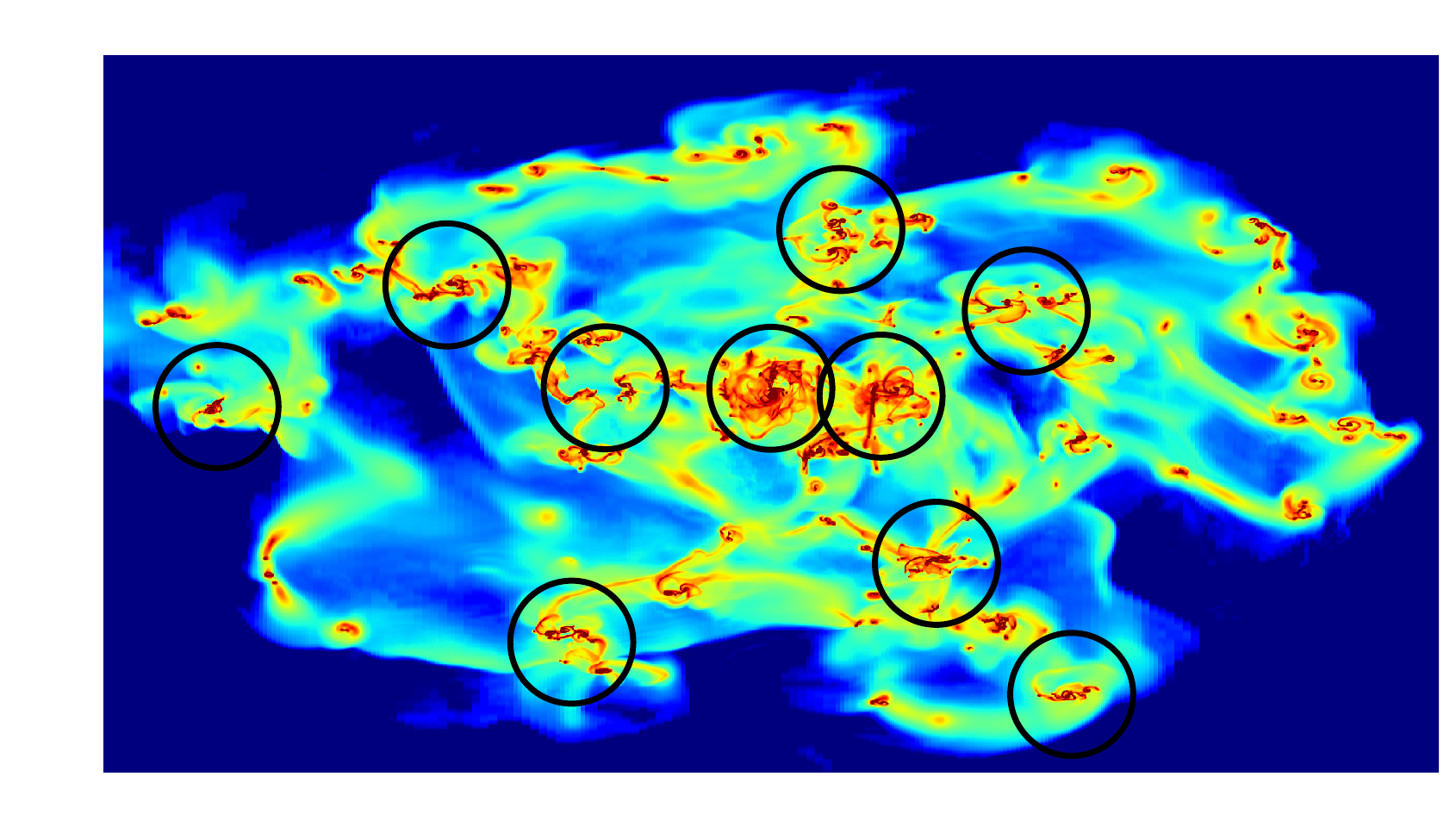}
  \includegraphics[width=0.49\linewidth]{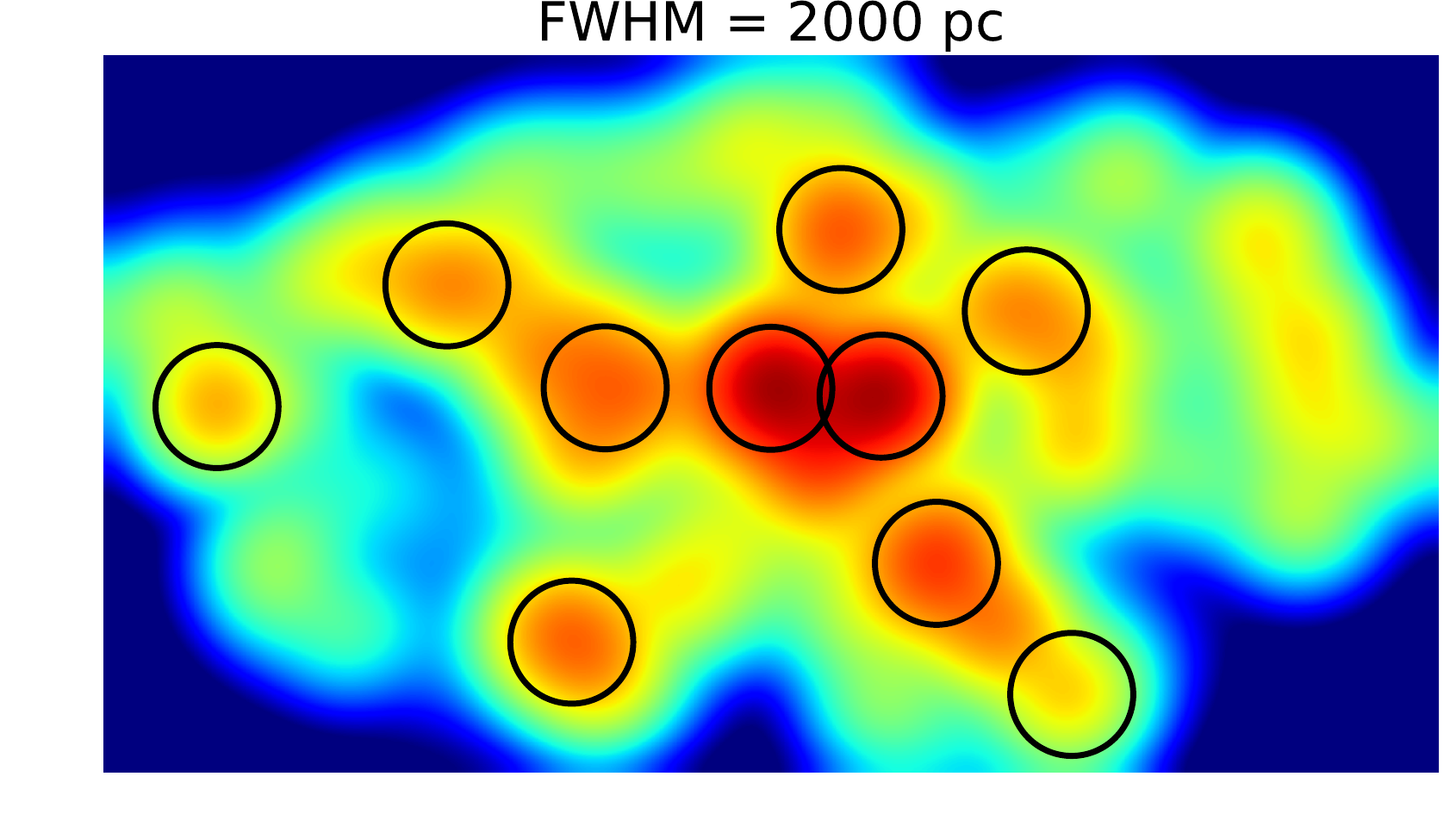}}\hfill

\includegraphics[width=.45 \linewidth]{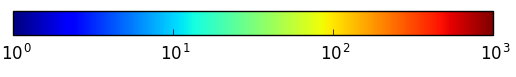} 

\caption{LOS observations of the surface density ($\mathrm{log_{10}} (\Sigma) \ \mathrm{[M_{\odot} \ pc^{-2}]}$) of the inclined galaxy ($60 \degr$) at timestep 655 Myr (same as in \citealp{2016ApJ...819L...2B}). Each panel shows a 22 x 12 $\mathrm{kpc^2}$ map centered on the disc. The left column is the original image which can be compared with different beam smeared versions in the right column. The identified groups or CCs are marked with a circle of the found structure size (FWHM).\label{fig:LOS_FWHM_set}}

\end{figure*}

  \begin{figure*}
\centering
\subfloat[ \label{fig:LOS_CC_Count_statistics}]
  {\includegraphics[width=0.47\linewidth]{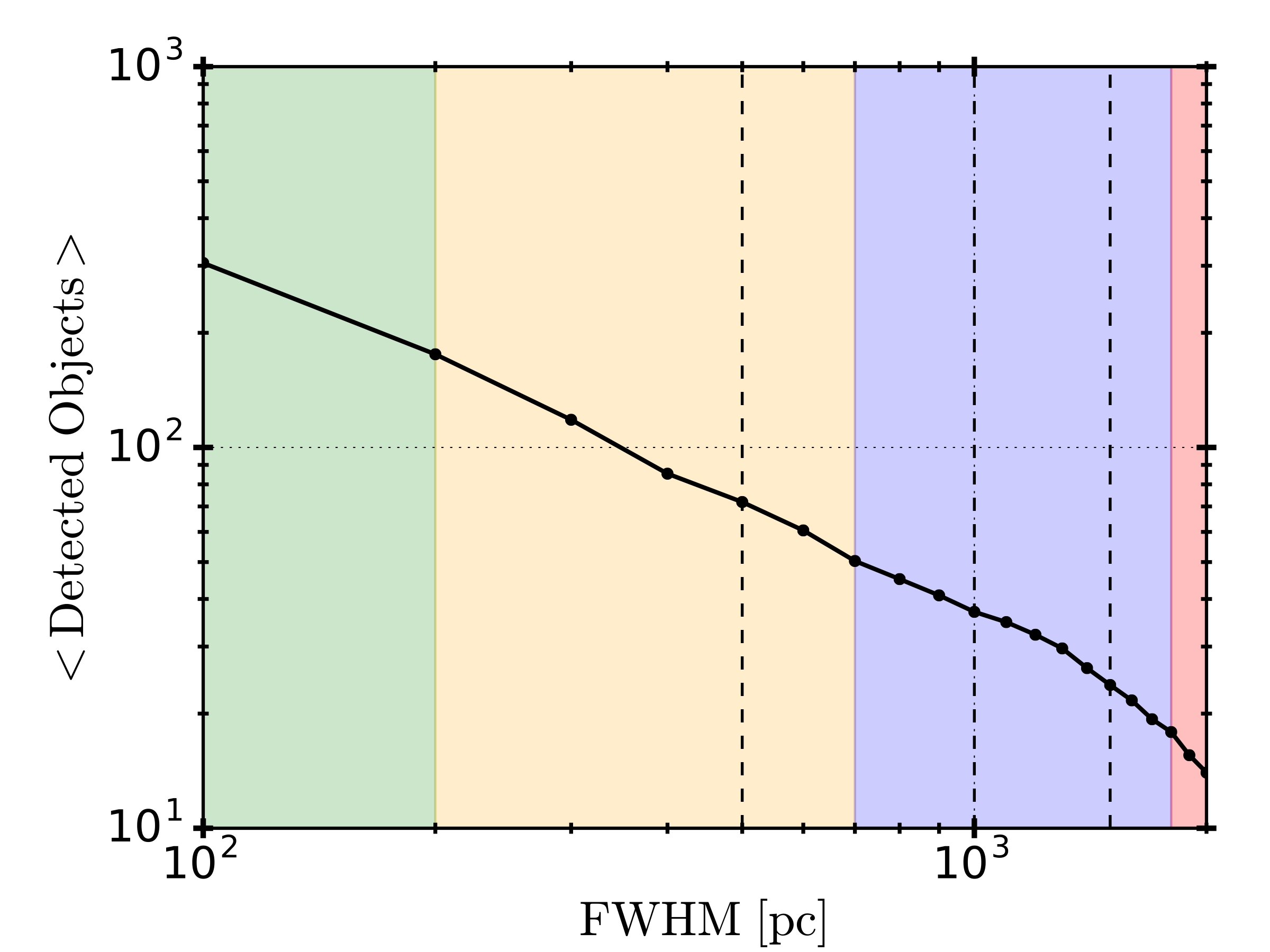}}\hfill
\subfloat[ \label{fig:LOS_CC_Mtot_statistics}]
  {\includegraphics[width=0.47\linewidth]{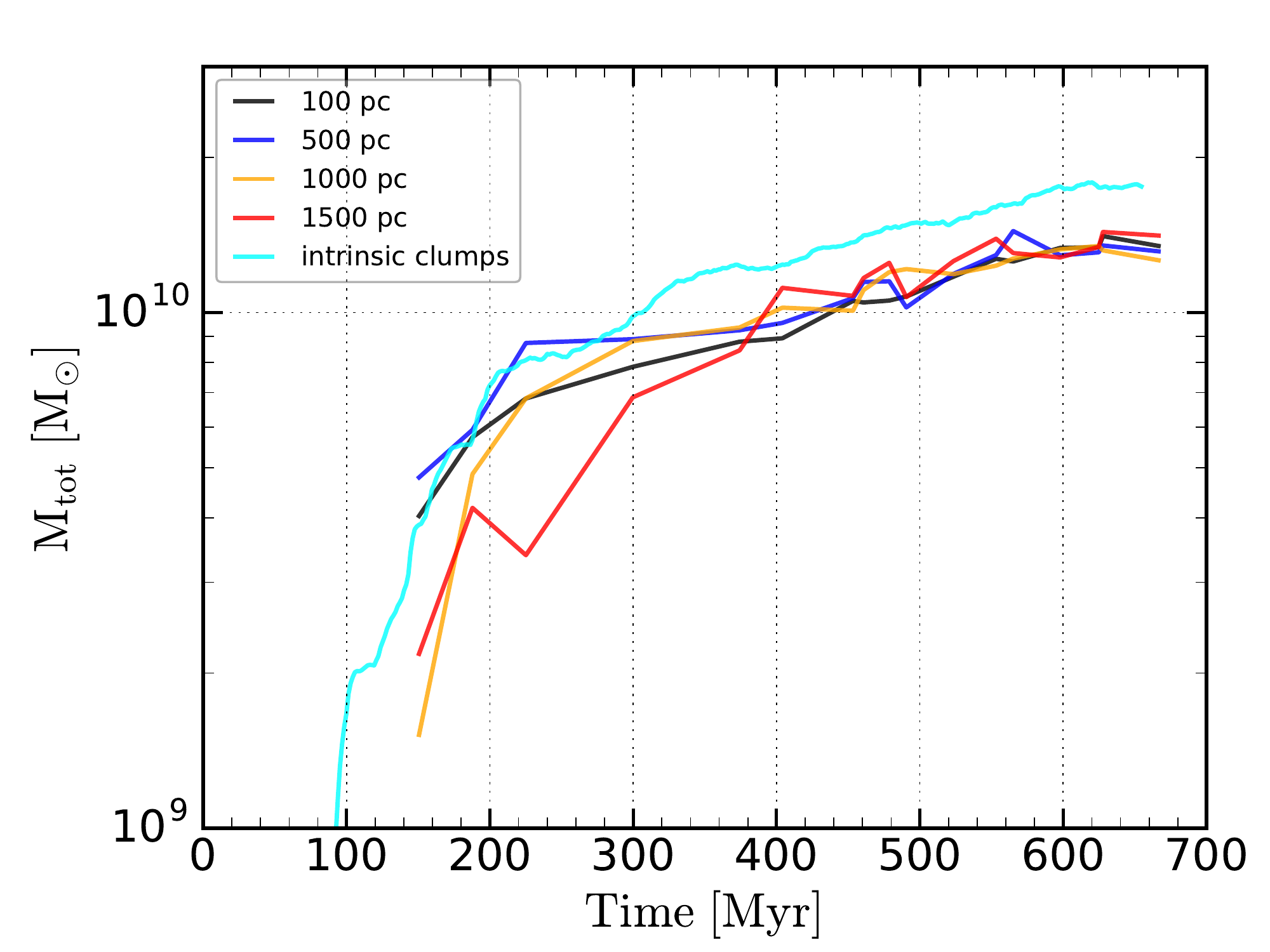}}

\caption{(a) The number of objects found by the blob finder dramatically decreases with the FWHM of the beam smearing (time average between 450 - 655 Myr). 
The background colors correspond to the ranges of the identified clump cluster scales (see Table \ref{tab:clump_cluster_definitions}). (b) Total mass of the identified objects over time for different FWHM of the beam smearing, compared to the total mass of the intrinsic clumps of run $MS$ (Section \ref{subsec:Detailed clump properties}).\label{fig:LOS_MccRcc_statistics_set}}
\end{figure*}

\section{The hierarchical clustering of clumps}
\label{sec:The effects of beam smearing on the main clump properties}
After the disc fragmentation to single clumps, they immediately group to clump clusters (CCs) on different spatial scales. In this section we identify these scales by using different beam-smearings (BS) and estimate the corresponding masses and sizes of the CCs. As in \citet{2016ApJ...819L...2B}, we interpolate the data on a uniform grid with pixel size 11.7 pc and incline the galaxy by 60 degrees. The surface density is observed along the line-of-sight and is spatially convolved by a Gaussian kernel of different full width at half maximum (FWHM = 100 pc - 2000 pc). Figure \ref{fig:LOS_FWHM_set} shows the evolved disc (same snapshot as in \citet{2016ApJ...819L...2B}) under different observed spatial resolutions, each compared to the highly resolved simulation picture. For every considered convolution we overplot the image with circles to indicate the positions and FWHM sizes of the detected structures (see Figure \ref{fig:LOS_FWHM_set} and for the size definition: Section \ref{subsubsec:2D images}). With larger FWHM of the beam smearing kernel the amount of visible clumps or clump clusters dramatically decreases (see also Figure \ref{fig:LOS_CC_Count_statistics}) since they are represented by a few giant objects on larger scales. The smooth relationship between detected objects and BS is caused by several reasons. Firstly, the CCs are not isolated entities, but in the convolution are also neighboring clumps or CCs involved. Secondly, isolated objects can disappear with larger BS due to a flat density profile and do not contribute to the statistics. With BS the mass is distributed over larger scales, expressed by lower densities in the maps. This is visible in the shift of the mass-fractions (see Figure \ref{fig:mass_factions}) of the density-regimes defined in the full resolution observation (Table \ref{tab:Surface density regions}). The highest density regime $H$ disappears with higher BS and the mass-fraction for the defined clump region ($> 10^2 \ \mathrm{M_{\odot} \ pc^{-2}}$) is declining, while the mass-fraction for the defined intermediate gas density ($10^1-10^2 \ \mathrm{M_{\odot} \ pc^{-2}}$), the arm-like features is increasing.

 \begin{table*}

  \caption{Overview of the hierarchical properties of the identified clumps and clump cluster mass and size scales. Derived from data between 450 - 670 Myrs. For range A we give only the identified clump masses and sizes at FWHM=100pc. Within the BS-ranges B,C,D,E we take their average masses or sizes.}
  \label{tab:clump_cluster_definitions}
  \begin{threeparttable}
  \begin{tabular}{cccccccc}
    \hline
    Name   & Definition 			   & Shading &	BS (FWHM)	& $M_{\mathrm{FWHM}}$  					  & $D_{\mathrm{FWHM}}$ 	& Detected Objects\tnote{a}  		& 	Description\\
 	Ranges &   & [color] & [pc]			& $[\mathrm{M_{\mathrm{\odot}}]}$ & $\mathrm{[pc]}$	& 	$\mathrm{[\%]}$ 		& 	\\
    \hline
    A 	 & clumps 	  		& \textcolor{ForestGreen}{green}   & 100-200		& $\sim  10^{7}$								  &		$\sim 125$	&  60-100 		& \makecell{intrinsic clumps} \\
    B 	 & smallest clusters  		& \textcolor{Dandelion}{yellow}  & 200-700		& $\sim 4.5 \times 10^{7}$								  & 	$\sim 450$		& 23-60 	& \makecell{dense groups} \\
    C 	 & intermediate clusters  		& \textcolor{blue}{blue}    & 700-1800 	& $\sim 3.2 \times 10^{8}	$								  & 	 $\sim 1500$ 	&  5-23	& \makecell{mainly dense groups, \\partly open CCs } \\
    D 	 & largest clusters	 	& \textcolor{red}{red}     & 1800-2000 	& $\sim 9 \times 10^{8}	$								  & 	$\sim 2000$		&  5-6  	&  \makecell{dense groups \\ and open CCs} \\
    E 	 & central clusters	& \textcolor{red}{(red)}   & 1800-2000 	& $\sim 3.2 \times  10^{9}$								  &		$\sim 3200$							&    			& \makecell{dense grouping \\ (sub-sample of D)}\\
    \hline
    
  \end{tabular}  
  \begin{tablenotes}
  \item[a] The relative percentage of each range to the number of clumps detected for a beam smearing of FWHM=100pc.
  \end{tablenotes}
  \end{threeparttable}
 \end{table*}

\begin{figure*}
\centering
\subfloat[ \label{fig:LOS_Mcc_distributions}]
  {\includegraphics[width=0.50\linewidth]{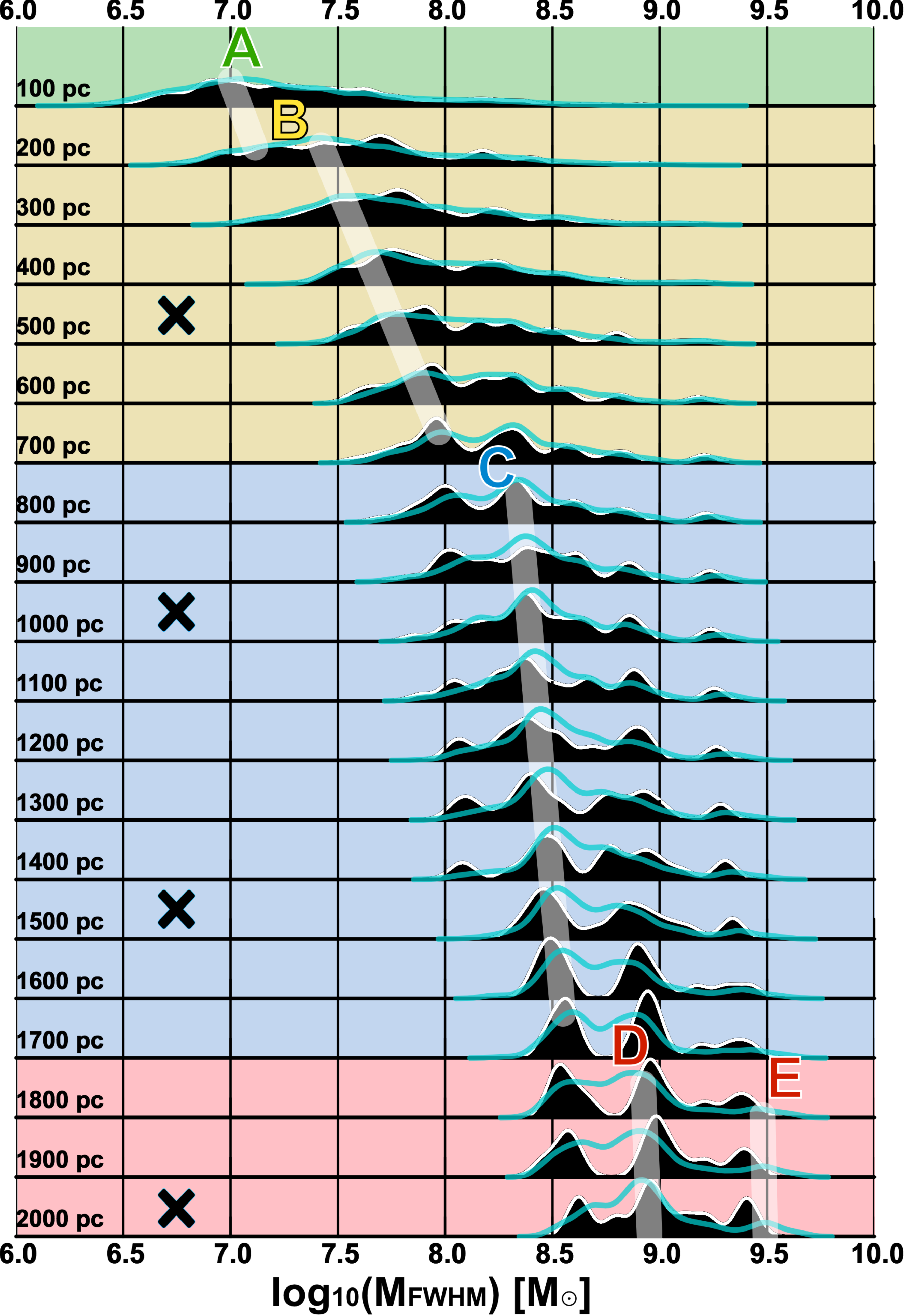}}\hfill
\subfloat[ \label{fig:LOS_Rcc_distributions}]
  {\includegraphics[width=0.49\linewidth]{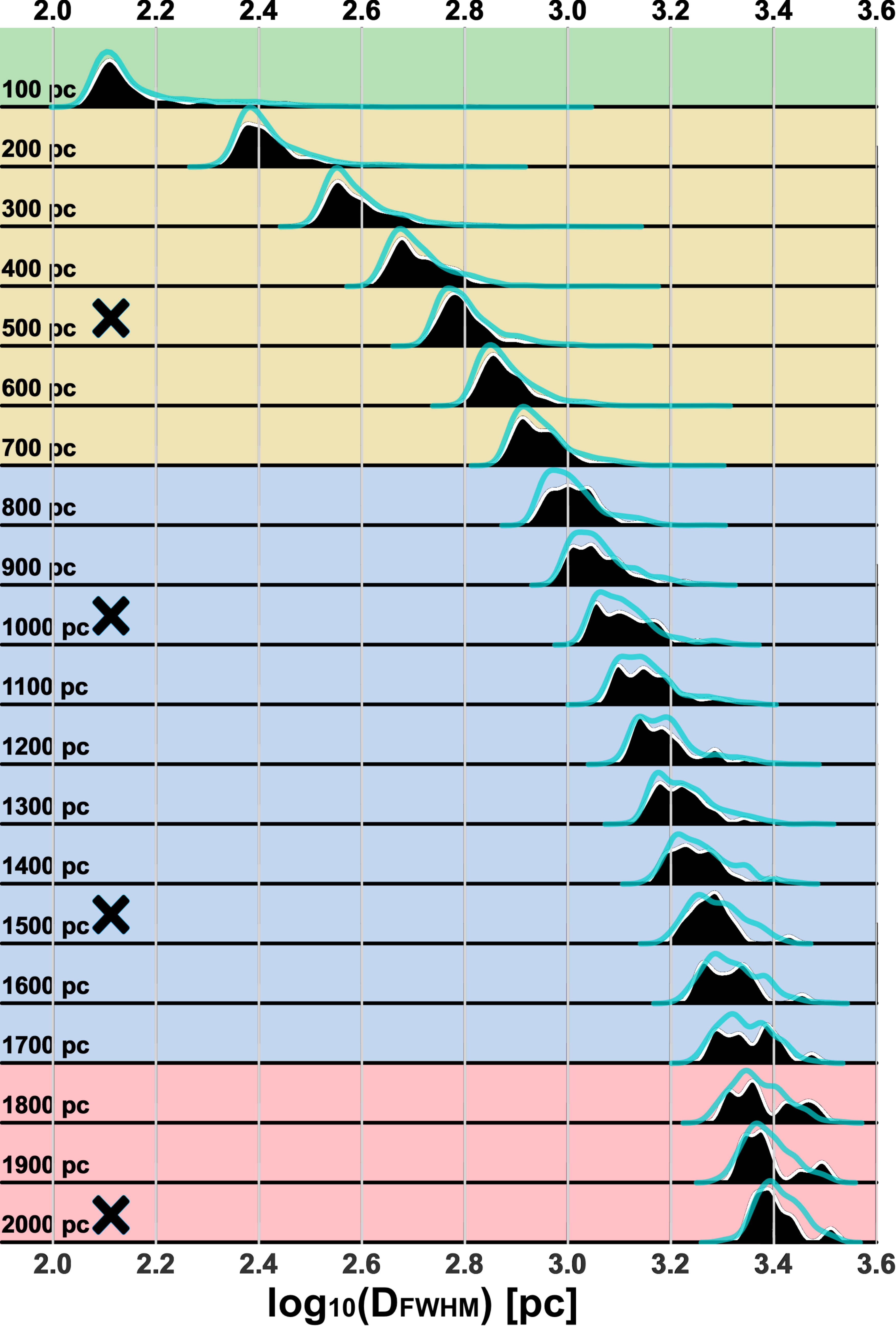}}

\caption{The dependence of the mass (a) and size (b) number density distributions of the identified clumps and clump clusters on the beam smearing in lin-log representation (y-axis: FWHM of the beam smearing kernel). The black shaded histograms correspond to the single time step at 655 Myr and is overlayed with the average distributions between the time 450 - 655 Myr (cyan line). The dominant number of clumps or clusters of the average distributions are highlighted by the white guidelines (A,B,C,D,E) and their identified ranges with color in the background (see Table \ref{tab:clump_cluster_definitions}). The line-of-sight maps in Figure \ref{fig:LOS_FWHM_set} are marked by the black crosses. \label{fig:LOS_MccRcc_statistics_set}}

\end{figure*}  

 The total mass of the identified objects is for the evolved disc $\sim 30$ percent of the total disc mass for all BS (Figure \ref{fig:LOS_CC_Mtot_statistics}). This is $\sim 20$ percent smaller than the clump mass obtained from the RAMSES clump finder (3D density approach) in Section \ref{subsec:Detailed clump properties}. If we increase the identified radius from the line-of-sight (LOS) clump detection by 50 percent we match the total mass of the intrinsic clumps. This means that the BS did not lead to fully Gaussian profiles since the FWHM should cover most of the clump mass within this area.

\begin{figure*}
\centering
\subfloat[ \label{fig:LOS_Mcc}]
  {\includegraphics[width=0.49\linewidth]{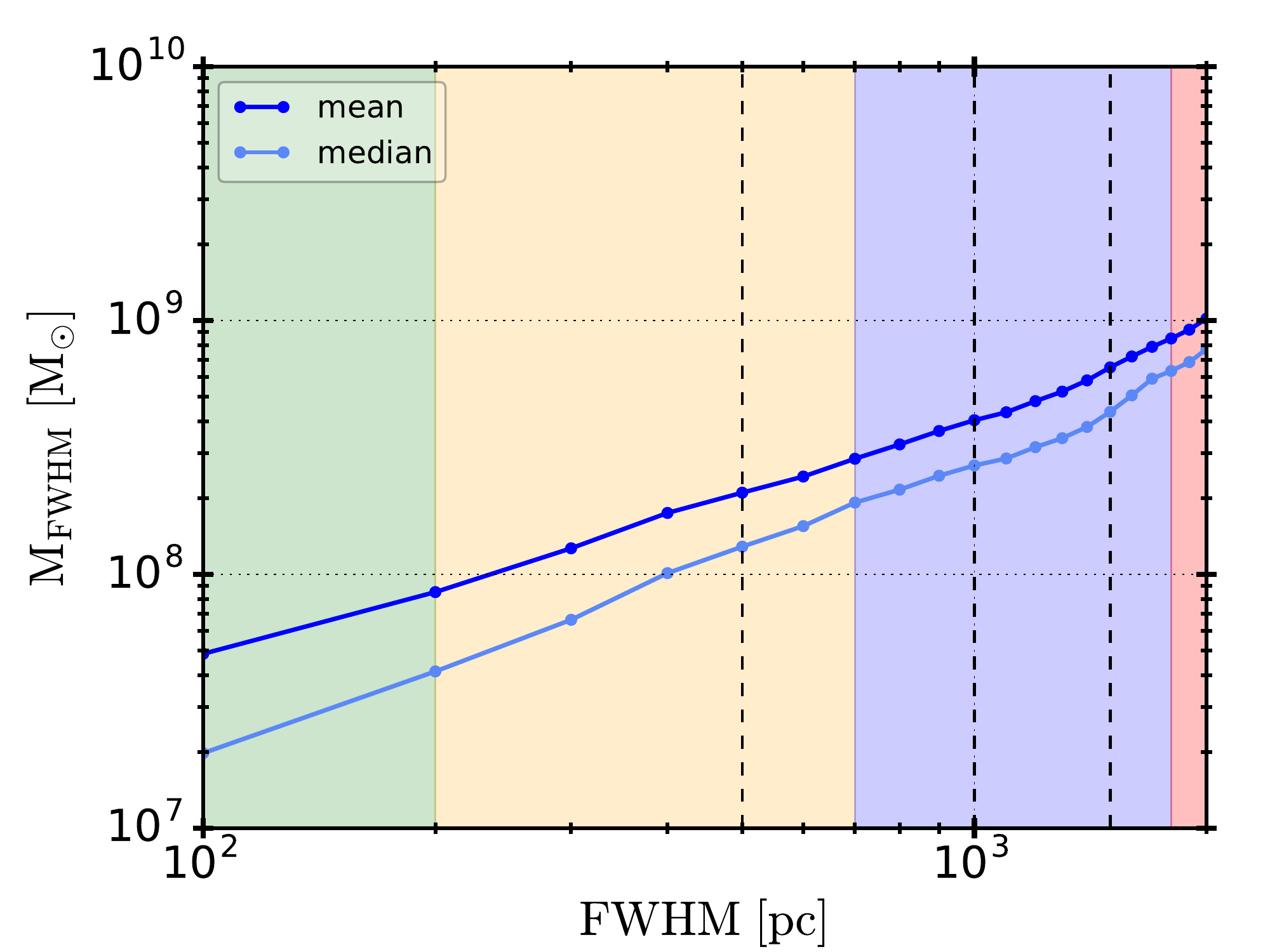}}\hfill
\subfloat[ \label{fig:LOS_Rcc}]
  {\includegraphics[width=0.49\linewidth]{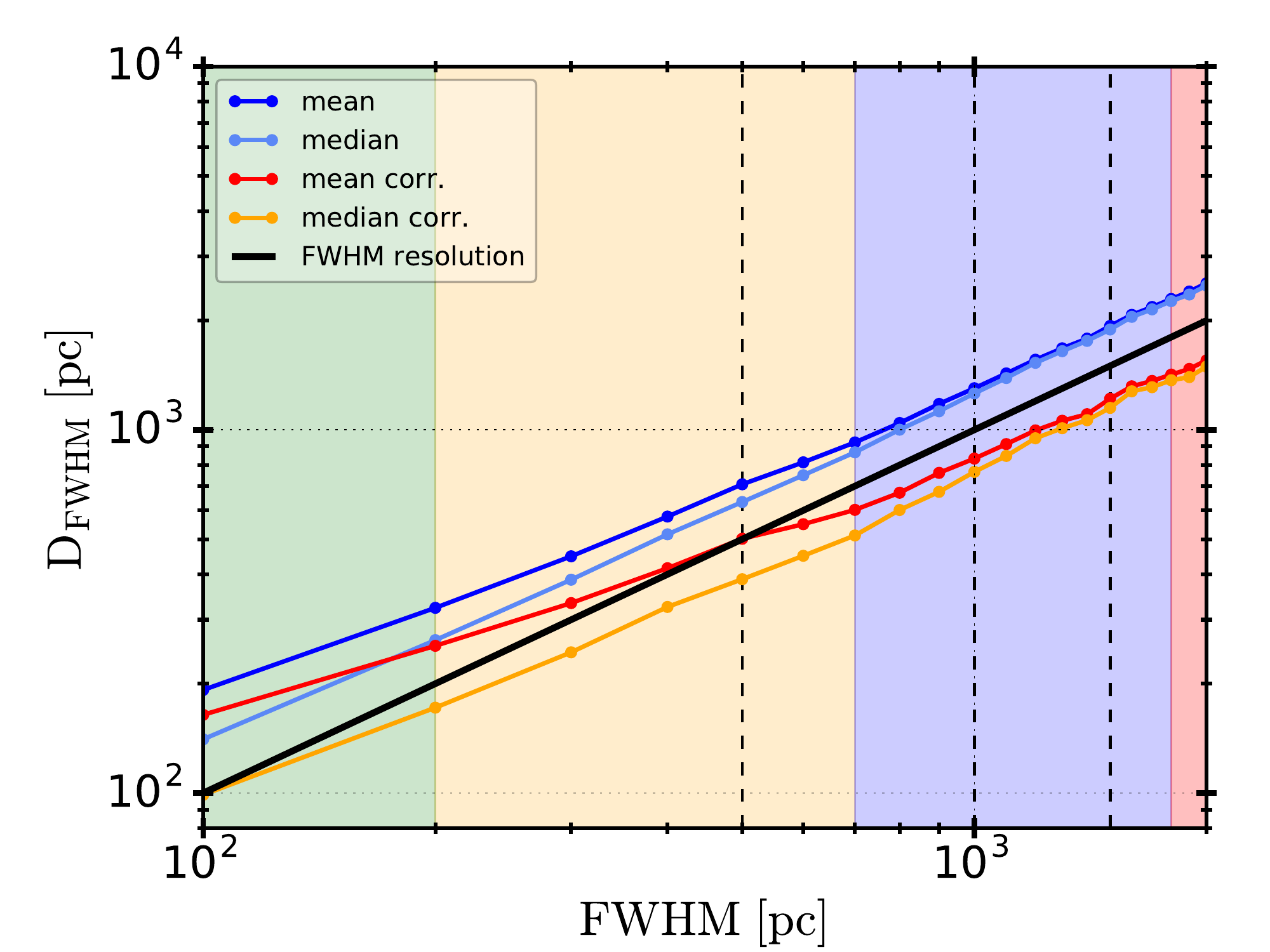}}

\caption{(a) The mean and the median of the mass of detected clumps and CCs is constantly rising with increasing FWHM of the beam smearing. (b) The linear relationship between the average and median size to the BS of detected clumps and CCs (blue and pale blue). The red and orange line result from quadratically subtracting the FWHM of the beam. 
    The background colors refer to the identified cluster ranges (see Table \ref{tab:clump_cluster_definitions}). The vertical dashed black lines correspond to the line-of-sight maps in Figure \ref{fig:LOS_FWHM_set}.
\label{fig:LOS_statistics_set}}

\end{figure*}

\subsection{Identification of characteristic CC scales}
\label{subsec:Identification of characteristic cluster scales}
For each beam smeared map (in FWHM=100 pc steps) we create mass and size histograms, compiled to the plots in Figure \ref{fig:LOS_MccRcc_statistics_set}. Each distribution is represented by a kernel density estimation (KDE) in log scale (using the \texttt{Seaborn} library \citealp{michael_waskom_2017_883859}) to make them better readable. To adequately represent the finite data by the smoothing of the KDE we find for the $D_{\mathrm{FWHM}}$ distribution the Gaussian kernel 0.014 dex and for the $M_{\mathrm{FWHM}}$ distribution 0.0525 dex. The single timestep at 655 Myr (as shown in Figure \ref{fig:LOS_FWHM_set}) is compared with the time average for the evolved disc (between 450-655 Myr, sampled every 15 Myrs). In general, we find a continuous rise for the mass and radius with increasing BS. The time averaged radius distributions have small widths and are mainly single peaked for all measured BS. The same peaks are visible in the single time step example as well, but break up into multiple sub-peaks. The mass distributions are much broader ($\sim 1 \ \mathrm{dex}$) with different features (single, double, triple peaked) that identify the typical mass of a cluster. The peaks of the mass distributions for the single timestep example are mainly close to the time average distributions but much more prominent. Furthermore, for smaller BS of FWHM 100-300 pc we find a smoother distribution for the time average, due to stronger deviations of the identified objects between the timesteps, but are still represented by the dominant peak for the time average. The features of the mass profiles can be explained as the following. With increasing BS more and more clumps are sufficiently close together to appear as larger and single objects. These groups appear as a bump in the mass distribution which stays the same with stronger BS while the size still increases. For an isolated group of clumps this works very well, but in a galaxy the groups are surrounded by more structures which leads to a small increase in mass. Therefore, we can identify a typical mass for CCs which stays quite similar in a certain range. These scales are dominating in different BS-ranges (colored background) which we highlight for the time-average data as guide lines in Figure \ref{fig:LOS_MccRcc_statistics_set} (summarized in Table \ref{tab:clump_cluster_definitions}). In the following we briefly describe each range and transition:

\begin{figure}
\includegraphics[width=84mm]{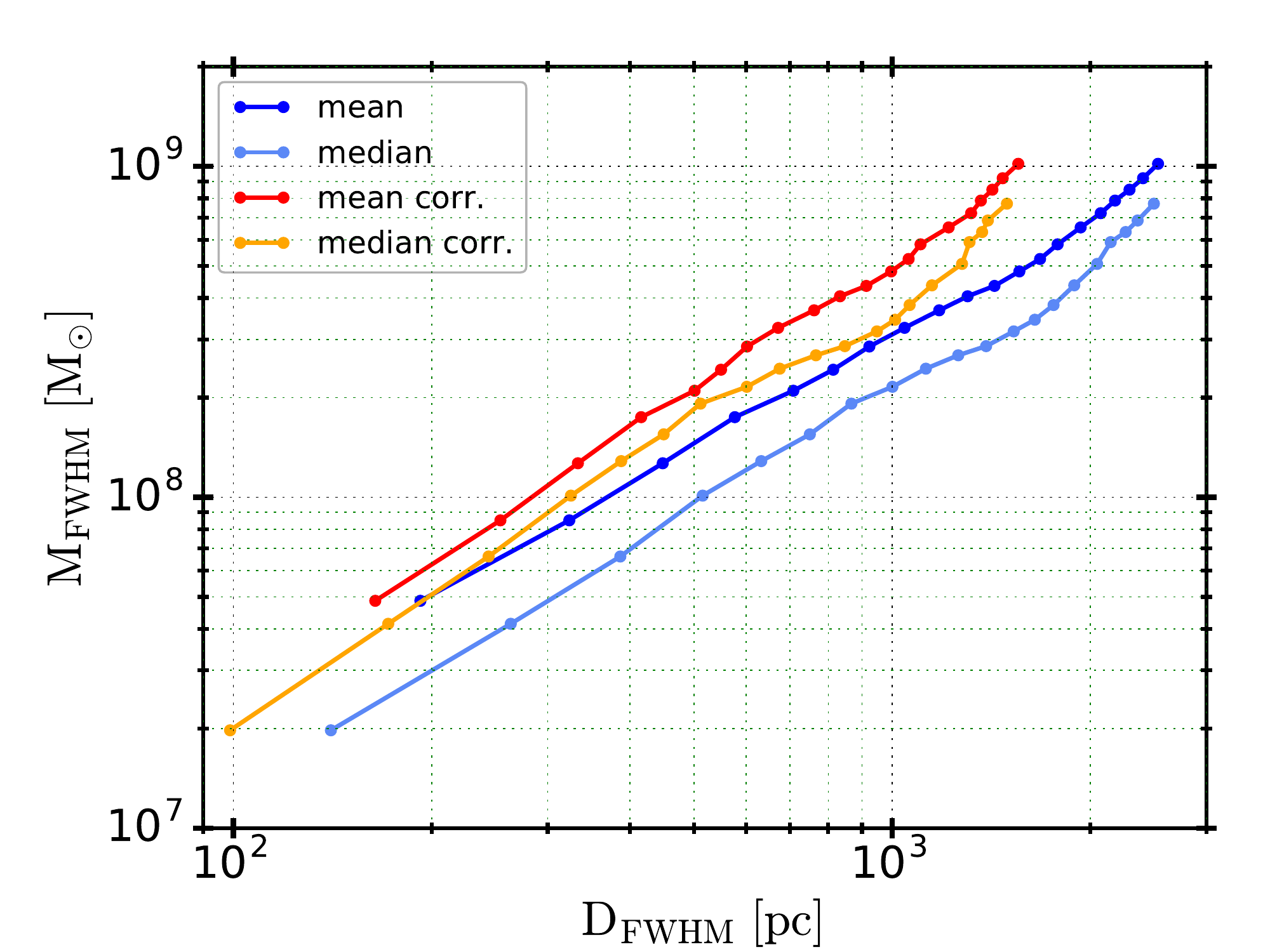} 
\caption{The relation between the mean and median mass and size of the detected clumps and CCs. The red and orange line result from quadratically subtracting the FWHM of the corresponding beam. 
\label{fig:LOS_Mcc_Rcc}} 
\end{figure}

\begin{enumerate}[label=\Alph*:]

\item Intrinsic clumps \\
At the smallest BS we find mainly single clumps. The estimated clump properties at a BS of FWHM = 100 pc  are close to the properties estimated from the intrinsic clump finder (Section \ref{subsec:Detailed clump properties}) for the average mass. We find agreement for the most frequent clumps in the distribution with $\mathrm{<M_{FWHM}> \ \simeq 10^7 \ M_{\odot}}$ and the size $\mathrm{<D_{FWHM}> \simeq 125}$ pc. The first transition from single clumps to groups is within the BS of 100 pc - 200 pc. The dominant peak is replaced by a new one, indicated by the semi-transparent guidelines. The identified fraction of objects decreases dramatically to 60 percent within this small increase of BS (Figure \ref{fig:LOS_CC_Count_statistics}).\\

\item Smallest clusters (FWHM 200-700 pc)\\
The smallest clusters dominate the distribution in this range at an average mass of $\mathrm{<M_{FWHM}>   \simeq 4.5 \times 10^7 \ M_{\odot}}$ and an average size of $\mathrm{<D_{FWHM}> \simeq 450}$ pc.The example with FWHM=500 pc in Figure \ref{fig:LOS_FWHM_set} shows how several clumps appear as single objects in dense groups on these scales and some of the remaining single and isolated clumps disappear, because the BS reduces the peak density below the detection threshold of the blob finder. With larger BS another peak begins to form, representing the next larger scale of typical clusters described in range C. The number of identified objects decrease from 60 to 18 percent. \\

\item Intermediate clusters (FWHM 700-1800 pc)\\
These CCs are mainly present on average for $\mathrm{<M_{FWHM}> \ \simeq 3.2 \times 10^8 \ M_{\odot}}$ on the average scale $\mathrm{<D_{FWHM}> \simeq 1500}$ pc and are formed  from groups of the previous range (compare in Figure \ref{fig:LOS_FWHM_set} the example with FWHM=500 pc and FWHM=1000 pc). The smallest isolated clusters are smeared to larger radii and are disappearing within this range (flatter surface density), especially at the galaxy edge. Groups with close partners combine to become larger and more mass rich entities (compare in Figure \ref{fig:LOS_FWHM_set} the example with FWHM=1000 pc and FWHM=1500 pc). Therefore, open clusters are also representing the giant BS objects. The dominant CCs of the following range D are already appearing (compare examples FWHM=1000 pc with FWHM=1500 pc in Figure \ref{fig:LOS_FWHM_set}) and in general represent the identified cluster scales in range C. A small fraction of 18-6 percent for the identified objects is left.\\

\item Largest clusters (FWHM 1800-2000 pc)\\
On these scales the dominant CCs have on average $\mathrm{<M_{FWHM}> \ \simeq 9 \times 10^8 \ M_{\odot}}$ and an average size of $\mathrm{D_{FWHM} \simeq 2000}$ pc. 
The clusters are the same and are represented on larger scales. This behaviour continuous with increasing BS (compare examples FWHM=1500 pc and FWHM=2000 pc in Figure \ref{fig:LOS_FWHM_set}). Some of the previous clusters cannot be identified because of their very low density contrast, again, mainly at larger disc radii. With 1.5 - 6 percent only a handful of objects can be identified. \\ 
\item Central clusters (FWHM 1800-2000 pc)\\
A peak at higher masses begins to evolve with $M_{\mathrm{FWHM}} \ \simeq 3.2 \times 10^9 \mathrm{M_{\odot}}$ in the center of the galaxy together with the peak described in range D, and typically consists of one or two distinguishable clusters. Since they are very dense groups, they are already present in the mass distributions from a BS of FWHM = 500 pc on, but increase in mass.
\end{enumerate}

\subsection{Average CC properties correlation with beam smearing}
\label{subsec:Average CC properties correlation with beam smearing}

The average mass and diameter profiles of the CCs are shown in Figure \ref{fig:LOS_statistics_set}. Both have a strong relation to the beam smearing and the  hierarchical grouping of the clumps is not visible. The positively skewed mass distributions are expressed by the smaller median and the higher mean values. We find
\begin{equation}
\mathrm{<M_{FWHM}> \ (M_{\odot}) \simeq (4 \times 10^5) \times FWHM \ (pc)}. 
\end{equation}

The CCs size $\mathrm{D_{FWHM}}$ distributions are very symmetrical, represented in the similar values for the mean and the median. We find for their relation to BS
\begin{equation}
\mathrm{<D_{FWHM}> \ (pc) \simeq (1.5-2) \times FWHM \ (pc)}. 
\end{equation}

Both relations together give the correlation between mass and sizes 
\begin{equation}
\mathrm{<M_{FWHM}> \ (M_{\odot}) \simeq (2-2.67) \times 10^5 \times D_{\mathrm{FWHM}} \ (pc)}
\end{equation}
If we correct for BS (quadratically subtract the corresponding BS), the relation is
\begin{equation}
\mathrm{<D_{FWHM}> \ (pc) \simeq (0.8-1) \times FWHM \ (pc)}. 
\end{equation}
and leads to
\begin{equation}
\mathrm{<M_{FWHM}> \ (M_{\odot}) \simeq (4-5) \times 10^5 \times D_{\mathrm{FWHM}} \ (pc)}. 
\end{equation}
We find strong deviations from the relationship below a BS of FWHM $<$ 100 pc and above  FWHM $>$ 2 kpc.

\newpage
\section{Summary and Discussion}
\label{sec:Summary an Discussion}

We simulated a massive gas disc with $\sim 3 \times 10^{10} \ \mathrm{M_{\odot}}$, Toomre unstable at different initial and maximum resolutions ($\sim 3-90$pc). All runs initially resolve the Toomre length by several resolution elements at all radii. Therefore, we can test the top-down hypothesis that giant clumps form directly on the Toomre scale and their sub-fragmentation on smaller scales within the clumps due to high resolution. Furthermore, we explore the relationship between the artificial pressure floor (APF) and the clump properties over time. Adding an APF is a numerical technique to prevent artificial fragmentation at maximum resolution. Finally, we investigate the clustering scales of the many clumps in the high resolution run and their properties related to the beam smearing between 0.1-2 kpc. \\

\begin{itemize}

	\item \textbf{Indication for a bottom-up scenario:}
	The simulations reveal significant differences concerning the clump properties in the fragmented disc. The runs considered for various maximum resolutions demonstrate that the initial fragmentation process changes from a few giant clumps of $10^8 - 10^{9} \ \mathrm{M_{\odot}}$ with a diameter of 0.6- 1 kpc ($\sim$ comparable to the initial Toomre length) to many and much smaller and less massive clumps with $\sim 10^7 \ \mathrm{M_{\odot}}$ with a diameter of $\sim 100$pc. A direct sub-fragmentation of the larger clumps with increasing resolution is absent. Instead, we see a transition from a direct formation of large clumps in the low resolution run to fully developing rings, which can further collapse until they fragment on much smaller scales than the initial Toomre length in the high resolution simulation. We find no strong relationship to the initial Toomre length in the evolved disc anymore. For this comparison study it is important to keep the minimum temperature the same at all times. It represents a minimum micro-turbulent pressure floor of the ISM which cannot be resolved in our simulations. Observations imply that the galaxies evolve with a Toomre parameter $Q \simeq 1 $ over cosmic time but can lie above or below this value within the uncertainties \citep{2015ApJ...799..209W}. We focus on the case of axisymmetric instabilities that arise for $\mathrm{Q < Q_{crit}} \simeq 0.7$, but are close to $\mathrm{Q_{crit}}$. The parameter-space in close proximity of the marginally stable case has to be explored further, e.g. for higher temperatures or small scale turbulence, to proof the bottom-up scenario and exclude the direct formation of clumps as large as the observed ones.\\

	\item \textbf{Main cause for the different outcome between the runs:} In general, we find that the high resolution runs can reach very high densities before the APF sets in (for $MS: \ n_{\mathrm{H}} \geq 5 \times 10^3 \ \mathrm{cm^{-3}}$ , for $SR: \ n_{\mathrm{H}} \geq 2 \times 10^3 \ \mathrm{cm^{-3}}$), which allows for further collapse and fragmentation into clumps on much smaller scales than the initial Toomre length. For the low resolution runs the APF already acts at very low densities preventing from further collapse (for $LR:$ $n_{\mathrm{H}} \geq 53 \ \mathrm{cm^{-3}}$, for $ULR: \ n_{\mathrm{H}} \geq 13 \ \mathrm{cm^{-3}}$). Furthermore, all clumps exhibit a hot core which is lower for smaller and higher for more massive ones and ranges between $c_{\mathrm{s}} \sim 30- 250 \ \mathrm{km \ s^{-1}}$. The APF sets in all runs the minimum thickness of the clumps. The APF has a different effect between the low and high resolution runs. The initial clumps for the main simulation $MS$ show less influence by the APF and are rotationally supported, even more the higher mass mergers. Smaller clumps are still supported by rotation but tend to be more influenced by the APF and/or the lower limit of $c_{\mathrm{s}} \sim 10 \ \mathrm{km \ s^{-1}}$. The initial clumps are three times larger than the minimum resolved Jeans length and the massive mergers 6 times. The initial clumps in the low resolution runs $LR$ and $ULR$ have a Jeans mass and an average density given by the APF and are only around 1.6 times larger than the minimum Jeans length. We conclude that these runs produce artificially given clumps that are in this case coincidental at mass scales similar to observed giant clumps at $\sim$kpc spatial resolution. Rotation might also play a role for the more massive mergers, which show lesser influence by the APF. \\

\end{itemize}

\begin{itemize}
  \item \textbf{General clump properties}: 
   For the high-resolution runs the initial clumps merge within $< 50$ Myr to $\sim 10^8 \ \mathrm{M_{\odot}}$ with a diameter of $\sim 120 \ \mathrm{pc}$ and are dominating the main clump mass. For all runs, typically 50 percent of the disc mass is within high densities and 30 percent in lower densities which directly surround the clumps. The definition of \textit{high} and \textit{low} densities is changing for the different runs. For the higher resolutions ($MS$, $SR$) the clumps are well defined above the densities $n_{\mathrm{H}} \geq 100\ \mathrm{cm^{-3}}$ and are shifted for run $LR$ to $n_{\mathrm{H}} \geq 10\ \mathrm{cm^{-3}}$ and for $ULR$ to $n_{\mathrm{H}} \geq 1\ \mathrm{cm^{-3}}$. The simulations from high to lower resolution experience shifts for several clump properties. The main mass of many clumps is transferred to much less clumps with $10^9 \ \mathrm{M_{\odot}}$ and smaller clumps are suppressed in $LR$ and $ULR$. E. g. the size of the $10^8 \ \mathrm{M_{\odot}}$ clumps is 3 times larger (diameter $\sim$ 360 pc) in run $LR$ and 6 times larger (diameter $\sim$ 720 pc)  in $ULR$ compared to the high resolution simulation $MS$. Also here, the clumps are closely surrounded by even lower densities with 30 percent of the disc mass. The properties we find for the runs $LR$ or $ULR$ are consistent in mass and size with clumpy galaxy simulations from various studies with a similar resolution and pressure floor.\\
    
	\item \textbf{Hierarchical scales of clump clusters:}
	At higher resolution, the vast number of clumps build larger groups from bottom-up within relatively short times $\ll 50 \ \mathrm{Myr}$ and over all the evolutionary time of 670 Myr. We identify clump clusters on several mass scales by spatially convolving the densities with a Gaussian between 100-2000 pc. The clusters appear in the mass distributions as single peaks with almost 1 dex difference for each beam smearing level if the bin sizes of the histograms are small enough. In our case, they describe the transition between the different cluster scales. Single clumps can only be identified with observations with spatial resolutions of at most 100 pc. This finding is equivalent to \cite{2017MNRAS.468.4792T}  (from $H\alpha$ mocks). The smallest CCs appear in dense groups on scales with a diameter of $\mathrm{D_{FWHM} \simeq 450}$ pc and a mass of $\mathrm{M_{FWHM} \simeq 4.5 \times 10^7 \ M_{\odot}}$. Furthermore, we find dense groups on the scale,  $\mathrm{D_{FWHM} \simeq 1.5}$ kpc with a mass of $\mathrm{M_{FWHM} \simeq 3.2 \times 10^8 \ M_{\odot}}$. On the next scale we identify more and more open clusters, which do not necessarily have a strong gravitational connection. They have on average $\mathrm{D_{FWHM} \simeq 2.5}$ kpc and a mass of $\mathrm{M_{FWHM} \simeq 9 \times 10^8 \ M_{\odot}}$. 
   The strong relationship to the beam smearing can be described on average by $\mathrm{<M_{FWHM}> \ (M_{\odot}) \simeq (4 \times 10^5) \times FWHM \ (pc)}. $ and $\mathrm{<D_{FWHM}> \ (pc) \simeq (0.8-1) \times FWHM \ (pc)}$ (resolution quadratically subtracted). The masses and sizes of the identified objects depend on the definition. Here, the convolution does not lead to perfect Gaussian profiles and the full mass for the FWHM is not captured. By increasing the measured sizes by 50 percent, we fully represent the total mass of the high density clumps (50 percent of the disc mass). In \cite{2017MNRAS.464..491F} the clump sizes are larger than the kernel of the convolved beam. In our case the sizes are larger on average for small beam smearing and with increasing convolution kernel smaller (resolution quadratically subtracted).
 It is difficult to compare the exact measured properties with observations. E.g. the high densities we take into account are shielded in H$\alpha$ observations. By considering a cut of the peak densities in the simulations leads to a larger FWHM of the clumps and consequently to a larger mass. Therefore, we refer here mainly to the similar trend of the clump clustering of the stars with observational resolution in the observations of \cite{2018NatAs...2...76C}. \\
With larger BS more and more massive clumps appear at the center of the galaxy while the isolated clusters at larger radii disappear in the environment. The clusters at FWHM=500 pc show similarities in mass and size to the low resolution runs ($LR$, $ULR$), but are much more numerous for the high resolution. Therefore, high and low resolution simulations are also in this case not directly comparable.\\

	\item \textbf{Final remarks and outlook:}
	For the presented bottom-up scenario, simulations with different galaxy properties have to be performed to explore the relationship to the clump clusters better and the convergence of the clump properties with resolution needs to be explored further. Overall, it is very useful to provide the effective resolution, since it is the minimum Jeans length that is resolved in a simulation and gives already information about the smallest possible structures. The maximum resolution alone is not sufficient enough to quantify how well clumps are resolved. \\
	The ring formation is predicted by the Toomre instability theory for axisymmetric conditions. In a disc with initially more local variations (as e.g. expected in a galaxy that is fed from outside) in density or velocity the rings may not fully develop, but we think that this is not affecting the main results in this study, namely that the resolution and artificial pressure floor are relevant to reach higher densities and allow for many structures on the sub-Toomre scale. Moreover, it should be further explored if reducing the APF with the cost of artificial fragmentation is more helpful in certain circumstances than producing inflated clumps. \\
	 The simulated model resembles galaxies, where the gas dominates the hydrodynamics and local gravity over the stars. Therefore, the lack of a stellar disc negligibly affects the fragmentation process we study if it is represented in the effective background potential as in our case. Nevertheless, the components of the galactic disc and a live dark matter halo can influence the dynamical friction and hence increase the migration of clumps to the centre where they can contribute to a stellar bulge component. The observed high-z galaxies typically have a bulge component, while bulge-less cases also exist. This causes a deeper potential well in the centre leading to additional rotation, which would have a stabilizing effect together with less fragmentation for the innermost fraction of the simulated galaxy and therefore, does not change the main findings. A follow-up paper will investigate the changes caused by star formation and stellar feedback. Interestingly, the isolated galaxy simulations in \cite{2012MNRAS.420.3490C} with an initial stellar disc component, stellar feedback and in \cite{2015arXiv150307660B} with added radiation pressure show visually similarities to our simulations.\\
	Furthermore, sophisticated comparisons with observations are necessary e.g. by using radiative transfer and by taking into account the sensitivity limit of the instruments \citep{2018NatAs...2...76C, 2017MNRAS.464..491F, 2017MNRAS.468.4792T}. The next generation of large telescopes like the Extremely-Large-Telescope (ELT) will have better resolutions than 100 pc and will hence allow to identify individual clumps and to proof the concept of hierarchical clustering. A part of the hierarchy can be investigated already by a spatial resolution better than 1 kpc.	
		
\end{itemize}

\section*{Acknowledgements} 
We thank the referee Fr\'ed\'eric Bournaud for helpful comments that improved the manuscript. We are grateful to Alessandro Ballone, Matias Bla\~na, Katharina Fierlinger, Guang-Xing Li, Go Ogiya, Michael Opitsch, for fruitful discussions. The computer simulations have been performed on the HPC system HYDRA at the Max Planck Computing \& Data Facility (MPCDF) center in Garching. MS acknowledges support by the Deutsche Forschungsgemeinschaft through grant no. BU 842/25-1. This work was supported by a grant from the Max-Planck Institute for Extraterrestrial Physics, by the Excellence-Cluster Universe, and by the German-Israeli Foundation through grant no. GIF 1341-303./2016.

\bibliographystyle{mn2e}
\bibliography{HierarchicalScales_Behrendt_revised}

\begin{thebibliography}{44}
\expandafter\ifx\csname natexlab\endcsname\relax\def\natexlab#1{#1}\fi

\bibitem[{{Agertz} {et~al}\mbox{.}(2009){Agertz}, {Lake}, {Teyssier}, {Moore},
  {Mayer}, \& {Romeo}}]{2009MNRAS.392..294A}
{Agertz} O., {Lake} G., {Teyssier} R., {Moore} B., {Mayer} L., {Romeo} A.~B.,
  2009, MNRAS, 392, 294

\bibitem[{{Behrendt}, {Burkert} \& {Schartmann}(2015){Behrendt}, {Burkert}, \&
  {Schartmann}}]{2015MNRAS.448.1007B}
{Behrendt} M., {Burkert} A., {Schartmann} M., 2015, MNRAS, 448, 1007

\bibitem[{{Behrendt}, {Burkert} \& {Schartmann}(2016){Behrendt}, {Burkert}, \&
  {Schartmann}}]{2016ApJ...819L...2B}
{Behrendt} M., {Burkert} A., {Schartmann} M., 2016, ApJ, 819, L2

\bibitem[{{Benincasa} {et~al}\mbox{.}(2018){Benincasa}, {Wadsley}, {Couchman},
  {Pettit}, \& {Tasker}}]{2018arXiv180802438B}
{Benincasa} S.~M., {Wadsley} J.~W., {Couchman} H.~M.~P., {Pettit} A.~R.,
  {Tasker} E.~J., 2018, ArXiv

\bibitem[{{Binney} \& {Tremaine}(2008)}]{2008gady.book.....B}
{Binney} J., {Tremaine} S., 2008, {Galactic Dynamics: Second Edition}.
  Princeton University Press

\bibitem[{{Bleuler} \& {Teyssier}(2014)}]{2014MNRAS.445.4015B}
{Bleuler} A., {Teyssier} R., 2014, MNRAS, 445, 4015

\bibitem[{{Bleuler} {et~al}\mbox{.}(2015){Bleuler}, {Teyssier}, {Carassou}, \&
  {Martizzi}}]{2015ComAC...2....5B}
{Bleuler} A., {Teyssier} R., {Carassou} S., {Martizzi} D., 2015, Computational
  Astrophysics and Cosmology, 2, 5

\bibitem[{Bournaud(2016)}]{Bournaud2016}
Bournaud F., 2016, Bulge Growth Through Disc Instabilities in High-Redshift
  Galaxies, Laurikainen E., Peletier R., Gadotti D., eds., Springer
  International Publishing, Cham, pp. 355--390

\bibitem[{{Bournaud}(2016)}]{2015arXiv150307660B}
{Bournaud} F., 2016, in Astrophysics and Space Science Library, Vol. 418,
  Galactic Bulges, {Laurikainen} E., {Peletier} R., {Gadotti} D., eds., p. 355

\bibitem[{{Bournaud} {et~al}\mbox{.}(2014){Bournaud}, {Perret}, {Renaud},
  {Dekel}, {Elmegreen}, {Elmegreen}, {Teyssier}, {Amram}, {Daddi}, {Duc},
  {Elbaz}, {Epinat}, {Gabor}, {Juneau}, {Kraljic}, \& {Le
  Floch'}}]{2014ApJ...780...57B}
{Bournaud} F. {et~al.}, 2014, ApJ, 780, 57

\bibitem[{{Burkert}(1995)}]{Burkert:1995jr}
{Burkert} A., 1995, ApJ, 447, L25

\bibitem[{{Cava} {et~al}\mbox{.}(2018){Cava}, {Schaerer}, {Richard},
  {P{\'e}rez-Gonz{\'a}lez}, {Dessauges-Zavadsky}, {Mayer}, \&
  {Tamburello}}]{2018NatAs...2...76C}
{Cava} A., {Schaerer} D., {Richard} J., {P{\'e}rez-Gonz{\'a}lez} P.~G.,
  {Dessauges-Zavadsky} M., {Mayer} L., {Tamburello} V., 2018, Nature Astronomy,
  2, 76

\bibitem[{{Ceverino}, {Dekel} \& {Bournaud}(2010){Ceverino}, {Dekel}, \&
  {Bournaud}}]{Ceverino:2010eh}
{Ceverino} D., {Dekel} A., {Bournaud} F., 2010, MNRAS, 404, 2151

\bibitem[{{Ceverino} {et~al}\mbox{.}(2012){Ceverino}, {Dekel}, {Mandelker},
  {Bournaud}, {Burkert}, {Genzel}, \& {Primack}}]{2012MNRAS.420.3490C}
{Ceverino} D., {Dekel} A., {Mandelker} N., {Bournaud} F., {Burkert} A.,
  {Genzel} R., {Primack} J., 2012, MNRAS, 420, 3490

\bibitem[{{Daddi} {et~al}\mbox{.}(2010){Daddi}, {Bournaud}, {Walter},
  {Dannerbauer}, {Carilli}, {Dickinson}, {Elbaz}, {Morrison}, {Riechers},
  {Onodera}, {Salmi}, {Krips}, \& {Stern}}]{2010ApJ...713..686D}
{Daddi} E. {et~al.}, 2010, ApJ, 713, 686

\bibitem[{{Dekel}, {Sari} \& {Ceverino}(2009{\natexlab{a}}){Dekel}, {Sari}, \&
  {Ceverino}}]{2009ApJ...703..785D}
{Dekel} A., {Sari} R., {Ceverino} D., 2009{\natexlab{a}}, ApJ, 703, 785

\bibitem[{{Dekel}, {Sari} \& {Ceverino}(2009{\natexlab{b}}){Dekel}, {Sari}, \&
  {Ceverino}}]{Dekel:2009bn}
{Dekel} A., {Sari} R., {Ceverino} D., 2009{\natexlab{b}}, ApJ, 703, 785

\bibitem[{{Elmegreen} \& {Elmegreen}(2005)}]{2005ApJ...627..632E}
{Elmegreen} B.~G., {Elmegreen} D.~M., 2005, ApJ, 627, 632

\bibitem[{{Elmegreen} {et~al}\mbox{.}(2005){Elmegreen}, {Elmegreen}, {Rubin},
  \& {Schaffer}}]{2005ApJ...631...85E}
{Elmegreen} D.~M., {Elmegreen} B.~G., {Rubin} D.~S., {Schaffer} M.~A., 2005,
  ApJ, 631, 85

\bibitem[{{Fisher} {et~al}\mbox{.}(2017){Fisher}, {Glazebrook}, {Damjanov},
  {Abraham}, {Obreschkow}, {Wisnioski}, {Bassett}, {Green}, \&
  {McGregor}}]{2017MNRAS.464..491F}
{Fisher} D.~B. {et~al.}, 2017, MNRAS, 464, 491

\bibitem[{{F{\"o}rster Schreiber} {et~al}\mbox{.}(2009){F{\"o}rster Schreiber},
  {Genzel}, {Bouch{\'e}}, {Cresci}, {Davies}, {Buschkamp}, {Shapiro},
  {Tacconi}, {Hicks}, {Genel}, {Shapley}, {Erb}, {Steidel}, {Lutz},
  {Eisenhauer}, {Gillessen}, {Sternberg}, {Renzini}, {Cimatti}, {Daddi},
  {Kurk}, {Lilly}, {Kong}, {Lehnert}, {Nesvadba}, {Verma}, {McCracken},
  {Arimoto}, {Mignoli}, \& {Onodera}}]{2009ApJ...706.1364F}
{F{\"o}rster Schreiber} N.~M. {et~al.}, 2009, ApJ, 706, 1364

\bibitem[{{F{\"o}rster Schreiber} {et~al}\mbox{.}(2011){F{\"o}rster Schreiber},
  {Shapley}, {Genzel}, {Bouch{\'e}}, {Cresci}, {Davies}, {Erb}, {Genel},
  {Lutz}, {Newman}, {Shapiro}, {Steidel}, {Sternberg}, \&
  {Tacconi}}]{2011ApJ...739...45F}
{F{\"o}rster Schreiber} N.~M. {et~al.}, 2011, ApJ, 739, 45

\bibitem[{{Genzel} {et~al}\mbox{.}(2008){Genzel}, {Burkert}, {Bouch{\'e}},
  {Cresci}, {F{\"o}rster Schreiber}, {Shapley}, {Shapiro}, {Tacconi},
  {Buschkamp}, {Cimatti}, {Daddi}, {Davies}, {Eisenhauer}, {Erb}, {Genel},
  {Gerhard}, {Hicks}, {Lutz}, {Naab}, {Ott}, {Rabien}, {Renzini}, {Steidel},
  {Sternberg}, \& {Lilly}}]{2008ApJ...687...59G}
{Genzel} R. {et~al.}, 2008, ApJ, 687, 59

\bibitem[{{Genzel} {et~al}\mbox{.}(2011){Genzel}, {Newman}, {Jones},
  {F{\"o}rster Schreiber}, {Shapiro}, {Genel}, {Lilly}, {Renzini}, {Tacconi},
  {Bouch{\'e}}, {Burkert}, {Cresci}, {Buschkamp}, {Carollo}, {Ceverino},
  {Davies}, {Dekel}, {Eisenhauer}, {Hicks}, {Kurk}, {Lutz}, {Mancini}, {Naab},
  {Peng}, {Sternberg}, {Vergani}, \& {Zamorani}}]{2011ApJ...733..101G}
{Genzel} R. {et~al.}, 2011, ApJ, 733, 101

\bibitem[{{Guo} {et~al}\mbox{.}(2018){Guo}, {Rafelski}, {Bell}, {Conselice},
  {Dekel}, {Faber}, {Giavalisco}, {Koekemoer}, {Koo}, {Lu}, {Mandelker},
  {Primack}, {Ceverino}, {de Mello}, {Ferguson}, {Hathi}, {Kocevski}, {Lucas},
  {P{\'e}rez-Gonz{\'a}lez}, {Ravindranath}, {Soto}, {Straughn}, \&
  {Wang}}]{2018ApJ...853..108G}
{Guo} Y. {et~al.}, 2018, ApJ, 853, 108

\bibitem[{{Harten}, {Lax} \& {van Leer}(1983){Harten}, {Lax}, \& {van
  Leer}}]{Harten_83}
{Harten} A., {Lax} P.~D., {van Leer} B., 1983, SIAM REVIEW, 25, 35

\bibitem[{{Hopkins} {et~al}\mbox{.}(2017){Hopkins}, {Wetzel}, {Keres},
  {Faucher-Giguere}, {Quataert}, {Boylan-Kolchin}, {Murray}, {Hayward},
  {Garrison-Kimmel}, {Hummels}, {Feldmann}, {Torrey}, {Ma}, {Angles-Alcazar},
  {Su}, {Orr}, {Schmitz}, {Escala}, {Sanderson}, {Grudic}, {Hafen}, {Kim},
  {Fitts}, {Bullock}, {Wheeler}, {Chan}, {Elbert}, \&
  {Narananan}}]{2017arXiv170206148H}
{Hopkins} P.~F. {et~al.}, 2017, ArXiv

\bibitem[{{Kim} {et~al}\mbox{.}(2016){Kim}, {Agertz}, {Teyssier}, {Butler},
  {Ceverino}, {Choi}, {Feldmann}, {Keller}, {Lupi}, {Quinn}, {Revaz},
  {Wallace}, {Gnedin}, {Leitner}, {Shen}, {Smith}, {Thompson}, {Turk}, {Abel},
  {Arraki}, {Benincasa}, {Chakrabarti}, {DeGraf}, {Dekel}, {Goldbaum},
  {Hopkins}, {Hummels}, {Klypin}, {Li}, {Madau}, {Mandelker}, {Mayer},
  {Nagamine}, {Nickerson}, {O'Shea}, {Primack}, {Roca-F{\`a}brega}, {Semenov},
  {Shimizu}, {Simpson}, {Todoroki}, {Wadsley}, {Wise}, \& {AGORA
  Collaboration}}]{2016ApJ...833..202K}
{Kim} J.-h. {et~al.}, 2016, ApJ, 833, 202

\bibitem[{{Oklop{\v c}i{\'c}} {et~al}\mbox{.}(2017){Oklop{\v c}i{\'c}},
  {Hopkins}, {Feldmann}, {Kere{\v s}}, {Faucher-Gigu{\`e}re}, \&
  {Murray}}]{2017MNRAS.465..952O}
{Oklop{\v c}i{\'c}} A., {Hopkins} P.~F., {Feldmann} R., {Kere{\v s}} D.,
  {Faucher-Gigu{\`e}re} C.-A., {Murray} N., 2017, MNRAS, 465, 952

\bibitem[{{Robertson} \& {Kravtsov}(2008)}]{2008ApJ...680.1083R}
{Robertson} B.~E., {Kravtsov} A.~V., 2008, ApJ, 680, 1083

\bibitem[{{Roe}(1986)}]{1986AnRFM..18..337R}
{Roe} P.~L., 1986, Annual Review of Fluid Mechanics, 18, 337

\bibitem[{{Swinbank} {et~al}\mbox{.}(2012){Swinbank}, {Smail}, {Sobral},
  {Theuns}, {Best}, \& {Geach}}]{2012ApJ...760..130S}
{Swinbank} A.~M., {Smail} I., {Sobral} D., {Theuns} T., {Best} P.~N., {Geach}
  J.~E., 2012, ApJ, 760, 130

\bibitem[{{Tacconi} {et~al}\mbox{.}(2013){Tacconi}, {Neri}, {Genzel}, {Combes},
  {Bolatto}, {Cooper}, {Wuyts}, {Bournaud}, {Burkert}, {Comerford}, {Cox},
  {Davis}, {F{\"o}rster Schreiber}, {Garc{\'{\i}}a-Burillo}, {Gracia-Carpio},
  {Lutz}, {Naab}, {Newman}, {Omont}, {Saintonge}, {Shapiro Griffin}, {Shapley},
  {Sternberg}, \& {Weiner}}]{2013ApJ...768...74T}
{Tacconi} L.~J. {et~al.}, 2013, ApJ, 768, 74

\bibitem[{{Tamburello} {et~al}\mbox{.}(2017){Tamburello}, {Rahmati}, {Mayer},
  {Cava}, {Dessauges-Zavadsky}, \& {Schaerer}}]{2017MNRAS.468.4792T}
{Tamburello} V., {Rahmati} A., {Mayer} L., {Cava} A., {Dessauges-Zavadsky} M.,
  {Schaerer} D., 2017, MNRAS, 468, 4792

\bibitem[{{Teyssier}(2002)}]{2002A&A...385..337T}
{Teyssier} R., 2002, A\&A, 385, 337

\bibitem[{{Teyssier}, {Chapon} \& {Bournaud}(2010){Teyssier}, {Chapon}, \&
  {Bournaud}}]{2010ApJ...720L.149T}
{Teyssier} R., {Chapon} D., {Bournaud} F., 2010, ApJ, 720, L149

\bibitem[{{Truelove} {et~al}\mbox{.}(1997){Truelove}, {Klein}, {McKee},
  {Holliman}, {Howell}, \& {Greenough}}]{1997ApJ...489L.179T}
{Truelove} J.~K., {Klein} R.~I., {McKee} C.~F., {Holliman}, II J.~H., {Howell}
  L.~H., {Greenough} J.~A., 1997, ApJ, 489, L179

\bibitem[{van~der Walt {et~al}\mbox{.}(2014)van~der Walt, {S}ch\"onberger,
  {Nunez-Iglesias}, {B}oulogne, {W}arner, {Y}ager, {G}ouillart, {Y}u, \& the
  scikit-image contributors}]{scikit-image}
van~der Walt S. {et~al.}, 2014, PeerJ, 2, e453

\bibitem[{{Wang} {et~al}\mbox{.}(2010){Wang}, {Klessen}, {Dullemond}, {van den
  Bosch}, \& {Fuchs}}]{Wang:2010wr}
{Wang} H.-H., {Klessen} R.~S., {Dullemond} C.~P., {van den Bosch} F.~C.,
  {Fuchs} B., 2010, MNRAS, 407, 705

\bibitem[{Waskom {et~al}\mbox{.}(2017)Waskom, Botvinnik, O'Kane, Hobson,
  Lukauskas, Gemperline, Augspurger, Halchenko, Cole, Warmenhoven, de~Ruiter,
  Pye, Hoyer, Vanderplas, Villalba, Kunter, Quintero, Bachant, Martin, Meyer,
  Miles, Ram, Yarkoni, Williams, Evans, Fitzgerald, Brian, Fonnesbeck, Lee, \&
  Qalieh}]{michael_waskom_2017_883859}
Waskom M. {et~al.}, 2017, mwaskom/seaborn: v0.8.1 (september 2017)

\bibitem[{{Wisnioski} {et~al}\mbox{.}(2015){Wisnioski}, {F{\"o}rster
  Schreiber}, {Wuyts}, {Wuyts}, {Bandara}, {Wilman}, {Genzel}, {Bender},
  {Davies}, {Fossati}, {Lang}, {Mendel}, {Beifiori}, {Brammer}, {Chan},
  {Fabricius}, {Fudamoto}, {Kulkarni}, {Kurk}, {Lutz}, {Nelson}, {Momcheva},
  {Rosario}, {Saglia}, {Seitz}, {Tacconi}, \& {van
  Dokkum}}]{2015ApJ...799..209W}
{Wisnioski} E. {et~al.}, 2015, ApJ, 799, 209

\bibitem[{{Wisnioski} {et~al}\mbox{.}(2012){Wisnioski}, {Glazebrook}, {Blake},
  {Poole}, {Green}, {Wyder}, \& {Martin}}]{2012MNRAS.422.3339W}
{Wisnioski} E., {Glazebrook} K., {Blake} C., {Poole} G.~B., {Green} A.~W.,
  {Wyder} T., {Martin} C., 2012, MNRAS, 422, 3339

\bibitem[{{Wuyts} {et~al}\mbox{.}(2013){Wuyts}, {F{\"o}rster Schreiber},
  {Nelson}, {van Dokkum}, {Brammer}, {Chang}, {Faber}, {Ferguson}, {Franx},
  {Fumagalli}, {Genzel}, {Grogin}, {Kocevski}, {Koekemoer}, {Lundgren}, {Lutz},
  {McGrath}, {Momcheva}, {Rosario}, {Skelton}, {Tacconi}, {van der Wel}, \&
  {Whitaker}}]{2013ApJ...779..135W}
{Wuyts} S. {et~al.}, 2013, ApJ, 779, 135

\bibitem[{{Zanella} {et~al}\mbox{.}(2015){Zanella}, {Daddi}, {Le Floc'h},
  {Bournaud}, {Gobat}, {Valentino}, {Strazzullo}, {Cibinel}, {Onodera},
  {Perret}, {Renaud}, \& {Vignali}}]{2015Natur.521...54Z}
{Zanella} A. {et~al.}, 2015, Nature Astronomy, 521, 54

\end{thebibliography}

\label{lastpage}

\end{document}